\documentclass[superscriptaddress, amsmath,amssymb, aps, twocolumn ]{revtex4-2}

\usepackage{graphicx}
\usepackage{dcolumn}
\usepackage{bm}
\usepackage{wasysym}
\usepackage{hyperref}
\usepackage[mathlines]{lineno}
\usepackage{xcolor}

\usepackage{gensymb}
\usepackage{enumitem}
\usepackage{multirow}
\usepackage{booktabs}
\usepackage{rotating}
\usepackage{array}
\newcolumntype{P}[1]{>{\centering\arraybackslash}p{#1}}
\usepackage{amssymb}
\usepackage{adjustbox}
\usepackage{orcidlink}
\usepackage{pifont}

\newcommand*{\defeq}{\mathrel{\vcenter{\baselineskip0.5ex \lineskiplimit0pt
                     \hbox{\scriptsize.}\hbox{\scriptsize.}}}
                     =}

\newcommand{\iu}{\mathrm{i}\mkern1mu}
\newcommand{\du}{\mathrm{d}}

\renewcommand{\checkmark}{\ding{51}}
\newcommand{\xmark}{\ding{55}}

\begin{document}

\preprint{APS/123-QED}

\title{Gravitational-Wave Data Analysis with High-Precision Numerical Relativity Simulations of Boson Star Mergers} 

\author{Tamara Evstafyeva \orcidlink{0000-0002-2818-701X}}
\email{te307@cam.ac.uk}
\affiliation{DAMTP, Centre for Mathematical Sciences, University of Cambridge, Wilberforce Road, Cambridge CB3 0WA, UK}

\author{Ulrich Sperhake 
\orcidlink{0000-0002-3134-7088}}
\email{U.Sperhake@damtp.cam.ac.uk}
\affiliation{DAMTP, Centre for Mathematical Sciences, University of Cambridge, Wilberforce Road, Cambridge CB3 0WA, UK}
\affiliation{Department of Physics and Astronomy, Johns Hopkins University, 3400 North Charles Street, Baltimore, Maryland 21218, USA}
\affiliation{{TAPIR 350-17, Caltech, 1200 E. California Boulevard, Pasadena, California 91125, USA}}

\author{Isobel M.~Romero-Shaw
\orcidlink{0000-0002-4181-8090}}
\email{ir346@cam.ac.uk}
\affiliation{DAMTP, Centre for Mathematical Sciences, University of Cambridge, Wilberforce Road, Cambridge CB3 0WA, UK}
\affiliation{Kavli Institute for Cosmology Cambridge, Madingley Road Cambridge CB3 0HA, United Kingdom}

\author{Michalis Agathos
\orcidlink{0000-0002-9072-1121}}
\email{m.agathos@qmul.ac.uk}
\affiliation{School of Mathematical Sciences, Queen Mary University of London, Mile End Road, London, E1 4NS, UK}
\affiliation{DAMTP, Centre for Mathematical Sciences, University of Cambridge, Wilberforce Road, Cambridge CB3 0WA, UK}
\affiliation{Kavli Institute for Cosmology Cambridge, Madingley Road Cambridge CB3 0HA, United Kingdom}

\date{\today}

\begin{abstract}
Gravitational-wave signals detected to date are commonly interpreted
under the paradigm that they originate from pairs of black holes
or neutron stars. Here, we explore the alternative scenario of
boson-star signals being present in the data stream. We perform
accurate and long ($\sim 20$ orbits) numerical simulations of boson-star
binaries and inject the resulting strain into LIGO noise. Our
Bayesian inference reveals that some boson-star signals 
are degenerate with current approximants, albeit with biased
parameters, while others exhibit smoking-gun signatures leaving
behind conspicuous residuals.
\end{abstract}

\maketitle

{\textit{\textbf{Introduction}}---}
Following the Nobel-Prize winning detection of
GW150914~\cite{Abbott:2016blz}, about 100 further gravitational-wave
(GW) events have been confidently detected by the Laser Interferometer
Gravitational-Wave Observatory (LIGO) and Virgo
\cite{AdvancedDetectorsLVK2018,KAGRA:2021vkt,Nitz:2021zwj}.  This
ever-increasing ensemble of GW events provides ample opportunities
to explore some of the deepest mysteries of the cosmos and test the
nature of compact objects: black holes (BHs), neutron stars (NSs)
and exotic compact objects (ECOs)~\cite{Cardoso:2019rvt}.  A research
program targeting such tests, however, faces several crucial
challenges: (i) Given a candidate class of ECOs, can we generate
gravitational waveforms of sufficient longevity and accuracy?  (ii)
Supposing ECO coalescences occur in the Universe, can we detect
them with our current search pipelines?  (iii) If yes, can we
distinguish them from traditional binary BH (BBH) or NS events?
(iv) Can we generate comprehensive GW template banks suitable as
alternatives to BH and NS approximants for parameter estimation
(PE)?  The main goal of this Letter is to explore the answers to
these questions for the case of scalar-field boson stars (BSs),
which are self-gravitating equilibrium solutions to the
Einstein-Klein-Gordon
equations~\cite{Kaup:1968zz,Ruffini:1969qy,Liebling:2012fv}.

BSs have attracted a great deal of interest over the years.  BSs
may account for part of the enigmatic dark-matter content of the
Universe~\cite{Sharma:2008sc, Marsh:2015xka}.  BS solutions cover
a wide range of compactness, thus forming a theoretical laboratory
for exploring extreme-gravity effects beyond BHs such as the geometry
of BS shadows~\cite{Vincent:2015xta,Rosa:2022tfv} or light
rings~\cite{Grandclement:2016eng, Cunha:2017qtt, Cunha:2022gde}.
BSs are an ideal proxy for a wide class of compact binaries
systematically deviating from BH systems, e.g.~through finite tidal
deformability~\cite{Sennett:2017etc}.  Combined with their high
amenability to numerical modeling, this makes BS binaries particularly
suitable for high-precision gravitational-wave (GW) source modeling
and template construction beyond BHs. Numerical studies of BS
binaries have made tremendous progress in the last decade, focusing
in particular on the merger
remnants~\cite{Palenzuela:2017kcg,Bezares:2022obu,Croft:2022bxq,
Siemonsen:2023age, Siemonsen:2023hko,Siemonsen:2024snb}, and analytic
approximations have been employed for performing PE of BS binary
signals~\cite{Pacilio:2020jza,Vaglio:2023lrd, Guo:2019sns} (see
also Refs.~\cite{CalderonBustillo:2020fyi,CalderonBustillo:2022cja}
for searches of Proca-star head-on collisions in GW data).  Here,
we numerically compute the first high-precision $\sim 20$ orbit
inspiral-merger-ringdown (IMR) waveforms for quasicircular BS
binaries and inject them into LIGO detector noise. We perform PE
using Bayesian inference and assess the ability of present BBH and
binary NS waveform templates to recover the injected signals.  We
use natural units $c=1=\hbar$, i.e.~$G=M_{\rm Pl}^{-2}$.
\\

{\textit{\textbf{Theory}}---}
BSs in general relativity are described by the action of a complex
scalar field $\varphi$ minimally coupled to gravity,
\begin{equation} \label{eq:action}
S = \int \frac{\sqrt{-g}}{2} \left\{\frac{R}{8 \pi G}- \left[g^{\mu \nu}
  \nabla_{\mu} \bar{\varphi} \nabla_{\nu} \varphi + V(\varphi) \right]
  \right \} \du ^4x,
\end{equation}
where $V(\varphi)$ is the potential, which we choose to be of
solitonic type \cite{Lee:1991ax, Lee:1986ts}, $V_{\rm
sol}=\mu^2|\varphi|^2(1-2|\varphi|^2/\sigma_0^2)^2$ with $\sigma_0=0.2$.
This choice allows us to construct BSs over a particularly wide
range of compactness. Through appropriate re-scaling of the
variables~\cite{Evstafyeva:2023kfg}, the mass of the scalar can be
set to $\mu=1$ which, henceforth, sets the length scale of our
units.  Varying the action~\eqref{eq:action} yields the
Einstein-Klein-Gordon equations and spherically symmetric solutions
are obtained by decomposing the scalar field into amplitude $A$ and
frequency $\omega$, $\varphi(t, r) = A(r)e^{\iu( \epsilon \omega t
+ \delta \phi)}$~\cite{Helfer:2021brt}.  Here we introduce the
parameter $\epsilon = \pm 1$, determining the rotation of the scalar
field in the complex plane, and a phase offset $\delta \phi$. Our
primary BS always has $\epsilon=1,~\delta\phi=0$ and we refer to
configurations with secondary parameters $(\delta \phi = 0,~\epsilon
= 1)$, $(\delta\phi=\tfrac{\pi}{2},~\epsilon=1)$, $(\delta \phi =
\pi,~\epsilon = 1)$, $(\delta \phi = 0,~\epsilon = -1)$ as
\textit{in-phase}, \textit{dephased}, \textit{antiphase} and
\textit{anti-BS} binaries respectively.  Given the central amplitude
$A_{\rm{ctr}} = A(0)$, we obtain a BS solution via a shooting
algorithm described in Ref.~\cite{Helfer:2021brt}.  In this work,
we consider two models: a compact ``\texttt{A17}'' star with $\sqrt{G}
A_{\rm{ctr}} = 0.17$, dimensionless tidal deformability $\Lambda
\sim 10$~\cite{Sennett:2017etc} and compactness $\mathcal{C}=0.2$
containing $0.99$ of its total mass $m$ within a radius $r_{99}=3.97$,
and a less compact ``\texttt{A147}'' star with $\sqrt{G} A_{\rm{ctr}}
= 0.147$, $\Lambda \sim 1000$, $\mathcal{C}=0.1$ and $r_{99}=4.48$.
Tidal deformability characterizes the stars' tidal interaction
-- the higher $\Lambda$, the more susceptible the BS is to tidal
deformations and corresponding changes in the GW phase.

{\textit{\textbf{NR Simulations}}---}
Our simulations have been performed using two codes, \textsc{GRChombo}
\begin{table}[t]
\caption{
Summary of the BS  binaries evolved in the center-of-mass
frame with initial boost velocity $v_{x, \rm{ini}}$ in the
$x$ direction, impact parameter $b$ in the $y$ direction and
separation $d$ in the $x$ direction. The dephased, antiphase and
anti-BS binaries are labeled as \texttt{p090}, \texttt{p180},
\texttt{e1}. $E_{l=2}$ is the GW energy contained in the $l=2$
modes.
}
\begin{tabular}{|c | c | c | c |c | c| c|} \hline
~~~Simulation~~~&~~~$v_{x, \rm{ini}}$~~~&~~~$b/M$~~~&~~~$d/M$~~~&~~~$E_{l=2}/M$~~~\\ \hline
\texttt{A17-d12} & 0.1671 & 12.283 & 0.2243 & 0.0333 \\ \hline
\texttt{A17-d12-p180} & 0.1671 & 12.283 & 0.2243 & 0.0353 \\ \hline
\texttt{A17-d12-e1} & 0.1671 & 12.283 & 0.2243 & 0.0299 \\ \hline
\texttt{A17-d14} & 0.1533 & 14.017 & 0.2102 & 0.0346 \\ \hline
\texttt{A17-d15} & 0.14625 & 15.087 & 0.2033 & 0.0367 \\ \hline
\texttt{A17-d15-p090} & 0.14625 & 15.087 & 0.2033 & 0.0377 \\ \hline
\texttt{A17-d15-p180} & 0.14625 & 15.087 & 0.2033 & 0.0387 \\ \hline
\texttt{A17-d15-e1} & 0.14625 & 15.087 & 0.2033 & 0.0352 \\ \hline
\texttt{A147-d17} & 0.1389 & 16.639 & 0.0693 & 0.0589 \\ \hline
\texttt{A147-d19} & 0.1256 & 19.412 & 0.1109 & 0.0691 \\ \hline
\end{tabular}
\label{tab:waveforms}
\end{table}
\cite{Clough:2015sqa, Radia:2021smk, Andrade:2021rbd} and
\textsc{Lean}~\cite{Sperhake:2006cy}.  Both codes evolve the Einstein
equations through fourth-order finite differencing of the CCZ4
formulation~\cite{Alic:2011gg}, employing the moving puncture
gauge~\cite{Campanelli:2005dd, Baker:2005vv}.  While \textsc{GRChombo}
is based on adaptive-mesh refinement provided by {\sc
Chombo}~\cite{chombo}, the \textsc{Lean} code, based on the {\sc
cactus} computational toolkit~\cite{Allen:1999}, utilizes mesh
refinement via {\sc carpet}~\cite{Schnetter:2003rb}. Apparent
horizons are computed using {\sc ahfinderdirect}~\cite{Thornburg:2003sf}.
We use a computational domain of length $L=1024$ with bitant symmetry,
8 nested refinement levels with resolutions $\Delta x=1/40$ to
$1/48$ on the innermost level and extract GWs at $R_{\rm ex}\in
[140,240]$.
\begin{figure}[t]
  \includegraphics[width=0.47\textwidth]{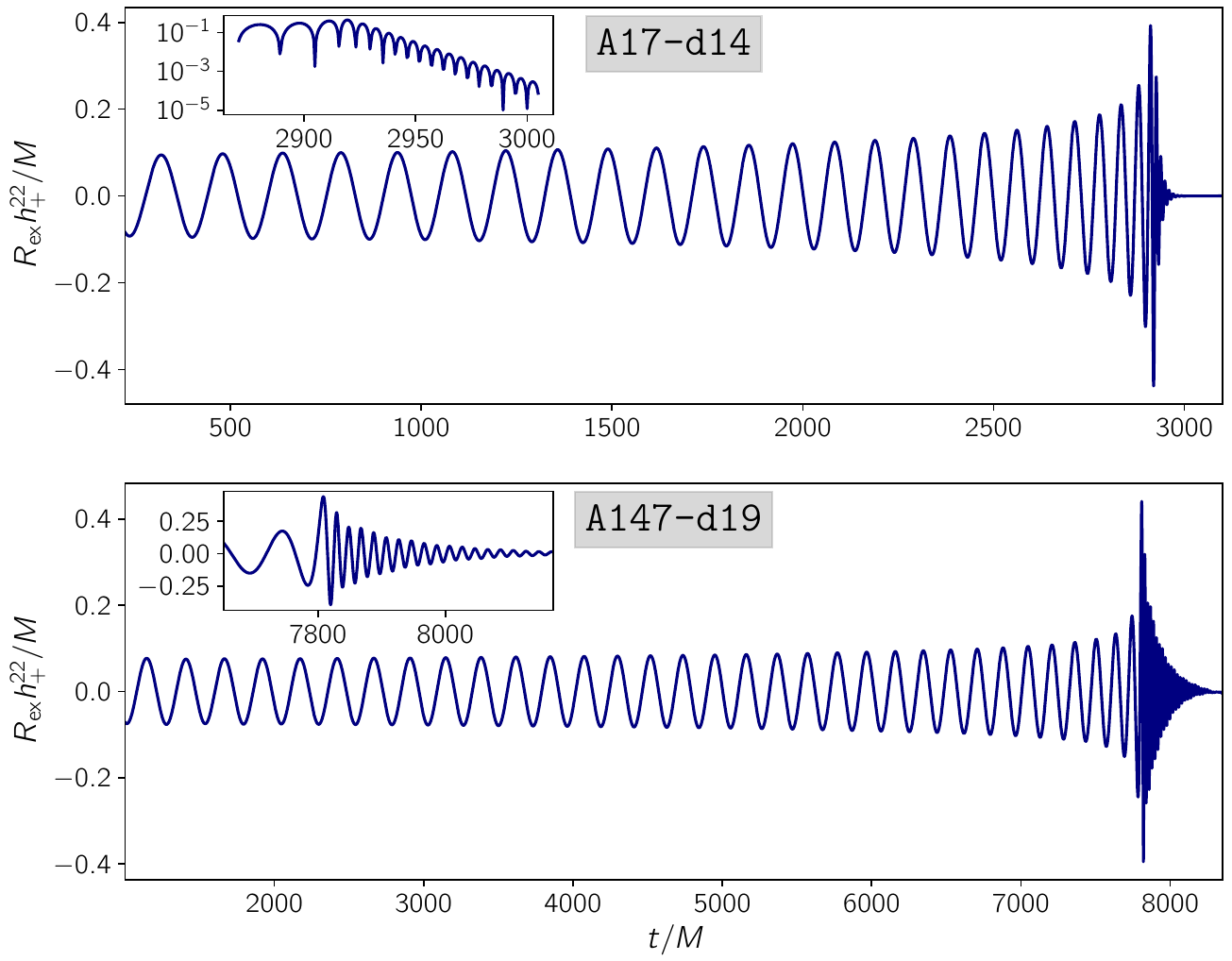}\\
  \caption{Plus polarizations of the $22$ IMR mode for the
  \texttt{A17-d14} and \texttt{A147-d19} binaries of Table~\ref{tab:waveforms}.
  }
  \label{fig:example_simulations}
\end{figure}

We focus on equal-mass nonspinning BS binaries of total mass $M=2m$,
listed in Table~\ref{tab:waveforms}, starting from initial data
constructed as in Refs.~\cite{Helfer:2021brt,Evstafyeva:2022bpr}.
By tuning the initial velocities, we reduce their orbital eccentricity,
estimated according to Eq.(17) of Ref.~\cite{Mroue:2010re}, as
$0.002-0.005$. By verifying convergence of the {\tt A17} and {\tt
A147} BS binaries and extrapolating GW signals to infinity, we
obtain an error budget of $(0.1,~4\,\%)$ for the GW phase and
amplitude for {\tt A17} binaries and $(0.2,~4\,\%)$ for {\tt A147}
binaries, similar to finite differencing production runs for BH
inspirals reported in Ref.~\cite{Hinder:2013oqa}; see the Supplemental
Material for details.

In physical terms, the most conspicuous difference between the
\texttt{A17} and \texttt{A147} binary families of Table~\ref{tab:waveforms}
is their ultimate fate. The compact \texttt{A17} coalescences
ubiquitously result in the formation of a BH remnant with dimensionless
final spin $a_{\rm fin} \sim 0.7$, similar to the end product of
nonspinning equal-mass BH mergers. By eye, the resulting GW form
looks indistinguishable from a BH signal;
cf.~Fig.~\ref{fig:example_simulations}.  In contrast, the ``fluffy''
\texttt{A147} binaries merge into a more compact, highly distorted
BS remnant with an oscillating central amplitude $0.156\le \sqrt{G}
A_{\rm{ctr}}\le 0.164$.  The GW signal emitted by these fluffy
binaries is characterized by a lower-amplitude inspiral part followed
by an immense GW burst at merger and a prolonged, rapidly pulsating
ringdown; cf.~Fig.~\ref{fig:example_simulations}.

For the \texttt{A17} family of stars we consider additional
configurations with nonzero phase offsets, $\delta \phi = \pi/2,
\pi$ and anti-BS binaries. All {\tt A17} binaries produce identical
signals in the early inspiral, but as the BSs get closer, the
scalar-field interaction~\cite{Palenzuela:2006wp, Sanchis-Gual:2022mkk}
leads to delayed mergers for $\delta \phi\ne 0$, $\epsilon=-1$ and,
especially, for $\delta \phi=\pi$.  A comparison with BH
waveforms~\cite{Radia:2021smk} furthermore suggests that the {\tt
A17} anti-BS signal resembles most closely that from a nonspinning,
equal-mass BH binary in terms of the GW amplitude in the late
inspiral and its phase evolution close to merger; cf.~Fig.~3 in the
Supplemental Material.

{\textit{\textbf{Parameter Estimation}}---}
We inject our numerically computed GW signals into Gaussian noise
colored by the O4 design sensitivities of the LIGO Hanford (H1)
and Livingston (L1) detectors~\cite{AdvancedDetectorsLVK2018} and
perform Bayesian inference using {\sc bilby}~\cite{Ashton:2018jfp,
Romero-Shaw:2020:Bilby}.  We focus on the dominant $l=2$ modes and
consider systems of total masses $M^{\rm{inj}}_{\rm tot}\in [5,
105]\,M_{\odot}$ in the source frame, corresponding to scalar-field
masses $\mu \in [10^{-13}, 10^{-12}]\,\rm{eV}$. We choose luminosity
distances $d_{\rm L}$ corresponding to (zero-noise) optimal injected
signal-to-noise ratios (SNRs) $\rho^{\rm H1}\in[15,60]$, $\rho^{\rm
L1}\in[10,45]$.  We recover injected BS signals using standard
waveform approximants, including effects of asymmetric mass ratio,
spin, orbital precession and tides, as used routinely by the
LIGO-Virgo-KAGRA Collaboration \cite{KAGRA:2021vkt}: \texttt{IMRPhenomD}
\cite{Husa:2015iqa, Khan:2015jqa}, \texttt{IMRPhenomPv3}
\cite{Khan:2018fmp}, \texttt{IMRPhenomXP}, \texttt{IMRPhenomXPHM}
\cite{Pratten:2020ceb}, \texttt{TaylorF2} \cite{Messina:2017yjg},
\texttt{IMRPhenomPv2\_NRTidal} \cite{Dietrich:2017aum, Dietrich:2019kaq}
and \texttt{TEOBResumS} \cite{Nagar:2018zoe}.  We obtain comparable
results for the various approximants, as reviewed in more detail
in the Supplemental Material, and focus the following discussion
on \texttt{IMRPhenomXP} and \texttt{IMRPhenomPv2\_NRTidal}. We
report the mass values in the source frame; quantities in the
detector frame will be denoted by ``det''.

We perform our inference using two sets of BBH-like spin priors:
(i) we fix the spins in the recovery templates to their zero injected
values; (ii) we explore the complete spin prior with all parameters
being free. In the process, we compute for each detector the optimal
SNR $\rho$ of the recovered and injected signals.  For each recovery
we compute the Bayes factor $\mathcal{B}^{\rm{S}}_{\rm{N}}$,
quantifying the evidence ratio between a signal being present in
the data relative to the hypothesis of pure Gaussian noise.  To
assess the quality of the recovery we additionally compute the
residual $r \defeq d - h_{\rm{max}}$ from the data stream $d$ and
the approximant's strain $h_{\rm max}$ evaluated at the
maximum-joint-log-likelihood source parameters across H1/L1.
Specifically, we construct a null distribution of the white-noise
optimal SNR for a signal duration $D$ by generating random noise
realizations via
\begin{equation}
    n_{\rm{w}} \defeq \frac{\tilde{n}(f)}{\sqrt{S_n(f)}} \sim
    \mathcal{N} \big(0, \tfrac{1}{2}\sqrt{D}\,\big)\,,
\end{equation}
where $S_n(f)$ is the detector's noise power-spectral density.  If
the residual's SNR falls above the 99th percentile of the noise
SNR, we regard the residual as incompatible with Gaussian noise.
\begin{table}[t]
\caption{Recovery results for a range of BS injections from Table
\ref{tab:waveforms} using the \texttt{IMRPhenomXP} approximant with
spins fixed to zero. We contrast the injected
($\rho_{\rm{net}}^{\rm{inj}}$) and recovered ($\rho_{\rm{net}}^{\rm{rec}}$)
network SNRs and report the recovered chirp-masses
$\mathcal{M}_{\rm{c}}^{\rm{rec}}$, mass-ratios $q^{\rm{rec}}$ at
maximum-joint-log-likelihood.
}
\begin{tabular}
{|P{0.8in}|P{0.45in}|P{0.45in}|P{0.45in}|P{0.45in}|P{0.45in}|} \hline
Run & $\rm{log}\mathcal{B}^{\rm{S}}_{\rm{N}}$&$\rho_{\rm{net}}^{\rm inj}$&$\rho_{\rm net}^{\rm rec}$&$\!\mathcal{M}_{\rm{c}}^{\rm rec}\!\!/M_{\odot}$&$q^{\rm rec}$\\ \hline
\multicolumn{6}{| c |}{$M^{\rm{inj}}_{\rm{tot}} = 72.4 M_{\odot}$, \quad $\mathcal{M}^{\rm{inj}}_{\rm{c}} = 31.5 M_{\odot}$, \quad $d^{\rm{inj}}_L=500\,{\rm Mpc}$} \\ \hline
\texttt{A17-d12}      &  717 & 40.37 & 38.63 & 26.60 & 0.32 \\ \hline
\texttt{A17-d15-p090} & 1019 & 45.17 & 46.74 & 31.63 & 0.65 \\ \hline
\texttt{A17-d12-p180} &  792 & 43.84 & 36.62 & 25.07 & 0.97 \\ \hline
\texttt{A17-d12-e1}   &  820 & 41.12 & 41.91 & 27.13 & 0.71 \\ \hline
\texttt{A17-d14}      &  678 & 41.41 & 38.26 & 26.72 & 0.30 \\ \hline
\texttt{A147-d19}     &  555 & 57.75 & 32.16 & 40.02 & 1.00 \\ \hline
\multicolumn{6}{| c |}{$M^{\rm{inj}}_{\rm{tot}} = 9.9 M_{\odot}$, \quad $\mathcal{M}^{\rm{inj}}_{\rm{c}} = 4.3 M_{\odot}$, \quad $d^{\rm{inj}}_L=62.5\,{\rm Mpc}$} \\ \hline
\texttt{A17-d12}       & 406 & 29.69 & 30.60  & 3.89 & 0.39 \\ \hline
\texttt{A17-d15-p090} & 698 & 39.15 & 38.73  & 4.07 & 0.93 \\ \hline
\texttt{A17-d12-p180}  & 328 & 30.91 & 28.28  & 3.94 & 1.00 \\ \hline
\texttt{A17-d12-e1}    & 407 & 30.13 & 29.07  & 4.03 & 0.92 \\ \hline
\texttt{A17-d14}      & 536 & 34.13 & 34.24  & 3.92 & 0.40 \\ \hline
\texttt{A147-d19}      & 286 & 37.73 & 24.45  & 4.41 & 0.13 \\ \hline
\end{tabular}
\label{tab:SNR_BHs}
\end{table}

{\textit{\textbf{Results}} ({\tt A17} families) ---
For the compact {\tt A17} binaries, which form a BH postmerger,
all waveform approximants enable confident recovery of the signals
with a residual compatible with Gaussian noise, regardless of the
analysis type; see Table~\ref{tab:SNR_BHs} for a representative set
of examples.  Approximants including tidal effects furthermore infer
tidal deformability parameters consistent with the injected
$\Lambda_{1,2} \sim \mathcal{O}(10)$, albeit with poor constraints
on $\Lambda_2$.
This corresponds to a well measured
tidal deformability $\delta \tilde{\Lambda}$ of the binary, but a
poorly constrained $\tilde{\Lambda}$ \cite{Wade:2014vqa}. All
waveform approximants, however, fail to correctly
infer some other injected parameters (component masses, spins and/or
luminosity distance) within the $90\%$ credible region;
cf.~Fig.~\ref{fig:corner_A17_d17_e1} for a typical example.
Provided the inspiral contributes significantly to
the SNR, i.e.~for $M^{\rm inj}_{\rm tot} \lesssim 80\,M_{\odot}$ in our case,
however, these
deviations are not random but exhibit clear systematics which we illustrate in Fig.~\ref{fig:families}
for a wide range of injections and now discuss
in more detail.
\begin{figure}[t!]
  \includegraphics[width=0.4\textwidth]{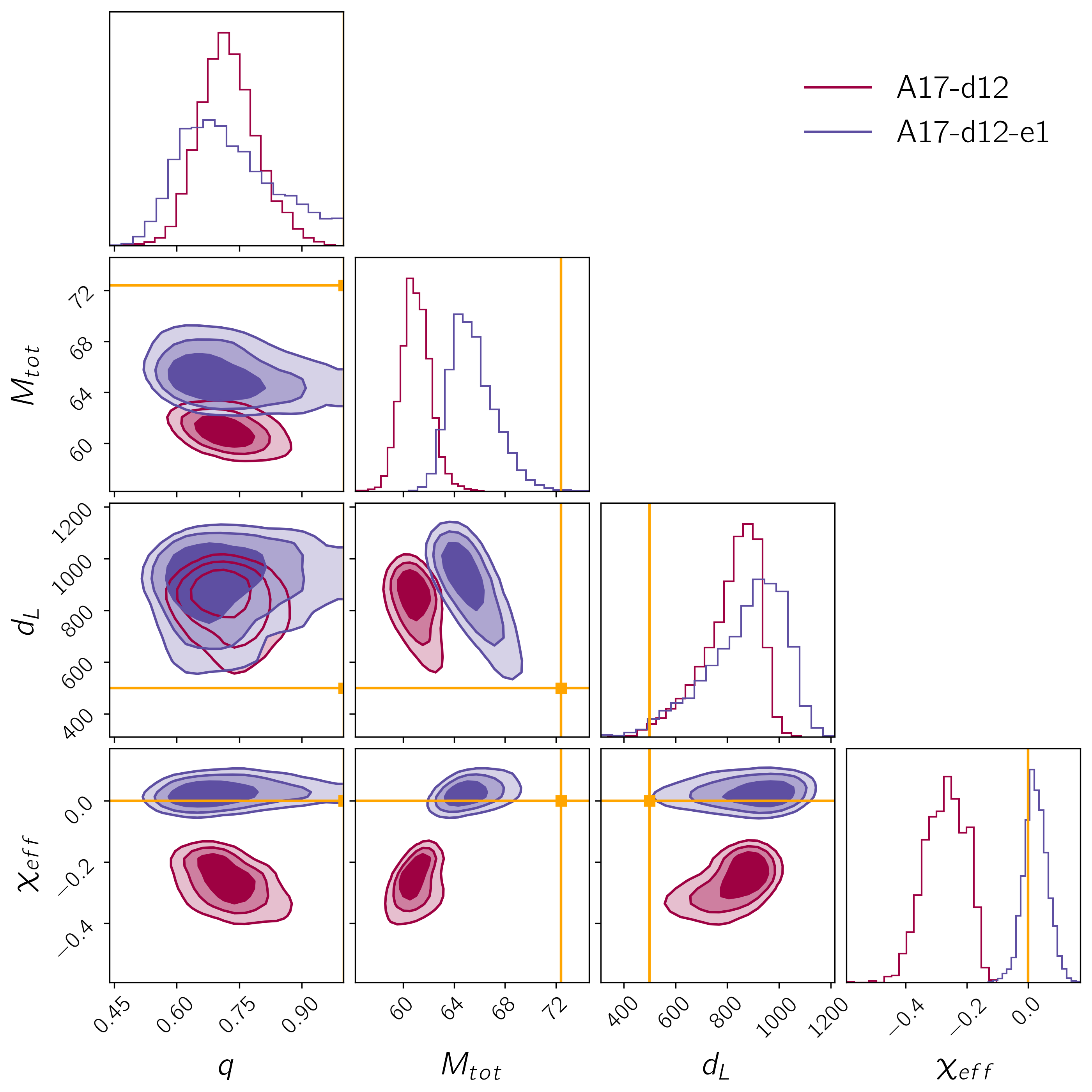}\\ 
  \caption{Comparison of key PE results for \texttt{A17-d12} and
  \texttt{A17-d12-e1} injections. We use \texttt{IMRPhenomXP} for
  recovery of a BS binary with $M^{\rm{inj}}_{\rm tot} = 72.4\,M_{\odot},~ d_{\rm{L}} = 500\,\rm{Mpc}$
  and allow all binary parameters to be sampled
  over. The 1, 1.5, 2-sigma contours are shown.
  }
  \label{fig:corner_A17_d17_e1}
\end{figure}

{\tt A17}, $\delta \phi=0$, $\epsilon=1$:
We obtain consistent results for using either
of the three waveform lengths employed, {\tt d12}, {\tt d14} and
{\tt d15}, indicating that they are sufficiently long for the mass
range considered. In general, the recovered luminosity distance is
overestimated,
likely compensating for amplitude effects arising from biases in the other
parameters.
Next, the posteriors from BH approximants
employed with spins fixed to zero systematically peak at total mass values within $\sim 10\,\%$ of the injected
$M^{\rm{inj}}_{\rm tot}$.  
For $M_{\rm tot}^{\rm inj}\lesssim 80\,M_{\odot}$, mass ratios are recovered with peaks in the range $0.3 \lesssim q\defeq m_2/m_1 \lesssim 0.7$, with no posteriors supporting the injected $q=1$ at $90\,\%$ confidence. This
picture changes considerably when we allow the spins to be sampled
over in the prior. Then we obtain more accurate estimates for the primary
mass $m_1$, often supporting the injected mass inside the $90\,\%$
confidence interval, and larger mass ratios
$0.5\lesssim q\lesssim 0.85$, albeit still well below the injected $q=1$.  
We furthermore infer significant spin magnitudes peaking at $0.3 \le a_{1,2} \le 0.98$ 
with clear preference for negatively aligned spins,
i.e.~$\chi_{\rm eff}<0$.

{\tt A17-p180}, $\delta \phi=\pi$, $\epsilon=1$:
As for $\delta \phi=0$, the recovered luminosity distance is often overestimated. Quite remarkably, however, these antiphase
binaries result in nearly opposite behavior in
PE results in every other regard.
For $M_{\rm tot}^{\rm inj}\lesssim 80\,M_{\odot}$,
we obtain good estimates 
for the primary mass $m_1$ and large mass
ratios $q\ge 0.8$ when the spins are {\it fixed} to their injected
values. Allowing the spins to vary, now results in significant
overestimates of $m_1$ and more unequal mass ratios $0.3\le q \le
0.8$. 
We typically obtain large spin magnitudes $a_{1,2}\ge 0.5$ but preferably
exhibiting partial alignment with the orbital angular momentum,
$0.1\le \chi_{\rm eff} \le 0.5$.

{\tt A17-d090}, $\delta \phi=\pi/2$, $\epsilon=1$:
This family of BS injections yields qualitatively similar behaviour
as {\tt A17-p180}, but with generally smaller departure from the
injected parameters.

{\tt A17-e1}, $\delta \phi=0$, $\epsilon=-1$:
The anti-BS family is the best recovered of all our injections,
supporting our conjecture that these binaries
most closely resemble waveforms from nonspinning BHs. We typically obtain mass ratios $q\gtrsim 0.75$, where 
the primary mass $m_1$ is either contained inside the $90\,\%$
confidence interval or close by, and,
for $M_{\rm tot}^{\rm inj}  \gtrsim 20 M_{\odot}$, spin values
close to zero. 
We illustrate this behavior by comparing in Fig.~\ref{fig:corner_A17_d17_e1} a representative
$M_{\rm tot}^{\rm inj}=72.4\,M_{\odot}$ example of this family with
its less accurately recovered {\tt A17-d12} counterpart.

\begin{figure}[t!]
  \includegraphics[width=0.48\textwidth]{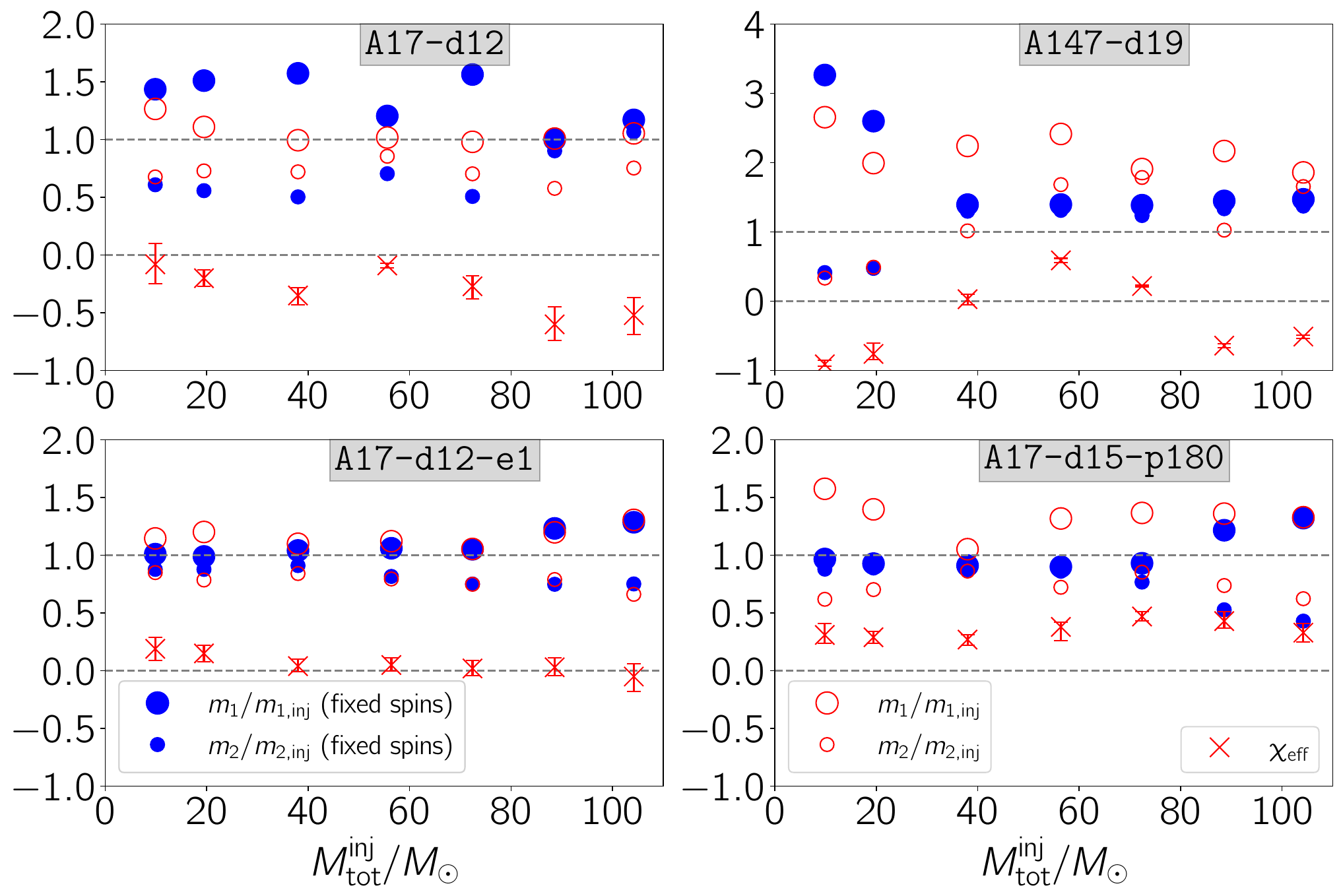}
  \caption{
  PE results obtained for four BS families and injections with
  total mass $M_{\rm tot}^{\rm inj} \in [5, 105] M_{\odot}$
  and luminosity distance $d_L=(M^{\rm{det}}_{\rm
  tot}/80\,M_{\odot}) 500\,{\rm Mpc}$.  For each injection, we display
  the median values of the
  recovered component masses $m_1,\,m_2$ normalized by their
  injected values (large and small circles) and the effective spin
  (crosses with $90\,\%$ confidence interval). The recoveries are
  obtained for fixed (blue) or variable spins (red symbols). The
  dashed lines mark the injected values, 1 for the masses and 0 for
  $\chi_{\rm eff}$. The recovered parameters deviate from these values
  non-randomly; cf.~the main text.
  By repeating selected injections with different $d_{\rm L}$, we
  have verified that the displayed trends are robust under variations
  of the SNR.
  }
  \label{fig:families}
\end{figure}

{\it Interpretation:}
The most pronounced difference between $\delta \phi=0$
BS binary signals, compared to their antiphase counterparts,
is the steep increase of the GW amplitude in the late
inspiral and merger whereas for $\delta \phi=\pi$, this
{\it chirp} is shallower; cf.~Fig.~3 in the Supplemental Material.
A closer inspection of GW signals from nonspinning BH binaries
in the frequency domain similarly yields a steeper increase in the GW amplitude close
to merger for unequal masses relative to $q=1$.
Unequal-mass BH chirps therefore resemble more closely
those from $\delta \phi=0$ BS binaries while $\delta \phi=\pi$
BS chirps are better approximated by equal-mass BH binaries. 

The inclusion of spins allows for an additional adjustment of the
chirp's steepness, typically boosting $\log (\mathcal{B}^{\rm{S}}_{\rm{N}})$
by $\sim 10\,\%$. BH binaries with aligned spins have an enhanced yet shallower inspiral than those with antialigned
spins; see e.g.~Figs.~1 and 3 in \cite{Campanelli:2006uy}. The
shallow inspiral-merger transition of $\delta\phi=\pi$ binaries,
therefore, favors the ``hang-up'' of aligned spins whereas the more
abrupt transition of $\delta\phi=0$ BS binaries favors partially
antialigned spins.
This effect also sheds light on the mass-ratio drifts
when we allow the spins to vary. For spins fixed
at zero, the steep chirp of $\delta \phi=0$ BS binaries can
{\it only} be reproduced through an unequal mass ratio in the
BBH approximant. For variable spins, in contrast, it can at least
partly be accommodated through antialigned spins
(the anti-hang-up effect); statistically we then expect
a drift towards $q=1$. For $\delta\phi=\pi$ BS binaries,
in turn, the inclusion of spins enables {\sc bilby} to reproduce the
shallow chirp in the BS signal through aligned spins rather than
resorting {\it exclusively} to a high ($q\approx 1$) mass ratio;
this results in a drift towards smaller $q$.
\begin{figure}[t!] 
  \includegraphics[width=0.47\textwidth]{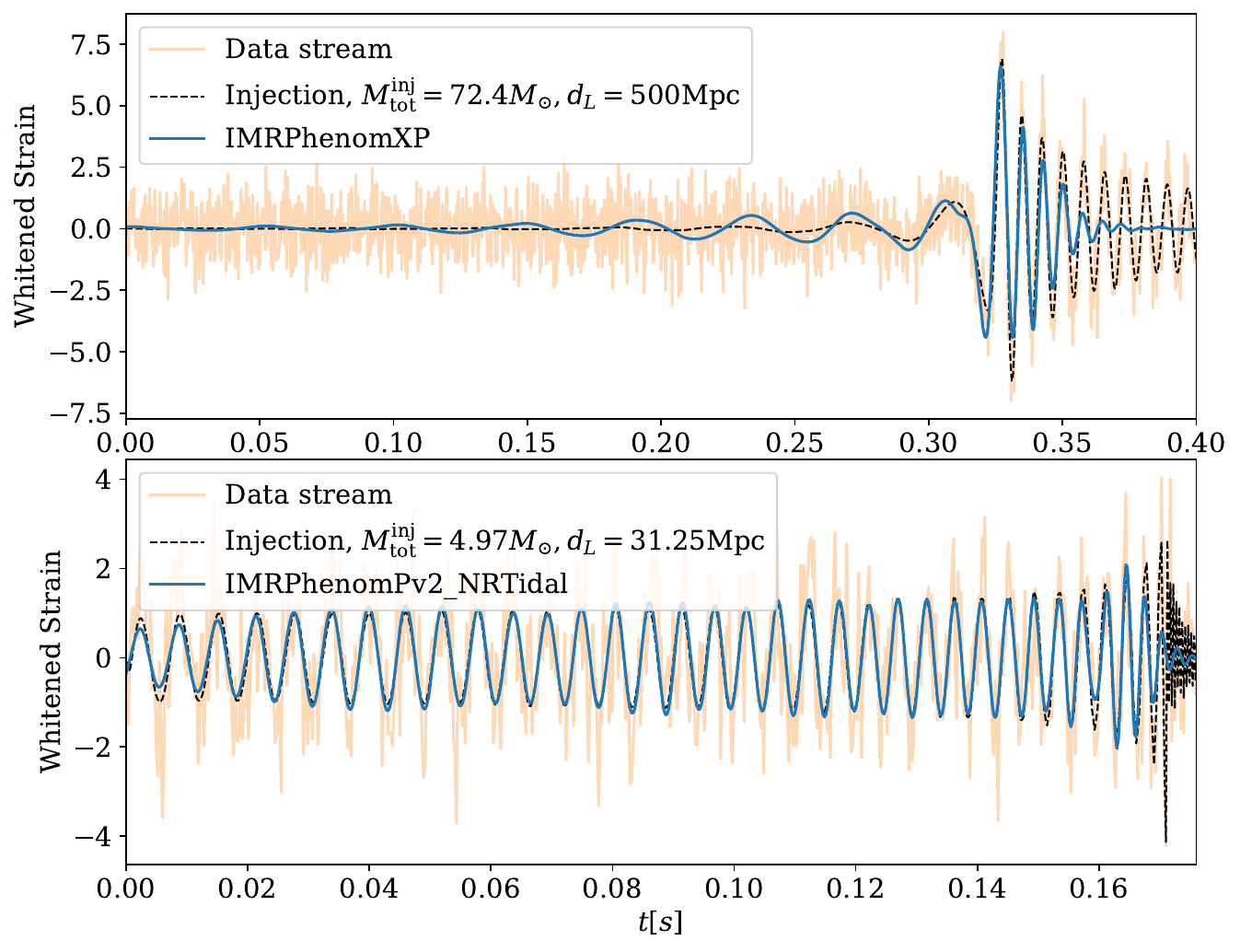}\\
  \caption{{\it Top:} The recovery of a high-mass \texttt{A147-d19}
  binary injection using \texttt{IMRPhenomXP} results in $m_1 \sim
  70 M_{\odot}$, $m_2 \sim 65 M_{\odot}$, $a_1 \sim 0.98$ and $a_2
  \sim 0.60$.
  {\it Bottom:} The recovery of a low-mass \texttt{A147-d19} binary
  injection using \texttt{IMRPhenomPv2\_NRTidal} yields
  maximum-joint-log-likelihood estimates $m_1 \sim 3.7 M_{\odot}$,
  $m_2 \sim 1.3 M_{\odot}$, $a_1 \sim 0.96$, $a_2 \sim 0.63$,
  $\Lambda_1 \sim 3000$ and $\Lambda_2 \sim 11\,000$.
  }
  \label{fig:A147}
\end{figure}

{\textit{\textbf{Results}} ({\tt A147} family)} ---
As illustrated in Fig.~\ref{fig:example_simulations}, GW signals
from our less compact {\tt A147} family of BS binaries differ more
pronouncedly from BH waveforms, especially around merger and ringdown.
These differences manifest themselves not only in the form of PE biases,
but also an incomplete recovery
of the injected signals. In particular, for high (low) total mass
$M_{\rm tot}^{\rm inj}$, the approximants are able to match the merger
(inspiral) part of the signals but never both and never the ringdown;
cf.~Fig.\ref{fig:A147}.  Overall, this leads to a reduction
in the recovered optimal SNR compared to the injection; cf.~Table
\ref{tab:SNR_BHs}. For injections with $\rho_{\rm inj}\lesssim
30$, the residual is still consistent with noise but for
$\rho_{\rm inj} \gtrsim 30$ we often obtain significant residuals. 

Furthermore, the late inspiral of less compact BS binaries occurs at lower frequencies than that of compact ones, so that BBH approximants systematically
overestimate the total mass; cf.~Fig.~\ref{fig:families}.
The addition of tidal effects improves
the approximant's ability to capture features in the inspiral portion
of the signal for small $M_{\rm tot}^{\rm inj}$; we thus obtain large tidal
parameters $\Lambda_{1,2}\sim 10^3$ to $10^4$ consistent with the
BS binary, albeit with large uncertainties in $\Lambda_2$.  Similar
to the compact BS injections, overall the inclusion of tidal effects
does not improve the estimation of other injection parameters.

{\textit{\textbf{Conclusions}}---}
By performing high-precision NR simulations of inspiralling and
merging BS binaries, we have investigated the capability of current
waveform models to recover BS GW signals with the following key results.
(i) Most BS systems considered here are detectable with current GW analysis pipelines
although current waveform templates systematically infer incorrect
source parameters. (ii) BS binaries can be numerically modeled with accuracy
comparable to BH binaries. (iii) BSs are excellent candidate sources
for constructing GW template banks alternative to BHs and NSs.

Our study suggests that, from the viewpoint of current GW
detectors, BS signals exhibit significant degeneracy with BBH waveforms obstructing the distinction between BS binaries, especially
those forming a BH upon merger, and NS or BH systems.
The most notable exception from this degeneracy
consists in the exceptionally strong inspiral-merger transition and a long-lived ringdown in binaries forming a BS postmerger; both these features are generally
poorly matched by present approximants leaving behind non-Gaussian residuals.

\begin{acknowledgments}
We thank Priscilla Canizares and Christopher J.~Moore for fruitful
discussions on GW analysis and the GRTL
collaboration\footnote{\href{https://github.com/GRTLCollaboration}{https://github.com/GRTLCollaboration}}.
T.E.~is supported by the Centre for Doctoral Training at the
University of Cambridge funded through STFC.
I.~M.~R.-S. acknowledges support from the Herchel Smith Postdoctoral Fellowship Fund.
This work has been supported by
STFC Research Grant No. ST/V005669/1.
We acknowledge support by the NSF Grants No.~PHY-090003, No.~PHY-1626190, and No.~PHY-2110594, DiRAC projects
ACTP284 and ACTP238, STFC capital Grants
No.~ST/P002307/1, No.~ST/R002452/1, No.~ST/I006285/1, and No.~ST/V005618/1, STFC operations Grant
No.~ST/R00689X/1. Computations were done on the CSD3 and Fawcett (Cambridge), 
Cosma (Durham), Hawk (Cardiff), CIT LIGO Lab (Pasadena), Stampede2 (TACC) and Expanse (SDSC) clusters.
\end{acknowledgments}

\vspace{0.2cm}
\textbf{\textit{Data availability}}--
Our data and animations are publicly
available~\cite{Evstafyeva_Boson_star_waveforms, video1, video2}.
\newpage

%


\onecolumngrid
\newpage
\begin{center}
  \textbf{\large{Supplemental Material}} \\
\end{center}
\twocolumngrid

\setcounter{equation}{0}
\setcounter{figure}{0}
\setcounter{table}{0}
\setcounter{page}{1}
\makeatletter
\renewcommand{\theequation}{S\arabic{equation}}
\renewcommand{\thefigure}{S\arabic{figure}}
\renewcommand{\bibnumfmt}[1]{[S#1]}
\renewcommand{\citenumfont}[1]{S#1}

\section*{Convergence tests}

\begin{figure}[b]
  \includegraphics[width=0.47\textwidth]{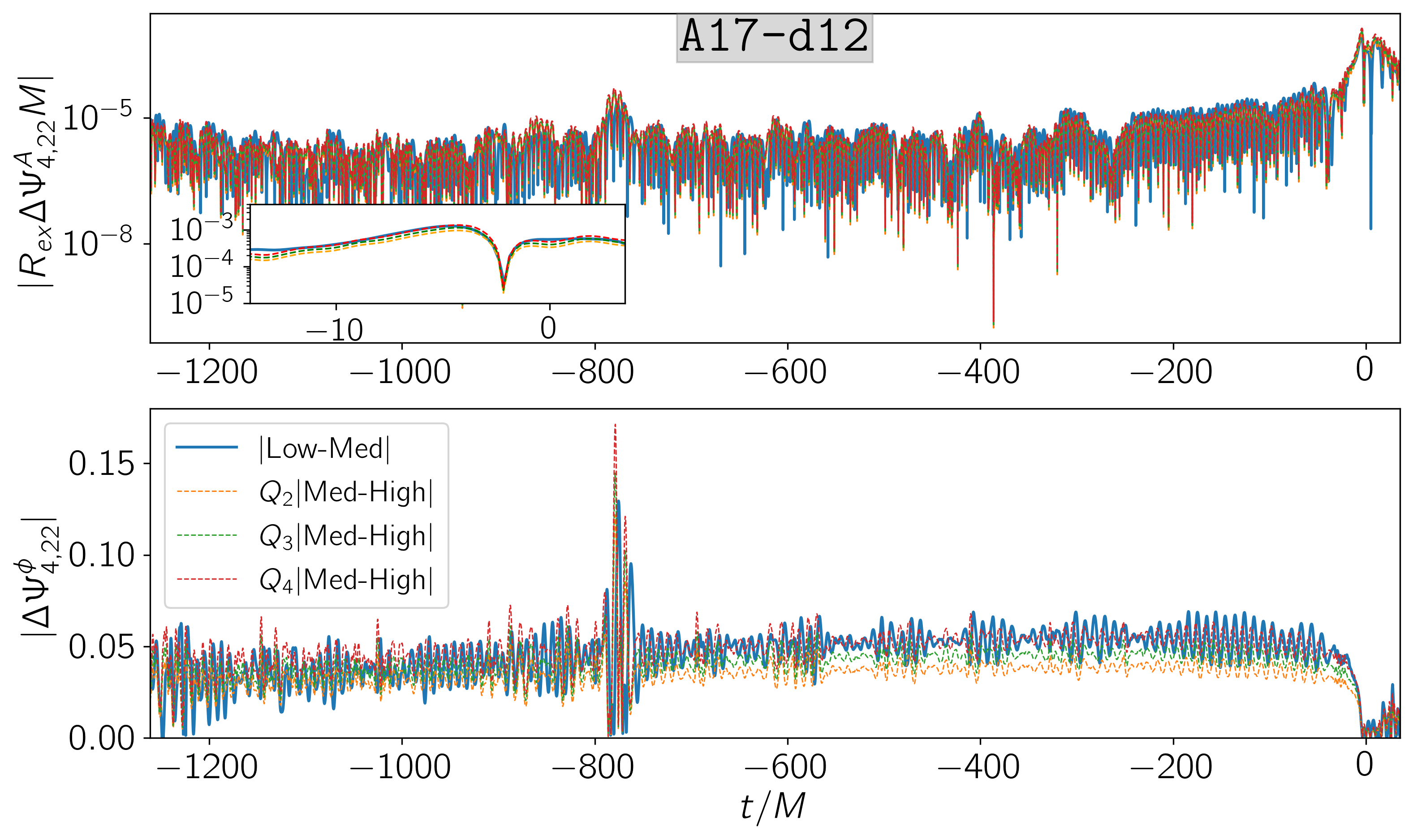}
  \includegraphics[width=0.47\textwidth]{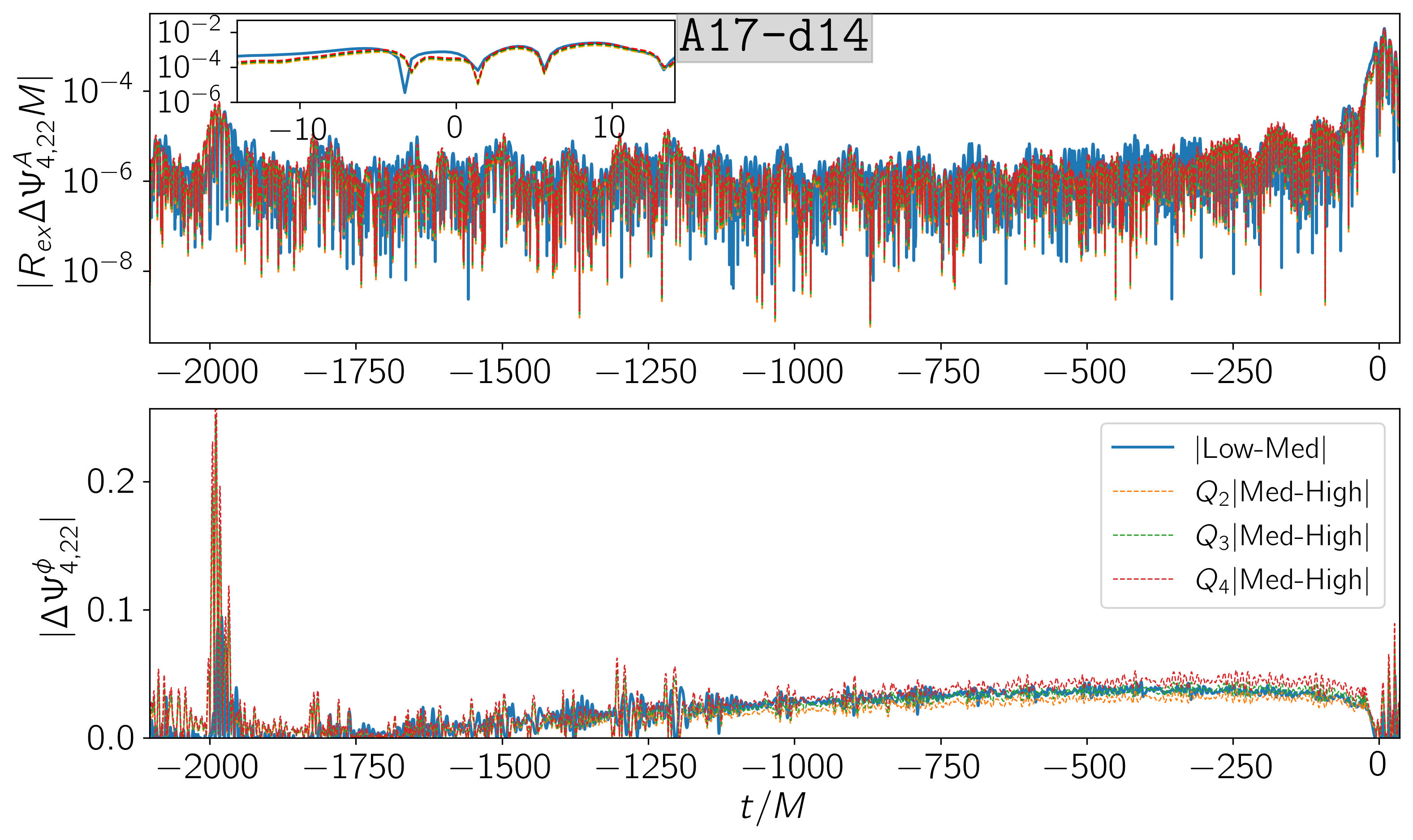}

  \caption{Convergence test of the compact configurations
  \texttt{A17-d12} and \texttt{A17-d14} of Table I in the main
  document.  The difference between high and medium resolutions has
  been re-scaled by factors $Q_2$ to $Q_4$, corresponding to
  convergence of second to fourth order.
  }
  \label{fig:soli_A017_convergence}
\end{figure}

To evaluate the numerical error incurred in our simulations, we
analyze their convergence properties for three representative
configurations. We start with the {\sc GRChombo} simulations of
binaries from the compact families \texttt{A17-d12} and
\begin{figure}[h]
  \includegraphics[width=0.47\textwidth]{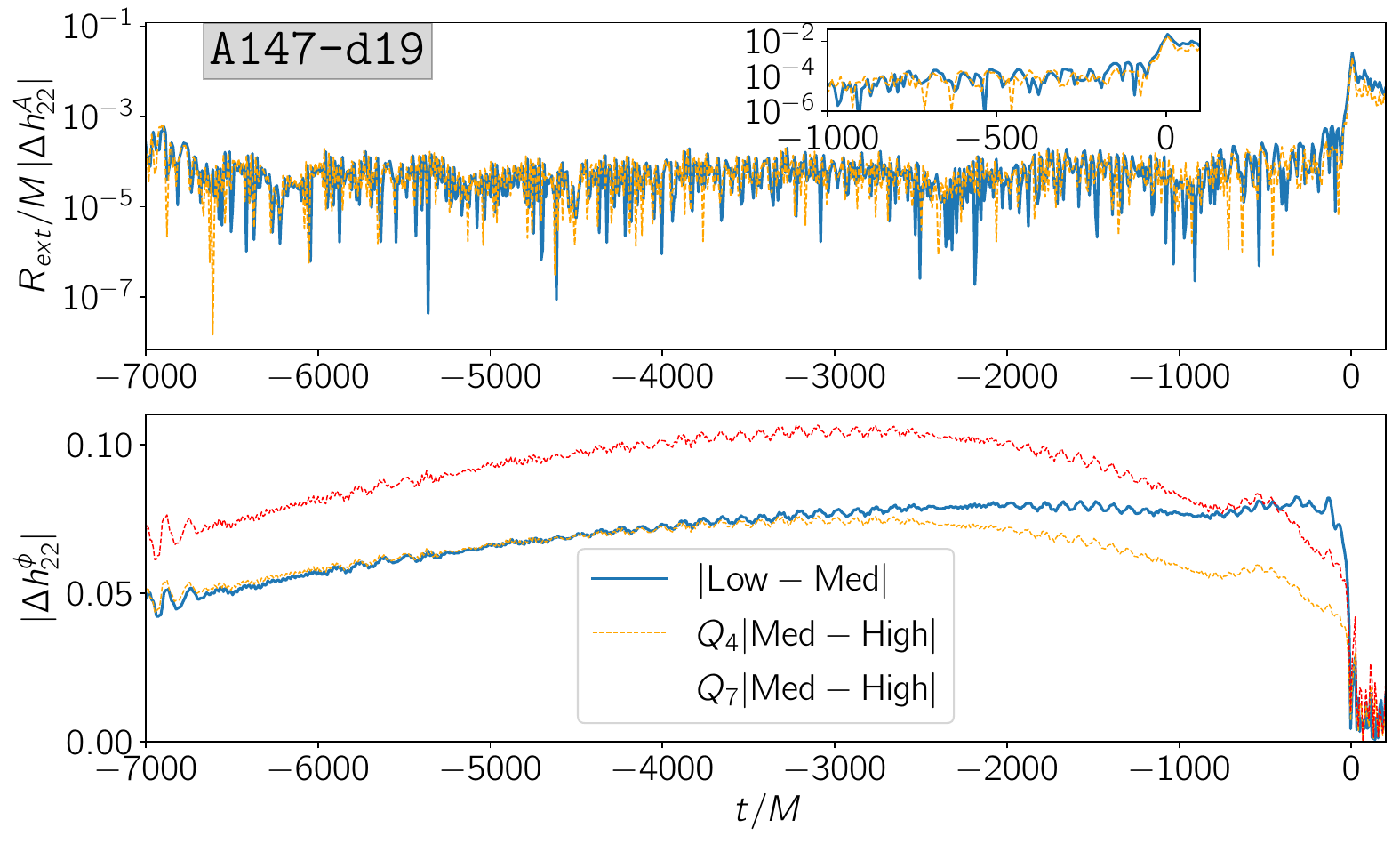}
  \caption{Convergence test of the configuration \texttt{A147-d19}
  of Table I in the main document for the GW strain (upper) and
  phase (lower panel). The difference between high and medium
  resolutions has been scaled by factors $Q_4$ and $Q_7$, corresponding
  to fourth-  and seventh-order convergence.}
  \label{fig:conv_A147-d14}
\end{figure}
\texttt{A17-d14}. Here we use $\Delta x_{\rm{low}} = 1/28$, $\Delta
x_{\rm{med}} = 1/32$, $\Delta x_{\rm{high}} = 1/40$ for the low,
medium and high resolutions, respectively.  The resulting convergence
analysis for the GW signals extracted at $R_{\rm ex}=140$ is shown
in Fig.~\ref{fig:soli_A017_convergence} and yields convergence
between third and fourth order.  For the shorter \texttt{A17-d12}
configuration, Richardson extrapolation yields an amplitude error
of 2.5\% and a phase error of 0.03 while we obtain 3\% in the
amplitude and 0.04 in the phase for the longer \texttt{A17-d14}
binary.  We quantify the error due to finite radius extraction
through extrapolation to infinity and thus find a phase error of
$0.06$ and an amplitude error of $1 \%$. Overall, this brings the
total numerical error budget for the \texttt{A17} family of runs
to 4\% in the amplitude and 0.1 in the phase.

For the less compact \texttt{A147-d19} configuration, we display
in Fig.~\ref{fig:conv_A147-d14} the convergence properties obtained
with the {\sc Lean} code for the GW multipole $h_{22}$ extracted
at $R_{\rm ex}=240$ for resolutions $\Delta x_{\rm low}=1/32$,
$\Delta x_{\rm med} =1/36$ and $\Delta x_{\rm high}=1/40$.  We
obtain fourth-order convergence for the amplitude and, during most
of the inspiral, for the phase. In the late inspiral and merger,
the phase evolution exhibits over convergence up to about seventh
order, likely due to fortuitous cancellations of different error
contributions. We conservatively estimate the numerical uncertainties
for this configuration assuming exclusively fourth-order convergence.
For the amplitude, we obtain $0.5\,\%$ in the early inspiral slowly
increasing to about $3\,\%$ around merger.  The phase error peaks
at about $0.09$ during the mid inspiral. The uncertainty due to
finite extraction radius is about $0.15$ or less in the GW phase
and decreases from $4\,\%$ (in the early inspiral) to $1\,\%$ (around
merger) in the amplitude. This results in an overall error budget
of $0.2$ in the phase and $4\,\%$ in the amplitude.

\section*{Inference set-up} \label{sec:analysis_setup}
In this Section, we outline the injection and analysis set-up
employed in our parameter estimation in {\sc bilby}.

\textit{Injections:} We utilise $\textsc{LALsuite}$ \cite{lalsuite,
swiglal} to perform reconstruction of our NR signals. Prior to
injecting, we remove junk radiation at the beginning of our waveforms
and prepare the NR data in line with the Format 1 standards of LIGO
Algorithms Library injections \cite{Schmidt2017numerical}. Unless
otherwise stated all of our injections use the same extrinsic
parameters. In particular, we use a right ascension $\alpha = 1.375$,
declination $\delta = -1.2108$, inclination angle $\iota
=\tfrac{\pi}{3}=60\degree$ and polarisation angle $\psi =
\tfrac{\pi}{2}=90\degree$. The total binary masses $M^{\rm
inj}_{\rm{tot}} \in [5, 105] M_{\odot}$ in the source frame considered
in this work correspond to $M^{\rm{det}}_{\rm{tot}} \in [5, 120]
M_{\odot}$ in the detector frame.

\textit{Parameter estimation:} In all analyses, we use a sampling
frequency of $4096$~Hz, a reference frequency of $50$~Hz and a
minimum frequency set by the starting frequency of our NR waveforms.
When the starting frequency of our injected waveforms drops below
20~Hz, we set the minimum frequency to 20~Hz. The NR signals are
injected into the Hanford and Livingston detectors, assuming noise
curves from the 4th observing run (O4)~\cite{AdvancedDetectorsLVK2018S}.
For all of the PE runs reported in the main text, we initialise the
sampler with 1000 points -- for some test runs included in Table
\ref{tab:runs} below, we have used 500 points instead -- and
marginalise over time and luminosity distance.  For approximants
including only the $(22)$-mode, we additionally marginalise over
the phase. We have used a range of nested samplers, such as
\texttt{dynesty} \cite{Speagle_2020}, \texttt{cpnest}
\cite{2022ascl.soft05021D} and \texttt{nessai}
\cite{nessai,Williams:2021qyt,Williams:2023ppp}, and have found
good agreement between them. The sampling space is determined by
{\sc bilby}'s default BBH and BNS priors (cf.~Tables 1 and 2 of
\citet{Ashton:S2018jfp}). However, we increase the spin magnitude
($a_1$, $a_2$) upper limits to $0.99$. When using nonprecessing
models, the prior on the spin magnitudes is a uniform distribution
$\chi_1, \chi_2 \in [-1, 1]$, where $\chi_1, \chi_2$ denote the
dimensionless spin projections onto the orbital angular momentum.
This range includes binaries with antialigned spins.  For BNS
models, we additionally increase the upper limits in the prior of
the dimensionless tidal deformability parameters ($\Lambda_1$,
$\Lambda_2$) to $20000$. We find this choice necessary for injections
of the less compact $\texttt{A147}$ family in order to circumvent
one of the tidal deformability numbers railing against the upper
prior bound.

We have tested the quality of our parameter estimation using
injections from different resolutions and extraction radii, and
have found consistent results.

\section*{Black-hole and boson-star signals} 
In this section, we compare the IMR signals of our BS families with
those from equal-mass nonspinning BH binaries and discuss the
varying degrees of deviation from the BH case which we observe. We
start this discussion by exploring first what behaviour we would
expect theoretically. For this purpose, we follow the analysis of
the scalar-field interaction outlined in Appendix B of
Ref.~\cite{Palenzuela:S2006wp} and in particular consider the energy
density of a scalar field $\varphi$ in the flat-space limit,
\begin{equation} \label{eq:indrho}
  \rho = \frac{1}{2} \left(|\Pi|^2 + |\partial_i \varphi|^2 +
  V(\varphi) \right)\,.
\end{equation}

During the plunge and merger phase of
\begin{figure*}[hbt!]
\includegraphics[width=0.45\textwidth]{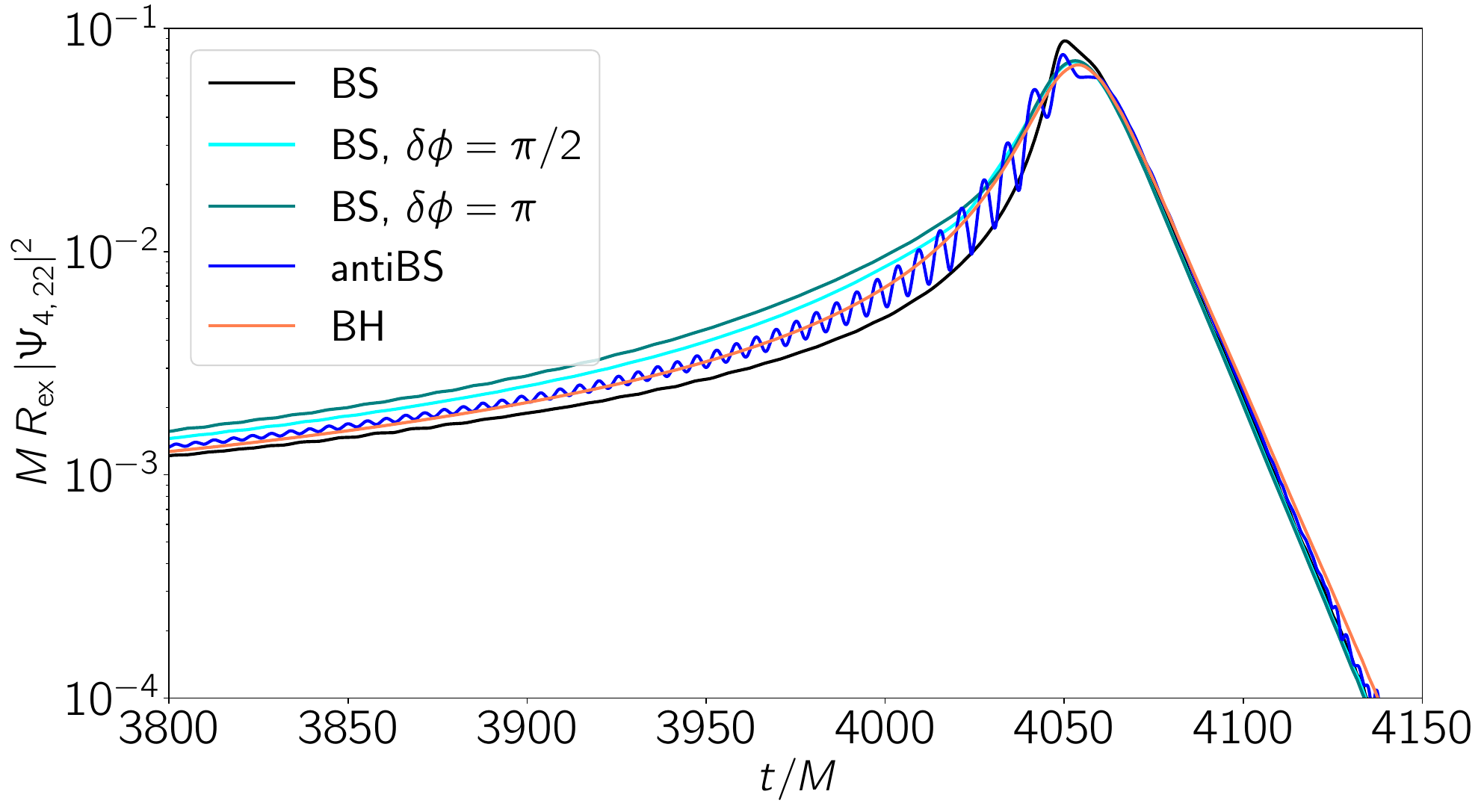} 
\includegraphics[width=0.45\textwidth]{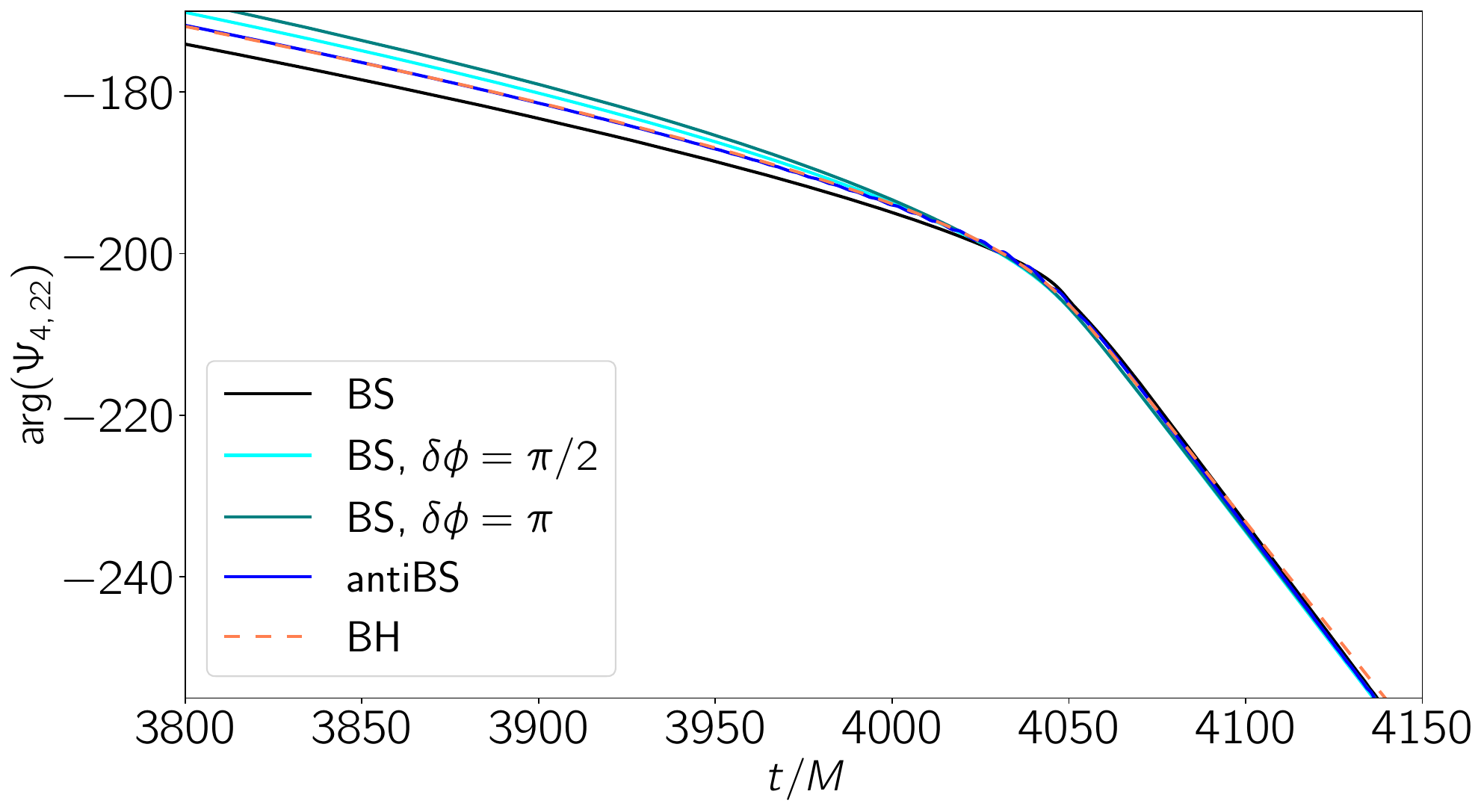}
  \caption{Comparison of amplitude (left) and phase (right) of the
  dominant (22)-modes obtained for the in-phase BS (\texttt{A17-d15}),
  dephased (\texttt{A17-d15-p090}), antiphase (\texttt{A17-d15-p180}),
  anti-BS (\texttt{A17-d15-e1}) and BH binaries.  Both amplitude
  and phase have been aligned such that the peak amplitudes coincide
  in time.  The anti-BS signal exhibits the best agreement with the
  BH waveform, whereas the in-phase and antiphase configurations
  result in the largest deviations from the BH signal.}
  \label{fig:BSBH_comp}
\end{figure*}
a binary the two BSs' scalar fields overlap and we can approximate
the resulting field and its time derivative as a superposition given
by $\varphi = \varphi_1 + \varphi_2$ and $\Pi = \Pi_1 + \Pi_2$.
This results in a combined energy density
\begin{equation}
  \rho = \rho_1 + \rho_2 + \Delta,
\end{equation}
where $\rho_1, \rho_2$ are the individual energy densities of the
stars as in Eq.~\eqref{eq:indrho} and $\Delta$ is the interaction
potential; cf.~Eq.(B3) of Ref~\cite{Palenzuela:S2006wp}. This
interaction term is composed of mixed products of the scalar field,
such as $\varphi_1 \bar{\varphi}_2$ or $\Pi_1 \bar{\Pi}_2$, and
therefore vanishes in the limit of large separations. We next recall
that the two BS scalar fields' phase offset and relative rotation
in the complex plane are determined by the parameters $\delta \phi$
and $\epsilon$, respectively. For an equal-mass binary system, the
scalar field profiles of the two stars forming a binary thus acquire
the following forms,
\begin{align}
  \varphi_1 &= A(t,r) e^{\iu\omega t}, \\
  \varphi_2 &= A(t,r) e^{\iu(\epsilon\omega t + \delta \phi)}.
\end{align}
We assume that the scalar field amplitude is a slowly varying
function of time so that the dynamical part is dominated by the
exponential factor $e^{\iu \omega t}$. Evaluating the interaction
potential $\Delta$ for the equal-mass (i) $\epsilon =1$, $\delta
\phi = 0$ (in-phase), (ii) $\epsilon =1$, $\delta \phi = \pi/2$
(dephased), (iii) $\epsilon =1$, $\delta \phi = \pi$ (antiphase),
and (iv) $\epsilon =-1$, $\delta \phi = 0$ (anti-BS) configurations,
we then find
\begin{align}
  \Delta_{\rm{BS}} &= \Delta_0, \\
  \Delta_{\delta \phi = \frac{\pi}{2}} &= 0, \\
  \Delta_{\delta \phi = \pi} &= -\Delta_0, \\
  \Delta_{\rm{anti-BS}} &= \Delta_0 \rm{cos}(2\omega t),
\end{align}
where $\Delta_0$ is strictly nonnegative. Note that the anti-BS
configuration results in an interaction potential with explicitly
harmonic time dependence that oscillates between the $\Delta$ values
of the in-phase and antiphase binaries. If the time-scale of the
late inspiral and plunge is much longer than the scalar oscillation
period of individual BSs, the interaction contribution to the energy
density averages out to zero. The interaction potential for the
$\delta \phi = \pi/2$ binary, in turn, vanishes completely without
any oscillating behaviour. Therefore, we may expect that the anti-BS
and $\delta \phi = \pi/2$ configurations will bear closest resemblance
to the BH case, where scalar self-interactions are absent by
construction. As we will see next, this expectation is largely, but
not completely, borne out by the empirical results.

For testing this hypothesis, we compare in Fig.~\ref{fig:BSBH_comp}
the aligned GW amplitude and phase obtained for the different BS
configurations with those obtained in Ref.~\cite{Radia:S2021smk}
for an equal-mass nonspinning BH binary. In accordance with our
above discussion, the anti-BS binary exhibits excellent agreement
with the BH results and displays an oscillatory pattern in the GW
amplitude commensurate with the expected oscillation of the interaction
term $\Delta_{\rm anti-BS}$. Its amplitude and phase furthermore
fall into the range between the in-phase and antiphase binaries.
The $\delta \phi = \pi/2$ binary, however, does not quite fulfill
the expectations from our theoretical analysis; even though its
amplitude and phase match the BH signal better than the in-phase
and antiphase binaries do, its deviations exceed those of the
(averaged) anti-BS signal.  This feature may simply mark the
limitations of the above flat-space analysis; and yet there remains
the question why the prediction works so well for the anti-BS
configuration but not for the $\delta\phi=\pi/2$ case. At present,
we do not have a fully satisfactory answer to this question.  A
more comprehensive exploration of the BS model parameter space may
shed more light on this puzzle.

\section*{Injections}
We present in Table~\ref{tab:runs} a list of injections of boson-star
(BS) waveforms together with the recovered parameters and consistency
checks of the recovered signals. We sort this table by the different
families of nonspinning BS binaries listed in Table I of the main
document. The varying injected parameters, namely the component
masses $m_1$, $m_2$ in the source frame and the luminosity distance
$d_{\rm L}$, are listed along with the resulting injected optimal
SNR values for the Hanford and Livingston detectors, $\rho^{\rm
H1}$ and $\rho^{\rm L1}$.  Each injection is analyzed using a
waveform approximant abbreviated as follows. \\

\begin{center}
\begin{tabular}{ccl}
  {\tt D} & = & {\tt IMRPhenomD} \\
  {\tt F2} & = & {\tt TaylorF2} \\
  {\tt Pv2T} & = & {\tt IMRPhenomPv2\_NRTidal} \\
  {\tt Pv3} & = & {\tt IMRPhenomPv3} \\
  {\tt TEOB} & = & {\tt TEOBResumS} \\
  {\tt XP} & = & {\tt IMRPhenomXP} \\
  {\tt XPHM} & = & {\tt IMRPhenomXPHM}
\end{tabular}
\end{center}
The recovery is performed using the listed samplers with $N_s$
points. Next, we report the recovered parameters which are obtained
from the posterior distributions with $90\,\%$ confidence intervals:
the component masses $m_1$ and $m_2$, the luminosity distance $d_L$,
the dimensionless spins \footnote{For nonprecessing models, we
report projections of the dimensionless spins onto the orbital
angular momentum, $\chi_1$, $\chi_2$, which are allowed to be
negative.} $a_1$ and $a_2$, the effective spin $\chi_{\rm eff}$,
the precession spin parameter $\chi_p$ and the tidal deformation
parameters $\Lambda_1$ and $\Lambda_2$. The recovered SNR is given
for the two detectors as $\rho^{\rm H1}$ and $\rho^{\rm L1}$ and
calculated from the inferred parameters maximising the
joint-log-likelihood. We flag with ticks or crosses whether the
residual is compatible with Gaussian noise and quantify the
significance of a detection with the Bayes factor
$\log(\mathcal{B}^{\rm{S}}_{\rm{N}})$.

\onecolumngrid
\newpage
\begingroup\squeezetable


\begingroup\squeezetable
\begin{sidewaystable}
\centering
\caption{
    List of all our injections for the BS families listed in Table
    I of the main document. We report the injected component masses,
    $m_1$ and $m_2$, luminosity distance $d_L$ and the corresponding
    SNR values for the H1 and L1 detectors. The recovery is performed
    using the specified approximants and samplers initialised with
    $N_s$ sampling points.  The recovered values for the masses,
    luminosity distance, the dimensionless spins $a_1$ and $a_2$,
    the effective and precession spin parameters $\chi_{\rm eff}$,
    $\chi_{\rm p}$ and the tidal deformability parameters $\Lambda_1$,
    $\Lambda_2$ are given in the form of median values with $90\,\%$
    confidence intervals.  We additionally list the recovered optimal
    SNR and Bayes evidence factors $\log(\mathcal{B}^{\rm{S}}_{\rm{N}})$
    and mark residuals consistent (inconsistent) with Gaussian noise
    by a \checkmark (\xmark) symbol.
    \label{tab:runs}
    }
    \begin{tabular}{l|ccccc|ccc|ccccccccccc|cc}

    \hline \hline
    \multicolumn{1}{c}{} &
    \multicolumn{5}{|c}{Injection} &
    \multicolumn{3}{|c}{Analysis} &
    \multicolumn{11}{|c}{Recovery} &
    \multicolumn{2}{|c}{Consistency} \\

    \hline
    Family & $\frac{m_1}{M_{\odot}}$ & $\frac{m_2}{M_{\odot}}$ & $\frac{d_L}{\rm Mpc}$ & $\rho^{\rm H1}$ & $\rho^{\rm L1}$ &
    Appr. & Sampler & $N_s$ &
    $\frac{m_1}{M_{\odot}}$ & $\frac{m_2}{M_{\odot}}$ & $\frac{d_L}{\rm Mpc}$ & $a_1$ & $a_2$ & $\chi_{\rm eff}$ & $\chi_{\rm p}$ & $\Lambda_1$ & $\Lambda_2$ & $\rho^{\rm H1}$ & $\rho^{\rm L1}$ &
    Res.(H,L) & $\log(\mathcal{B}^{\rm{S}}_{\rm{N}})$ \\
    \hline
    %
    {\tt A147-d19     } & 2.48& 2.48& 30 & 24.5 & 17.4 &
    {\tt Pv2T} & {\tt nessai} & 1000 &
    $ 3.70^{+0.12}_{-0.10}$ & $1.75^{+0.05}_{-0.06}$ & $51.2^{+24.3}_{-16.9}$ &
    -- & -- &
    -- & -- &
    $10407^{+1915}_{-1795}$ & $10002^{+8973}_{-8897}$ & 21.7 & 14.7 &
    (\xmark,\xmark) &  316 \\
    %
    {\tt A147-d19     } & 2.48& 2.48& 30 & 24.5 & 17.4 &
    {\tt Pv2T} & {\tt nessai} & 1000 &
    $3.64^{+0.15}_{-0.13}$ & $1.70^{+0.15}_{-0.21}$ & $49.8^{+9.9 }_{-11.3}$ &
    $0.78^{+0.18}_{-0.29}$ & $0.47^{+0.44}_{-0.41}$ &
    $-0.33^{+0.27}_{-0.28}$ & $0.53^{+0.26}_{-0.20}$ &
    $6102^{+2519}_{-2352}$ & $10302^{+8602}_{-8881}$ &
    22.1 & 17.0 &
    (\checkmark,\checkmark) & 341 \\
    {\tt A147-d19} & 2.48 & 2.48 & 31.25 & 23.8 & 17.1 &
    {\tt Pv2T} & {\tt nessai} & 1000 & 
    $3.57_{-0.11}^{+0.20}$ & $1.59_{-0.19}^{+0.16}$ & 
    $54.3_{-13.1}^{+8.6}$ &
    $0.79_{-0.38}^{+0.18}$ & $0.52_{-0.46}^{+0.42}$ & $-0.40_{-0.25}^{+0.34}$ & $0.47_{-0.22}^{+0.30}$ & $7894_{-3089}^{+2791}$ & $9998_{-8867}^{+8902}$ & 22.5 & 16.0 &
    (\checkmark,\checkmark) & 350 \\
    %
    {\tt A147-d19} & 2.48& 2.48& 31.25 & 23.8 & 17.1 &
    {\tt XP} & {\tt nessai} & 1000 &
    $6.92^{+0.56}_{-0.33}$ & $1.07^{+0.07}_{-0.08}$ & $ 74.0^{+23.9}_{-28.5}$ &
    -- & -- & -- & -- & -- & -- & 17.2 & 12.6 &
    (\checkmark,\checkmark) & 214  \\
    %
    %
    {\tt A147-d19} & 2.48& 2.48& 31.25 & 23.8 & 17.1 &
    {\tt XP} & {\tt nessai} & 1000 &
    $5.31^{+1.23}_{-0.72}$ & $1.37^{+0.18}_{-0.21}$ & $  432^{+63  }_{-368 }$ &
    $0.69^{+0.16}_{-0.16}$ & $0.80^{+0.17}_{-0.51}$ &
    $0.00^{+0.14}_{-0.12}$ & $0.68^{+0.13}_{-0.16}$ &
    -- & -- & 3.14 & 1.66 &
    (\xmark,\xmark) & 237  \\
    {\tt A147-d19} & 4.98 & 4.98 & 20 & 96.2 & 68.1 &
    {\tt XP} & {\tt nessai} & 1000 &
    $13.7_{-0.06}^{+0.05}$ & $1.72_{-0.01}^{+0.01}$ &
    $26.0_{-0.84}^{+0.96}$ & $0.86_{-0.01}^{+0.01}$ & $0.98_{-0.05}^{+0.01}$ & $-0.74_{-0.02}^{+0.02}$ & $0.46_{-0.01}^{+0.01}$ & -- & -- & 75.6 & 51.6 & 
    (\xmark,\xmark) & 4221 \\ 
    {\tt A147-d19} & 4.94 & 4.94 & 50 & 38.5 & 27.2 &
    {\tt XP} & {\tt nessai} & 1000 &
    $15.0_{-0.4}^{+0.2}$ & $1.90_{-0.02}^{+0.06}$ & $56.1_{-3.8}^{+8.2}$ &
    $0.90_{-0.02}^{+0.04}$ & $0.87_{-0.39}^{+0.11}$ & $-0.57_{-0.03}^{+0.05}$ & $0.71_{-0.02}^{+0.03}$ & -- & -- & 26.6 & 17.3 & 
    (\checkmark,\checkmark) & 674 \\
    {\tt A147-d19} & 4.93 & 4.93 & 62.5 & 30.8 & 21.8 &
    {\tt XP} & {\tt nessai} & 1000 &
    $15.8_{-0.2}^{+0.2}$ & $1.99_{-0.02}^{+0.03}$ & $140_{-68}^{+35}$ &
    -- & -- & -- & -- & -- & -- & 19.9 & 14.2 & 
    (\checkmark,\checkmark) & 286 \\ 
    %
    {\tt A147-d19     } & 4.93& 4.93&62.5& 30.8 & 21.8 &
    {\tt XP} & {\tt nessai} & 1000 &
    $16.1^{+0.2 }_{-0.7 }$ & $2.04^{+0.04}_{-0.07}$ & $ 110^{+190 }_{-47  }$ &
    -- & -- & -- & -- & -- & -- & 18.8 & 14.1 &
    (\checkmark,\checkmark) & 253 \\
    %
    {\tt A147-d19     } & 4.93& 4.93&62.5& 30.8 & 21.8 &
    {\tt XP} & {\tt nessai} & 1000 &
    $13.1^{+0.1 }_{-0.1 }$ & $1.65^{+0.02}_{-0.02}$ & $75.4^{+17.3}_{-10.0}$ &
    $0.99^{+0.00}_{-0.01}$ & $0.87^{+0.11}_{-0.39}$ &
    $-0.91^{+0.06}_{-0.03}$ & $0.32^{+0.05}_{-0.08}$ &
    -- & -- & 25.2 & 14.7 &
    (\checkmark,\checkmark) & 378 \\
    {\tt A147-d19} & 4.93 & 4.93 & 62.5 & 30.8 & 21.8 &
    {\tt F2} & {\tt nessai} & 1000 &
    $9.83_{-1.13}^{+3.11}$ & $2.06_{-0.29}^{+0.23}$ & $171_{-59}^{+42}$ &
    $-0.83_{-0.11}^{+0.31}$ & $-0.05_{-0.52}^{+0.41}$ & $-0.69_{-0.11}^{+0.23}$ & -- & $511_{-279}^{+235}$ & $2544_{-2221}^{+2136}$ & 21.4 & 15.6 &
    (\checkmark,\checkmark) & 342 \\ 
    {\tt A147-d19} & 4.93 & 4.93 & 62.5 & 30.8 & 21.8 &
    {\tt TEOB} & {\tt nessai} & 1000 &
    $12.9_{-0.13}^{+0.14}$ & $1.62_{-0.02}^{+0.02}$ & $98.4_{-40.6}^{+37.2}$ &
    $-0.98_{-0.01}^{+0.02}$ & $-0.84_{-0.12}^{+0.27}$ & $-0.96_{-0.02}^{+0.03}$ & -- & -- & -- & 21.0 & 14.1 &
    (\checkmark,\checkmark) & 337 \\ 
    %
    %
    %
    %
    {\tt A147-d19     } & 4.89& 4.89&100 & 19.2 & 13.6 &
    {\tt F2} & {\tt nessai} &1000 &
    $13.2^{+1.0 }_{-1.3 }$ & $1.93^{+0.20}_{-0.15}$ & $ 273^{+113}_{-101}$ &
    -- & -- &
    -- & -- &
    $1666^{+601}_{-344}$ & $2514^{+2161}_{-2201}$ &
    14.1 & 11.6 &
    (\checkmark,\checkmark) & 111 \\
    %
    {\tt A147-d19     } & 4.89& 4.89&100 & 19.2 & 13.6 &
    {\tt F2} & {\tt nessai} &1000 &
    $10.2^{+2.9 }_{-2.3 }$ & $2.18^{+0.50}_{-0.39}$ & $ 294^{+89 }_{-110}$ &
    $-0.61^{+0.22}_{-0.25}$ & $-0.03^{+0.39}_{-0.50}$ &
    $-0.51^{+0.16}_{-0.17}$ & -- &
    $ 688^{+724}_{-356}$  & $2518^{+2186}_{-2192}$ &
    14.8 & 10.6 &
    (\checkmark,\checkmark) & 114 \\
    %
    {\tt A147-d19     } & 9.73& 9.73&125 & 38.7 & 27.2 &
    {\tt XPHM} & {\tt cpnest} & 1000 &
    $23.6 ^{+0.8 }_{-1.0 }$ & $4.60^{+0.41}_{-0.11}$ & $ 477^{+112}_{-166}$ &
    -- & -- &
    -- & -- &
    -- & -- &
    18.0 & 12.1 &
    (\xmark,\checkmark) &  193 \\
    %
    {\tt A147-d19     } & 9.73& 9.73&125 & 38.7 & 27.2 &
    {\tt XPHM} & {\tt cpnest} & 1000 &
    $20.3 ^{+1.4 }_{-3.6 }$ & $4.97^{+0.86}_{-0.20}$ & $ 227^{+32 }_{-31 }$ &
    $0.70^{+0.25}_{-0.14}$ & $0.63^{+0.34}_{-0.56}$ &
    $-0.55^{+0.15}_{-0.22}$ & $0.37^{+0.11}_{-0.06}$ &
    -- & -- &
    22.6 & 15.1 &
    (\xmark,\xmark) &  318 \\
    %
    {\tt A147-d19     } & 9.73& 9.73&125 & 38.7 & 27.2 &
    {\tt XP} & {\tt nessai} & 1000 &
    $25.3 ^{+1.8 }_{-0.9 }$ & $4.62^{+0.20}_{-0.15}$ & $ 473^{+156}_{-184}$ &
    -- & -- &
    -- & -- &
    -- & -- &
    18.6 & 11.6 &
    (\checkmark,\checkmark) &  181 \\ 
    %
    {\tt A147-d19     } & 9.73& 9.73&125 & 38.7 & 27.2 &
    {\tt XP} & {\tt nessai} & 1000 &
    $19.4^{+0.8 }_{-3.1 }$ & $4.79^{+1.31}_{-0.06}$ & $ 247^{+28 }_{-46 }$ &
    $0.96^{+0.03}_{-0.21}$ & $0.84^{+0.14}_{-0.37}$ &.
    $-0.76^{+0.16}_{-0.08}$ & $0.38^{+0.10}_{-0.06}$ &
    -- & -- & 22.6 & 16.1 &
    (\xmark,\checkmark) & 340 \\
    %
    %
    %
    {\tt A147-d19     } & 19.0& 19.0&250 & 43.4 & 30.4 &
    {\tt XP} & {\tt nessai} & 1000 &
    $26.5 ^{+2.5 }_{-1.6 }$ & $24.7^{+2.1 }_{-1.7 }$ & $1119^{+324}_{-463}$ &
    -- & -- &
    -- & -- &
    -- & -- &
    21.0 & 15.1 &
    (\xmark,\checkmark) &  284 \\
    %
    {\tt A147-d19     } & 19.0& 19.0&250 & 43.4 & 30.4 &
    {\tt XP} & {\tt nessai} & 1000 &
    $42.6^{+1.3 }_{-1.3 }$ & $19.3^{+0.7 }_{-0.7 }$ & $ 101^{+4  }_{-1  }$ &
    $0.33^{+0.05}_{-0.05}$ & $0.66^{+0.15}_{-0.12}$ &
    $0.03^{+0.07}_{-0.08}$ & $0.06^{+0.01}_{-0.01}$ &
    -- & -- & 32.3 & 19.7 &
    (\checkmark,\checkmark) & 660 \\
    %
    {\tt A147-d19     } &28.2 & 28.2&300 & 57.0 & 39.9 &
    {\tt XP} & {\tt nessai} & 1000 &
    $39.4^{+3.0}_{-2.5}$    & $37.0^{+2.7}_{-2.8}$  & $889^{+376}_{-362}$  &
    -- & -- &
    -- & -- &
    -- & -- &
    30.2 & 21.2 &
    (\xmark,\xmark) &  642 \\
    %
    {\tt A147-d19     } & 28.2& 28.2&300 & 57.0 & 39.9 &
    {\tt XP} & {\tt nessai} & 1000 &
    $68.1^{+2.7 }_{-2.4 }$ & $47.5^{+1.1 }_{-1.7 }$ & $ 103^{+9  }_{-3  }$ &
    $0.96^{+0.02}_{-0.07}$ & $0.65^{+0.05}_{-0.04}$ &
    $0.59^{+0.03}_{-0.03}$ & $0.42^{+0.03}_{-0.03}$ &
    -- & -- & 42.8 & 25.4 &
    (\checkmark,\checkmark) &1243 \\
    {\tt A147-d19} & 37.9 & 37.9 & 250 & 94.7 & 66.2 &
    {\tt XP} & {\tt nessai} & 1000 &
    $53.9_{-2.4}^{+3.4}$ & $51.4_{-2.6}^{+3.2}$ & $721_{-300}^{+198}$ &
    -- & -- & -- & -- & -- & -- & 52.6 & 35.7 &
    (\xmark,\xmark) & 2002 \\
    {\tt A147-d19} & 37.9 & 37.9 & 250 & 94.7 & 66.2 &
    {\tt XP} & {\tt nessai} & 1000 &
    $67.4_{-0.38}^{+0.42}$ & $60.2_{-0.33}^{+0.35}$ & $124_{-6}^{+5}$ &
    $0.83_{-0.01}^{+0.00}$ & $0.99_{-0.01}^{+0.00}$ & $-0.02_{-0.00}^{+0.00}$ & $0.71_{-0.01}^{+0.01}$ & -- & -- & 78.0 & 44.9 &
    (\xmark,\xmark) & 4041 \\
    %
    %
    {\tt A147-d19     } & 36.9& 36.9&400 & 61.0 & 42.3 &
    {\tt XPHM} & {\tt cpnest} & 500 &
    $83.9^{+3.5 }_{-2.7 }$ & $57.6^{+4.2 }_{-3.5 }$ & $  103^{+10  }_{-2   }$ &
    $0.57^{+0.10}_{-0.09}$ & $0.98^{+0.01}_{-0.02}$ &
    $0.49^{+0.06}_{-0.05}$ & $0.47^{+0.06}_{-0.05}$ &
    -- & -- & 47.6 & 26.8 &
    (\checkmark,\checkmark) &1560 \\
    %
    %
    %
    %
    {\tt A147-d19} & 36.2 & 36.2 & 500 & 47.3 & 33.1 &
    {\tt XP} & {\tt nessai} &1000 &
    $50.2_{-3.9}^{+5.6}$ & $44.8_{-4.8}^{+4.6}$ & $ 1350_{-533}^{+373}$ &
    -- & -- &
    -- & -- &
    -- & -- & 26.5 & 18.2 &
    (\xmark,\checkmark) & 457 \\
    {\tt A147-d19} & 36.2 & 36.2 & 500 & 47.3 & 33.1 &
    {\tt XP} & {\tt nessai} & 1000 &
    $69.1_{-0.9}^{+0.8}$ & $64.7_{-0.9}^{+0.8}$ & $116_{-12}^{+15}$ &
    $0.97_{-0.05}^{+0.02}$ & $0.59_{-0.03}^{+0.02}$ & $0.22_{-0.01}^{+0.01}$ & $0.80_{-0.05}^{+0.04}$ & -- & -- & 38.1 & 23.7 &
    (\checkmark,\checkmark) & 919 \\
    {\tt A147-d19} & 36.2 & 36.2 & 500 & 47.3 & 33.1 &
    {\tt D} & {\tt nessai} & 1000 &
    $45.6_{-2.3}^{+3.3}$ & $43.2_{-2.4}^{+3.0}$ & $826_{-358}^{+245}$ &
    $-0.97_{-0.01}^{+0.03}$ & $-0.97_{-0.02}^{+0.03}$ &
    $-0.97_{-0.01}^{+0.02}$ & -- &
    -- & -- & 31.5 & 19.8 &
    (\checkmark,\checkmark) & 671 \\
    %
    {\tt A147-d19     } & 36.2& 36.2&500 & 47.3 & 33.1 &
    {\tt XPHM} & {\tt cpnest} &1000 &
    $64.7^{+1.0 }_{-1.0 }$ & $36.8^{+1.7 }_{-1.6 }$ & $  247^{+35  }_{-36  }$ &
    $ 0.98^{+0.01}_{-0.02}$ & $0.96^{+0.02}_{-0.08}$ &
    $-0.42^{+0.04}_{-0.05}$ & $0.41^{+0.06}_{-0.06}$ &
    -- & -- & 41.0 & 21.7 &
    (\checkmark,\checkmark) & 955 \\
    %
    {\tt A147-d19     } & 44.3& 44.3&625 & 48.4 & 33.8 &
    {\tt XP} & {\tt nessai} &1000 &
    $64.2^{+6.4}_{-5.0}$   & $59.2^{+5.6}_{-5.2}$   & $1345^{+509}_{-539}$    &
    -- & -- &
    -- & -- &
    -- & -- & 30.5 & 21.0 &
    (\checkmark,\checkmark) & 644 \\
    %
    {\tt A147-d19     } & 44.3& 44.3&625 & 48.4 & 33.8 &
    {\tt XP} & {\tt nessai} &1000 &
    $96.1^{+1.4 }_{-2.4 }$ & $45.5^{+2.4 }_{-1.8 }$ & $  106^{+13  }_{-6   }$ &
    $0.98^{+0.01}_{-0.04}$ & $0.96^{+0.02}_{-0.07}$ &
    $-0.64^{0.03}_{-0.03}$ & $0.53^{+0.06}_{-0.06}$ &
    -- & -- & 41.0 & 20.7 &
    (\checkmark,\checkmark) &1017 \\
    %
    {\tt A147-d19     } & 52.1& 52.1&750 & 48.6 & 34.0 &
    {\tt XP} & {\tt nessai} &1000 &
    $76.7^{+6.6}_{-5.8}$   & $71.6^{+6.3}_{-5.6}$   & $1305^{+560}_{-492}$    &
    -- & -- &
    -- & -- &
    -- & -- & 33.1 & 23.0 &
    (\checkmark,\checkmark) & 753 \\
    %
    {\tt A147-d19     } & 52.1 & 52.1 &750 & 48.6 & 34.0 &
    {\tt XP} & {\tt nessai} &1000 &
    $96.9^{+0.8 }_{-1.3 }$ & $86.3^{+1.1 }_{-1.3 }$ & $  102^{+6   }_{-2   }$ &
    $0.98^{+0.01}_{-0.04}$ & $0.97^{+0.02}_{-0.04}$ &
    $-0.51^{0.02}_{-0.03}$ & $0.85^{+0.02}_{-0.03}$ &
    -- & -- & 40.5 & 24.1 &
    (\checkmark,\checkmark) &1047 \\
    \hline
    \hline \hline
    \end{tabular}
\end{sidewaystable}
\endgroup
\begin{sidewaystable}[t]
\centering
    \begin{tabular}{l|ccccc|ccc|ccccccccccc|cc}

    \hline \hline
    \multicolumn{1}{c}{} &
    \multicolumn{5}{|c}{Injection} &
    \multicolumn{3}{|c}{Analysis} &
    \multicolumn{11}{|c}{Recovery} &
    \multicolumn{2}{|c}{Consistency} \\

    \hline
    Family & $\frac{m_1}{M_{\odot}}$ & $\frac{m_2}{M_{\odot}}$ & $\frac{d_L}{\rm Mpc}$ & $\rho^{\rm H1}$ & $\rho^{\rm L1}$ &
    Appr. & Sampler & $N_s$ &
    $\frac{m_1}{M_{\odot}}$ & $\frac{m_2}{M_{\odot}}$ & $\frac{d_L}{\rm Mpc}$ & $a_1$ & $a_2$ & $\chi_{\rm eff}$ & $\chi_{\rm p}$ & $\Lambda_1$ & $\Lambda_2$ & $\rho^{\rm H1}$ & $\rho^{\rm L1}$ &
    Res.(H,L) & $\log(\mathcal{B}^{\rm{S}}_{\rm{N}})$ \\
    \hline
    %
    %
    %
    %
    {\tt A17-d12} & 2.48 & 2.48 & 31.25 & 16.5 & 11.8 &
    {\tt Pv2T} & {\tt nessai} & 1000 &
    $2.75_{-0.31}^{+0.67}$ & $2.19_{-0.41}^{+0.25}$ & $55.0_{-21.6}^{+20.2}$ &
    $0.57_{-0.48}^{+0.37}$ & $0.53_{-0.47}^{+0.41}$ & $0.30_{-0.24}^{+0.23}$ & $0.44_{-0.29}^{+0.34}$ & $79.8_{-73.2}^{+257}$ & $175_{-158}^{+357}$ & 16.3 & 12.9 &
    (\checkmark,\checkmark) & 154 \\
    {\tt A17-d12} & 2.48 & 2.48 & 31.25 & 16.3 & 11.5 &
    {\tt XP} & {\tt nessai} & 1000 &
    $2.80_{-0.39}^{+0.70}$ & $1.98_{-0.46}^{+0.35}$ & $52.2_{-22.2}^{+17.0}$ &
    -- & -- & -- & -- & -- & -- & 16.4 & 11.2 &
    (\checkmark,\checkmark) & 151 \\
    {\tt A17-d12} & 2.48 & 2.48 & 31.25 & 16.3 & 11.5 &
    {\tt XP} & {\tt nessai} & 1000 &
    $3.33_{-0.85}^{+1.48}$ & $1.76_{-0.49}^{+0.58}$ & $48.2_{-16.9}^{+19.4}$ &
    $0.52_{-0.43}^{+0.41}$ & $0.52_{-0.45}^{+0.42}$ & $0.13_{-0.20}^{+0.22}$ & $0.49_{-0.34}^{+0.41}$ & -- & -- & 15.2 & 10.4 &
    (\checkmark,\checkmark) & 153 \\ 
    {\tt A17-d12} & 4.93 & 4.93 & 62.5 & 24.2 & 17.2 &
    {\tt F2} & {\tt nessai} & 1000 &
    $10.7_{-4.9}^{+3.0}$ & $2.44_{-0.56}^{+2.01}$ & $126_{-41}^{+35}$ &
    $0.41_{-0.09}^{+0.17}$ & $0.04_{-0.45}^{+0.54}$ & $0.35_{-0.06}^{+0.07}$ & -- 
    & $0.80_{-0.75}^{+11.3}$ & $94.0_{-88.2}^{+618}$ & 22.3 & 17.7 &
    (\checkmark,\checkmark) & 357 \\
    {\tt A17-d12} & 4.93 & 4.93 & 62.5 & 24.2 & 17.2 &
    {\tt XP} & {\tt nessai} & 1000 &
    $7.08_{-0.79}^{+0.59}$ & $3.00_{-0.31}^{+0.48}$ & $98.7_{-38.8}^{+24.4}$ &
    -- & -- & -- & -- & -- & -- & 24.6 & 18.2 &
    (\checkmark,\checkmark) & 406 \\
    {\tt A17-d12} & 4.93 & 4.93 & 62.5 & 24.2 & 17.2 &
    {\tt XP} & {\tt nessai} & 1000 &
    $6.24_{-1.55}^{+1.84}$ & $3.34_{-0.69}^{+0.95}$ & $103.6_{-33.7}^{+19.9}$ &
    $0.21_{-0.19}^{+0.38}$ & $0.42_{-0.37}^{+0.46}$ & $-0.08_{-0.17}^{+0.18}$ & $0.23_{-0.17}^{+0.32}$ & -- & -- & 21.6 & 16.2 &
    (\checkmark,\checkmark) & 406 \\
    {\tt A17-d12} & 9.73 & 9.73 & 125 & 31.0 & 22.0 &
    {\tt XP} & {\tt nessai} & 1000 &
    $14.7_{-0.9}^{+0.9}$ & $5.43_{-0.42}^{+0.49}$ & $197_{-74}^{+45}$ &
    -- & -- & -- & -- & -- & -- & 31.7 & 22.2 & 
    (\checkmark,\checkmark) & 685 \\
    {\tt A17-d12} & 9.73 & 9.73 & 125 & 31.0 & 22.0 &
    {\tt XP} & {\tt nessai} & 1000 &
    $10.8_{-1.2}^{+1.0}$ & $7.09_{-0.59}^{+0.84}$ & $125_{-40}^{+84}$ &
    $0.51_{-0.22}^{+0.26}$ & $0.57_{-0.36}^{+0.26}$ & $-0.20_{-0.07}^{+0.07}$ & $0.40_{-0.16}^{+0.18}$ & -- & -- & 33.1 & 21.4 &
    (\checkmark,\checkmark) & 697 \\
    {\tt A17-d12} & 19.0 & 19.0 & 250 & 35.0 & 24.8 &
    {\tt TEOB} & {\tt nessai} & 1000 &
    $17.7_{-1.5}^{+2.3}$ & $14.8_{-1.8}^{+1.3}$ & $376_{-166}^{+95}$ &
    $-0.41_{-0.36}^{+0.41}$ & $-0.31_{-0.51}^{+0.41}$ & $-0.37_{-0.06}^{+0.06}$ & -- & -- & -- & 34.1 & 22.8 &
    (\checkmark,\checkmark) & 806 \\
    {\tt A17-d12} & 19.0 & 19.0 & 250 & 35.0 & 24.8 & 
    {\tt D} & {\tt nessai} & 1000 &
    $17.7_{-1.6}^{+2.5}$ & $14.5_{-1.8}^{+1.4}$ & $373_{-168}^{+101}$ &
    $-0.41_{-0.37}^{+0.37}$ & $-0.39_{-0.44}^{+0.43}$ & $-0.40_{-0.06}^{+0.06}$ & -- & -- & -- & 33.5 & 22.3 &
    (\checkmark,\checkmark) & 792 \\
    {\tt A17-d12} & 19.0 & 19.0 & 250 & 35.0 & 24.8 &
    {\tt F2} & {\tt nessai} & 1000 &
    $50.9_{-4.6}^{+3.1}$ & $6.86_{-0.37}^{+0.64}$ & $523_{-243}^{+152}$ &
    $0.39_{-0.03}^{+0.04}$ & $-0.01_{-0.48}^{+0.46}$ & $0.34_{-0.03}^{+0.03}$ & -- & $0.20_{-0.18}^{+0.35}$ & $203_{-180}^{+401}$ & 34.1 & 22.5 &
    (\checkmark,\checkmark) & 765 \\
    {\tt A17-d12} & 19.0 & 19.0 & 250 & 35.0 & 24.8 &
    {\tt XP} & {\tt nessai} & 1000 &
    $29.9_{-4.1}^{+1.9}$ & $9.57_{-1.34}^{+0.92}$ & $410_{-213}^{+917}$ &
    -- & -- & -- & -- & -- & -- & 32.2 & 21.7 &
    (\checkmark,\checkmark) & 762 \\
    {\tt A17-d12} & 19.0 & 19.0 & 250 & 35.0 & 24.8 &
    {\tt XP} & {\tt nessai} & 1000 &
    $18.9_{-5.85}^{+2.04}$ & $13.7_{-4.4}^{+1.5}$ & $302_{-125}^{+3203}$ &
    $0.56_{-0.24}^{+0.28}$ & $0.56_{-0.33}^{+0.28}$ & $-0.35_{-0.08}^{+0.07}$ & $0.35_{-0.15}^{+0.16}$ & -- & -- & 33.5 & 23.5 &
    (\checkmark,\checkmark) & 796 \\
    {\tt A17-d12} & 29.5 & 29.5 & 75 & 171.6 & 122.0 &
    {\tt XPHM} & {\tt dynesty} & 500 &
    $30.5_{-0.07}^{+0.06}$ & $26.4_{-0.05}^{+0.08}$ & $58.9_{-2.7}^{+2.9}$ &
    $0.63_{-0.01}^{+0.01}$ & $0.98_{-0.02}^{+0.01}$ & $-0.12_{-0.00}^{+0.00}$ & $0.15_{-0.01}^{+0.01}$ & -- & -- & 170.0 & 115.9 &
    (\checkmark,\checkmark) & 21637 \\
    {\tt A17-d12} & 29.0 & 29.0 & 150 & 85.8 & 61.0 &
    {\tt XP} & {\tt nessai} & 1000 & 
    $30.2_{-0.14}^{+0.13}$ & $26.0_{-0.09}^{+0.14}$ & $116_{-10}^{+11}$ &
    $0.61_{-0.02}^{+0.02}$ & $0.97_{-0.04}^{+0.02}$ & $-0.12_{-0.01}^{+0.01}$ & $0.15_{-0.01}^{+0.02}$ & -- & -- & 86.2 & 60.5 &
    (\checkmark,\checkmark) & 5413 \\
    {\tt A17-d12} & 28.7 & 28.7 & 200 & 64.3 & 45.7 &
    {\tt XP} & {\tt nessai} & 1000 & 
    $30.0_{-0.23}^{+0.24}$ & $25.9_{-0.18}^{+0.25}$ & $147_{-23}^{+25}$ &
    $0.61_{-0.03}^{+0.02}$ & $0.96_{-0.06}^{+0.02}$ & $-0.12_{-0.01}^{+0.01}$ & $0.15_{-0.02}^{+0.02}$ & -- & -- & 64.0 & 45.2 &
    (\checkmark,\checkmark) & 3032 \\
    {\tt A17-d12} & 27.8 & 27.8 & 375 & 34.3 & 24.4 &
    {\tt XP} & {\tt nessai} & 1000 &
    $33.5_{-3.5}^{+1.9}$ & $19.6_{-1.0}^{+2.9}$ & $665_{-237}^{+159}$ & 
    -- & -- & -- & -- & -- & -- & 32.9 & 23.7 & 
    (\checkmark,\checkmark) & 794 \\
    {\tt A17-d12} & 27.8 & 27.8 & 375 & 34.3 & 24.4 &
    {\tt XP} & {\tt nessai} & 1000 & 
    $28.3_{-0.9}^{+1.1}$ & $23.8_{-1.02}^{+0.87}$ & $525_{-128}^{+185}$ &
    $0.81_{-0.19}^{+0.15}$ & $0.91_{-0.17}^{+0.07}$ & $-0.09_{-0.02}^{+0.02}$ & $0.74_{-0.11}^{+0.12}$ & -- & -- & 33.8 & 24.4 & 
    (\checkmark,\checkmark) & 834 \\
    {\tt A17-d12} & 27.8 & 27.8 & 375 & 34.3 & 24.4 &
    {\tt XPHM} & {\tt dynesty} & 500 &
    $32.8_{-2.2}^{+1.5}$ & $19.8_{-1.2}^{+2.2}$ & $690_{-154}^{+122}$ &
    -- & -- & -- & -- & -- & -- & 32.5 & 23.2 &
    (\checkmark,\checkmark) & 794 \\ 
    {\tt A17-d12} & 27.8 & 27.8 & 375 & 34.3 & 24.4 &
    {\tt XPHM} & {\tt dynesty} & 500 &
    $28.8_{-0.8}^{+0.8}$ & $24.5_{-0.8}^{+0.8}$ & $403_{-84}^{+117}$ &
    $0.86_{-0.21}^{+0.12}$ & $0.89_{-0.19}^{+0.09}$ & $-0.12_{-0.03}^{+0.03}$ & $0.70_{-0.13}^{+0.13}$ & -- & -- & 32.3 & 23.2 &
    (\checkmark,\checkmark) & 738 \\
    {\tt A17-d12} & 27.2 & 27.2 & 500 & 25.7 & 18.3 &
    {\tt XP} & {\tt nessai} & 1000 & 
    $27.9_{-1.5}^{+1.5}$ & $23.4_{-1.4}^{+1.3}$ & $633_{-224}^{+304}$ &
    $0.72_{-0.18}^{+0.23}$ & $0.89_{-0.26}^{+0.09}$ & $-0.09_{-0.03}^{+0.04}$ & $0.61_{-0.39}^{+0.26}$ & -- & -- & 24.7 & 17.4 &
    (\checkmark,\checkmark) & 459 \\
    {\tt A17-d12} & 26.2 & 26.2 & 700 & 18.4 & 13.1 &
    {\tt XP} & {\tt nessai} & 1000 & 
    $26.1_{-2.2}^{+3.6}$ & $22.3_{-2.6}^{+2.1}$ & $985_{-310}^{+435}$ &
    $0.67_{-0.54}^{+0.29}$ & $0.61_{-0.51}^{+0.34}$ & $-0.07_{-0.04}^{+0.05}$ & $0.67_{-0.35}^{+0.25}$ & -- & -- & 19.4 & 15.1 &
    (\checkmark,\checkmark) & 224 \\
    {\tt A17-d12} & 25.8 & 25.8 & 800 & 16.1 & 11.4 &
    {\tt XP} & {\tt nessai} & 1000 &
    $25.5_{-3.0}^{+3.9}$ & $21.7_{-3.4}^{+2.2}$ & $1176_{-381}^{+719}$ &
    $0.61_{-0.51}^{+0.34}$ & $0.52_{-0.46}^{+0.41}$ & $-0.07_{-0.05}^{+0.06}$ & $0.64_{-0.36}^{+0.28}$ & -- & -- & 15.6 & 11.2 &
    (\checkmark,\checkmark) & 167 \\
    {\tt A17-d12} & 36.2 & 36.2 & 500 & 32.9 & 23.4 &
    {\tt XPHM} & {\tt cpnest} & 1000 &
    $53.2_{-8.7}^{+5.1}$ & $19.1_{-2.6}^{+6.4}$ & $896_{-123}^{+138}$ &
    -- & -- & -- & -- & -- & -- & 28.7 & 20.8 &
    (\checkmark,\checkmark) & 692 \\
    {\tt A17-d12} & 36.2 & 36.2 & 500 & 32.9 & 23.4 &
    {\tt XPHM} & {\tt cpnest} & 500 &
    $34.3_{-4.5}^{+3.6}$ & $24.5_{-3.3}^{+3.9}$ & $692_{-290}^{+142}$ &
    $0.77_{-0.34}^{+0.20}$ & $0.64_{-0.53}^{+0.32}$ & $-0.53_{-0.13}^{+0.12}$ & $0.37_{-0.24}^{+0.25}$ & -- & -- & 32.8 & 24.0 &
    (\checkmark,\checkmark) & 764 \\
    {\tt A17-d12} & 36.2 & 36.2 & 500 & 32.9 & 23.4 &
    {\tt XP} & {\tt nessai} & 1000 &
    $56.6_{-3.8}^{+3.9}$ & $18.4_{-1.4}^{+1.9}$ & $740_{-281}^{+188}$ &
    -- & -- & -- & -- & -- & -- & 31.4 & 22.5 &
    (\checkmark,\checkmark) & 717 \\
    {\tt A17-d12} & 36.2 & 36.2 & 500 & 32.9 & 23.4 &
    {\tt XP} & {\tt nessai} & 1000 &
    $35.5^{+2.8}_{-3.0}$ & $25.5^{+2.3}_{-2.0}$ & $838_{-240}^{+113}$ &
    $0.33_{-0.18}^{+0.26}$ & $0.78_{-0.24}^{+0.14}$ & $-0.27_{-0.11}^{+0.09}$ & $0.46_{-0.18}^{+0.15}$ & -- & -- & 32.7 & 22.5 &
    (\checkmark,\checkmark) & 792 \\
    {\tt A17-d12} & 36.2 & 36.2 & 500 & 32.9 & 23.4 &
    {\tt Pv3} & {\tt nessai} & 1000 &
    $58.3_{-3.7}^{+3.9}$ & $17.8_{-1.4}^{+1.7}$ & $736_{-284}^{+185}$ &
    -- & -- & -- & -- & -- & -- & 31.4 & 21.9 &
    (\checkmark,\checkmark) & 706 \\
    {\tt A17-d12} & 36.2 & 36.2 & 500 & 32.9 & 23.4 &
    {\tt Pv3} & {\tt nessai} & 1000 &
    $35.3_{-2.5}^{+2.8}$ & $26.3_{-2.0}^{+2.1}$ & $811_{-214}^{+119}$ &
    $0.29_{-0.14}^{+0.22}$ & $0.42_{-0.25}^{+0.38}$ & $-0.21_{-0.14}^{+0.07}$ & $0.27_{-0.17}^{+0.23}$ & -- & -- & 34.8 & 24.1 &
    (\checkmark,\checkmark) & 792 \\
    {\tt A17-d12} & 44.3 & 44.3 & 625 & 30.9 & 22.1 &
    {\tt XP} & {\tt nessai} & 1000 &
    $44.3_{-3.1}^{+4.7}$ & $40.0_{-3.8}^{+3.6}$ & $1192_{-469}^{+303}$ &
    -- & -- & -- & -- & -- & -- & 27.2 & 19.6 &
    (\checkmark,\checkmark) & 625 \\
    {\tt A17-d12} & 44.3 & 44.3 & 625 & 30.9 & 22.1 &
    {\tt XP} & {\tt nessai} & 1000 &
    $44.7_{-5.3}^{+3.9}$ & $25.6_{-4.1}^{+7.2}$ & $780_{-250}^{+199}$ &
    $0.88_{-0.29}^{+0.10}$ & $0.60_{-0.54}^{+0.37}$ & $-0.60_{-0.14}^{+0.15}$ & $0.26_{-0.18}^{+0.34}$ & -- & -- & 29.2 & 22.2 &
    (\checkmark,\checkmark) & 680 \\
    {\tt A17-d12} & 44.3 & 44.3 & 625 & 30.9 & 22.1 &
    {\tt XPHM} & {\tt dynesty} & 500 &
    $50.4_{-3.6}^{+3.5}$ & $35.5_{-3.2}^{+3.5}$ & $1041_{-257}^{+253}$ &
    -- & -- & -- & -- & -- & -- & 27.2 & 19.6 &
    (\checkmark,\checkmark) & 633 \\
    {\tt A17-d12} & 44.3 & 44.3 & 625 & 30.9 & 22.1 &
    {\tt XPHM} & {\tt dynesty} & 500 &
    $44.3_{-3.1}^{+3.3}$ & $26.1_{-2.9}^{+4.0}$ & $830_{-152}^{+146}$ &
    $0.88_{-0.22}^{+0.10}$ & $0.33_{-0.30}^{+0.59}$ & $-0.56_{-0.11}^{+0.11}$ & $0.28_{-0.19}^{+0.32}$ & -- & -- & 28.8 & 20.4 &
    (\checkmark,\checkmark) & 681 \\
    {\tt A17-d12} & 52.1 & 52.1 & 750 & 29.3 & 20.9 &
    {\tt XP} & {\tt nessai} & 1000 &
    $61.0_{-2.9}^{+4.6}$ & $55.6_{-4.8}^{+2.8}$ & $398_{-81}^{+79}$ &
    -- & -- & -- & -- & -- & -- & 28.1 & 21.7 &
    (\checkmark,\checkmark) & 543 \\
    {\tt A17-d12} & 52.1 & 52.1 & 750 & 29.3 & 20.9 &
    {\tt XP} & {\tt nessai} & 1000 &
    $54.9_{-6.0}^{+5.6}$ & $39.3_{-7.5}^{+6.6}$ & $260_{-56}^{+61}$ &
    $0.81_{-0.37}^{+0.16}$ & $0.45_{-0.38}^{+0.41}$ & $-0.52_{-0.17}^{+0.15}$ & $0.26_{-0.17}^{+0.23}$ & -- & -- & 28.4 & 22.4 &
    (\checkmark,\checkmark) & 592 \\
    \hline
    %
    %
    %
    %
    {\tt A17-d14} & 2.48 & 2.48 & 31.25 & 19.6 & 13.9 &
    {\tt Pv2T} & {\tt nessai} & 1000 &
    $2.68_{-0.31}^{+0.73}$ & $2.05_{-0.42}^{+0.27}$ &
    $54.3_{-17.7}^{+15.0}$ &
    $0.44_{-0.39}^{+0.45}$ & $0.52_{-0.46}^{+0.41}$ & $0.12_{-0.17}^{+0.19}$ & $0.48_{-0.30}^{+0.36}$ & $84.8_{-78.9}^{+296}$ & $233_{-210}^{+482}$ & 17.9 & 13.4 &
    (\checkmark,\checkmark) & 205 \\
    %
    %
    %
    %
    {\tt A17-d14} & 4.93 & 4.93 & 62.5 & 27.8 & 19.8 &
    {\tt F2} & {\tt nessai} & 1000 &
    $11.2_{-3.3}^{+1.7}$ & $2.27_{-0.29}^{+0.88}$ &
    $111_{-34}^{+31}$ &
    $0.38_{-0.07}^{+0.06}$ & $0.02_{-0.43}^{+0.50}$ & $0.32_{-0.06}^{+0.05}$ & -- & $0.39_{-0.35}^{+1.28}$ & $74.9_{-70.4}^{+239.9}$ & 26.1 & 21.4 &
    (\checkmark,\checkmark) & 528 \\
    {\tt A17-d14} & 4.93 & 4.93 & 62.5 & 27.8 & 19.8 &
    {\tt XP} & {\tt nessai} & 1000 &
    $7.19_{-0.44}^{+0.42}$ & $2.98_{-0.21}^{+0.24}$ &
    $95.9_{-33.6}^{+23.5}$ &
    -- & -- & -- & -- & -- & -- & 27.5 & 20.4 &
    (\checkmark,\checkmark) & 536 \\ 
    {\tt A17-d14} & 4.93 & 4.93 & 62.5 & 27.8 & 19.8 &
    {\tt XP} & {\tt nessai} & 1000 &
    $5.00_{-0.30}^{+1.45}$ & $4.15_{-0.89}^{+0.20}$ &
    $97.0_{-22.6}^{+27.7}$ &
    $0.61_{-0.47}^{+0.30}$ & $0.83_{-0.71}^{+0.15}$ & $-0.19_{-0.07}^{+0.12}$ & $0.66_{-0.51}^{+0.16}$ & -- & -- & 22.0 & 17.7 &
    (\checkmark,\checkmark) & 532 \\
    {\tt A17-d14} & 19.0 & 19.0 & 250 & 37.8 & 27.0 &
    {\tt F2} & {\tt nessai} & 1000 &
    $49.1_{-2.1}^{+2.4}$ & $7.09_{-0.35}^{+0.34}$ &
    $479_{-146}^{+109}$ &
    $0.37_{-0.04}^{+0.04}$ & $0.01_{-0.51}^{+0.55}$ & $0.32_{-0.03}^{+0.04}$ & -- & $0.03_{-0.03}^{+0.05}$ & $21.0_{-18.7}^{+48.2}$ & 36.7 & 27.5 &
    (\checkmark,\checkmark) & 1004 \\
    {\tt A17-d14} & 19.0 & 19.0 & 250 & 37.8 & 27.0 &
    {\tt XP} & {\tt nessai} & 1000 &
    $29.8_{-1.1}^{+1.1}$ & 
    $10.1_{-0.4}^{+0.5}$ & $362_{-120}^{+85}$ & -- & -- & -- & -- & -- & -- & 34.5 & 26.7 &
    (\checkmark,\checkmark) & 994 \\
    {\tt A17-d14} & 19.0 & 19.0 & 250 & 37.8 & 27.0 &
    {\tt XP} & {\tt nessai} & 1000 &
    $18.3_{-2.4}^{+1.0}$ & $15.4_{-0.8}^{+2.0}$ & $415_{-61}^{+48}$ &
    $0.33_{-0.16}^{+0.30}$ & $0.65_{-0.32}^{+0.27}$ & $-0.20_{-0.08}^{+0.04}$ & $0.49_{-0.24}^{+0.24}$ & -- & -- & 36.8 & 28.0 &
    (\checkmark,\checkmark) & 1038 \\
    {\tt A17-d14} & 36.2 & 36.2 & 500 & 33.6 & 24.2 &
    {\tt XPHM} & {\tt dynesty} & 750 &
    $51.2_{-3.8}^{+3.9}$ & $21.0_{-2.3}^{+2.6}$ & $884_{-113}^{+100}$ &
    -- & -- & -- & -- & -- & -- & 32.5 & 22.7 &
    (\checkmark,\checkmark) & 737 \\
    {\tt A17-d14} & 36.2 & 36.2 & 500 & 33.6 & 24.2 &
    {\tt XPHM} & {\tt dynesty} & 750 &
    $37.0_{-2.6}^{+2.9}$ & $22.3_{-2.3}^{+2.8}$ & $698_{-100}^{+85}$ &
    $0.88_{-0.20}^{+0.10}$ & $0.53_{-0.49}^{+0.44}$ & $-0.53_{-0.09}^{+0.10}$ & $0.33_{-0.23}^{+0.24}$ & -- & -- & 34.3 & 24.1 &
    (\checkmark,\checkmark) & 805 \\
    {\tt A17-d14} & 36.2 & 36.2 & 500 & 33.6 & 24.2 &
    {\tt XP} & {\tt nessai} & 1000 &
    $58.8_{-8.1}^{+4.9}$ & $17.8_{-2.0}^{+4.2}$ & $710_{-255}^{+208}$ &
    -- & -- & -- & -- & -- & -- & 30.5 & 23.1 &
    (\checkmark,\checkmark) & 678 \\
    {\tt A17-d14} & 36.2 & 36.2 & 500 & 33.6 & 24.2 &
    {\tt XP} & {\tt nessai} & 1000 &
    $35.3_{-2.7}^{+2.3}$ & $25.7_{-1.9}^{+2.3}$ & $727_{-218}^{+116}$ &
    $0.5_{-0.21}^{+0.24}$ & $0.72_{-0.30}^{+0.23}$ & $-0.33_{-0.13}^{+0.08}$ & $0.47_{-0.21}^{+0.18}$ & -- & -- & 34.1 & 24.2 &
    (\checkmark,\checkmark) & 866 \\
    {\tt A17-d14} & 36.2 & 36.2 & 500 & 33.9 & 23.7 &
    {\tt TEOB} & {\tt nessai} & 1000 &
    $35.6_{-3.9}^{+3.3}$ & $24.0_{-2.3}^{+3.5}$ & $662_{-235}^{+150}$ &
    $-0.73_{-0.19}^{+0.36}$ & $-0.24_{-0.56}^{+0.29}$ & $-0.53_{-0.05}^{+0.06}$ & -- & -- & -- & 34.1 & 24.2 &
    (\checkmark,\checkmark) & 808 \\
    \hline \hline
    \end{tabular}
\end{sidewaystable}
\endgroup

\begingroup
\squeezetable
\begin{sidewaystable}
\centering

    \begin{tabular}{l|ccccc|ccc|ccccccccccc|cc}

    \hline \hline
    \multicolumn{1}{c}{} &
    \multicolumn{5}{|c}{Injection} &
    \multicolumn{3}{|c}{Analysis} &
    \multicolumn{11}{|c}{Recovery} &
    \multicolumn{2}{|c}{Consistency} \\

    \hline
    Family & $\frac{m_1}{M_{\odot}}$ & $\frac{m_2}{M_{\odot}}$ & $\frac{d_L}{\rm Mpc}$ & $\rho^{\rm H1}$ & $\rho^{\rm L1}$ &
    Appr. & Sampler & $N_s$ &
    $\frac{m_1}{M_{\odot}}$ & $\frac{m_2}{M_{\odot}}$ & $\frac{d_L}{\rm Mpc}$ & $a_1$ & $a_2$ & $\chi_{\rm eff}$ & $\chi_{\rm p}$ & $\Lambda_1$ & $\Lambda_2$ & $\rho^{\rm H1}$ & $\rho^{\rm L1}$ &
    Res.(H,L) & $\log(\mathcal{B}^{\rm{S}}_{\rm{N}})$ \\
    \hline
    %
    %
    {\tt A17-d15      } & 4.94& 4.94& 50 & 39.2 & 27.8 &
    {\tt F2} & {\tt nessai} &1000 &
    $4.76^{+0.33}_{-0.15}$ & $4.44^{+0.15}_{-0.29}$ & $71.9^{+34.6}_{-23.5}$ &
     -- & -- &
     -- & -- &
    $1.77^{+17.93}_{-1.63}$  & $2.20^{+26.27}_{-2.02}$ &
    36.6 & 29.2 &
    (\checkmark,\checkmark) &1023 \\
    %
    {\tt A17-d15      } & 4.94& 4.94& 50 & 38.0 & 27.8 &
    {\tt F2} & {\tt nessai} &1000 &
    $13.8^{+0.2 }_{-0.5 }$ & $1.77^{+0.07}_{-0.03}$ & $ 95.5^{+18.1}_{-31.2}$ &
    $ 0.36^{+0.04}_{-0.03}$ & $-0.04^{+0.47}_{-0.56}$ &
    $0.31^{+0.03}_{-0.03}$ & -- &
    $0.10^{+0.26}_{-0.09}$ &
    $ 107^{+302 }_{-99  }$ &
    37.4 & 26.4 &
    (\checkmark,\checkmark) & 977 \\
    %
    {\tt A17-d15      } & 4.89& 4.89& 100 & 19.6 & 13.9 &
    {\tt F2} & {\tt dynesty} & 500 &
    $4.65^{+0.17}_{-0.17}$ & $4.30^{+0.17}_{-0.15}$ & $  158^{+63  }_{-57  }$ &
    -- & -- &
    -- & -- &
    $168^{+153 }_{-151 }$ & $ 208^{+206 }_{-186 }$ &
    20.1 & 13.1 &
    (\checkmark,\checkmark) & 235 \\
    %
    {\tt A17-d15      } & 4.89& 4.89& 100 & 19.6 & 13.9 &
    {\tt F2} & {\tt dynesty} & 500 &
    $11.7^{+1.8 }_{-6.0 }$ & $1.96^{+1.70}_{-0.24}$ & $  182^{+44  }_{-66  }$ &
    $0.34^{+0.06}_{-0.25}$ & $0.01^{+0.48}_{-0.45}$ &
    $0.29^{+0.06}_{-0.20}$ & -- &
    $1.29^{+5.97}_{-1.17}$ & $ 352^{+1581}_{-340 }$ &
    19.4 & 14.0 &
    (\checkmark,\checkmark) & 235 \\
    %
    {\tt A17-d15      } & 4.89& 4.89& 100 & 19.6 & 13.9 &
    {\tt F2} & {\tt nessai} &1000 &
    $4.80^{+0.68}_{-0.29}$ & $4.18^{+0.28}_{-0.53}$ & $  171^{+55  }_{-65  }$ &
    -- & -- &
    -- & -- &
    $11.4^{+50.9}_{-10.5}$ & $ 19.6^{+86.6}_{-18.0}$ &
    18.3 & 14.8 &
    (\checkmark,\checkmark) & 225 \\
    %
    %
    {\tt A17-d15      } & 19.0& 19.0&250 & 40.8 & 28.9 &
    {\tt XPHM} & {\tt cpnest} & 500 &
    $17.9^{+2.4 }_{-1.5 }$ & $14.4^{+1.5 }_{-1.8 }$ & $ 342^{+82 }_{-118}$ &
    $0.75^{+0.21}_{-0.45}$ & $0.51^{+0.44}_{-0.44}$ &
    $-0.45^{+0.08}_{-0.06}$ & $0.34^{+0.33}_{-0.21}$ &
    -- & -- & 38.8 & 28.5 &
    (\checkmark,\checkmark) & 1103 \\
    %
    {\tt A17-d15      } & 39.1& 39.1&100 &172   & 123  &
    {\tt XPHM} & {\tt cpnest} &1000 &
    $38.9^{+1.2 }_{-1.3 }$ & $25.4^{+1.6 }_{-1.2 }$ & $ 135^{+7  }_{-10 }$ &
    $0.97^{+0.02}_{-0.10}$ & $0.94^{+0.05}_{-0.88}$ &
    $-0.62^{+0.04}_{-0.03}$ & $0.52^{+0.12}_{-0.46}$ &
    -- & -- &172   & 123  &
    (\checkmark,\checkmark) &22229\\
    %
    {\tt A17-d15      } & 38.3& 38.3&200 & 86.1 & 61.6 &
    {\tt XPHM} & {\tt cpnest} &1000 &
    $37.8^{+1.1 }_{-1.1 }$ & $25.4^{+1.1 }_{-1.2 }$ & $ 236^{+23 }_{-22 }$ &
    $0.95^{+0.03}_{-0.13}$ & $0.13^{+0.72}_{-0.12}$ &
    $-0.59^{+0.04}_{-0.05}$ & $0.14^{+0.32}_{-0.08}$ &
    -- & -- & 83.1 & 64.0 &
    (\checkmark,\checkmark) &5392 \\
    %
    {\tt A17-d15      } & 37.6& 37.6&300 & 57.4 & 41.1 &
    {\tt XPHM} & {\tt cpnest} &1000 &
    $36.5^{+2.1 }_{-1.8 }$ & $26.3^{+3.5 }_{-2.2 }$ & $ 393^{+48 }_{-52 }$ &
    $0.87^{+0.10}_{-0.23}$ & $0.57^{+0.39}_{-0.50}$ &
    $-0.51^{+0.12}_{-0.08}$ & $0.46^{+0.33}_{-0.29}$ &
    -- & -- & 54.2 & 42.7 &
    (\checkmark,\checkmark) &2356 \\
    %
    {\tt A17-d15      } & 36.9& 36.9&400 & 43.0 & 30.8 &
    {\tt XPHM} & {\tt cpnest} & 500 &
    $35.6^{+3.3 }_{-3.4 }$ & $24.7^{+3.5 }_{-2.9 }$ & $ 515^{+90 }_{-127}$ &
    $0.86^{+0.12}_{-0.30}$ & $0.55^{+0.39}_{-0.51}$ &
    $-0.57^{+0.09}_{-0.11}$ & $0.33^{+0.27}_{-0.22}$ &
    -- & -- & 42.9 & 31.3 &
    (\checkmark,\checkmark) &1294 \\
    %
    {\tt A17-d15      } & 36.9& 36.9&400 & 43.0 & 30.8 &
    {\tt XPHM} & {\tt cpnest} &1000 &
    $36.8^{+1.9 }_{-2.1 }$ & $23.9^{+2.4 }_{-1.9 }$ & $ 473^{+62 }_{-93 }$ &
    $0.90^{+0.08}_{-0.21}$ & $0.45^{+0.49}_{-0.41}$ &
    $-0.60^{+0.08}_{-0.09}$ & $0.25^{+0.30}_{-0.18}$ &
    -- & -- & 43.8 & 30.3 &
    (\checkmark,\checkmark) &1402 \\
    %
    %
    {\tt A17-d15      } & 36.2& 36.2&500 & 35.6 & 25.1 &
    {\tt XPHM} & {\tt cpnest} & 500 &
    $35.1^{+2.9 }_{-2.9 }$ & $26.0^{+3.5 }_{-3.4 }$ & $ 538^{+207}_{-172}$ &
    $0.77^{+0.19}_{-0.35}$ & $0.57^{+0.37}_{-0.50}$ &
    $-0.54^{+0.10}_{-0.10}$ & $0.32^{+0.30}_{-0.21}$ &
    -- & -- & 34.7 & 25.7 &
    (\checkmark,\checkmark) & 896 \\
    %
    {\tt A17-d15      } & 36.2& 36.2&500 & 35.6 & 25.1 &
    {\tt XPHM} & {\tt cpnest} &1000 &
    $37.7^{+4.2 }_{-3.5 }$ & $26.2^{+4.4 }_{-5.2 }$ & $ 507^{+215}_{-229}$ &
    $0.80^{+0.17}_{-0.29}$ & $0.56^{+0.40}_{-0.50}$ &
    $-0.44^{+0.11}_{-0.14}$ & $0.43^{+0.35}_{-0.27}$ &
    -- & -- & 37.3 & 27.5 &
    (\checkmark,\checkmark) & 895 \\
    %
    {\tt A17-d15      } & 29.3& 29.3&2000& 8.61 & 6.16 &
    {\tt XPHM} & {\tt cpnest} &1000 &
    $31.1^{+12.9}_{-5.8 }$ & $22.8^{+6.0 }_{-7.6 }$ & $2689^{+1372}_{-1247}$ &
    $0.50^{+0.44}_{-0.45}$ & $0.52^{+0.42}_{-0.47}$ &
    $-0.15^{+0.23}_{-0.31}$ & $0.48^{+0.40}_{-0.36}$ &
    -- & -- & 8.44 & 7.66 &
    (\checkmark,\checkmark) &  32 \\
    \hline \hline
    %
    %
    %
    %
    {\tt A17-d12-p180} & 2.48 & 2.48 & 31.25 & 16.8 & 11.8 &
    {\tt Pv2T} & {\tt nessai} & 1000 &
    $2.68_{-0.30}^{+0.77}$ & $2.14_{-0.41}^{+0.24}$ &
    $63.0_{-25.1}^{+19.1}$ &
    $0.74_{-0.50}^{+0.23}$ & $0.65_{-0.55}^{+0.31}$ & $0.45_{-0.25}^{+0.23}$ & $0.48_{-0.30}^{+0.32}$ & $109_{-101}^{+397}$ & $283_{-255}^{+574}$ & 17.9 & 11.1 &
    (\checkmark,\checkmark) & 146 \\
    {\tt A17-d12-p180} & 4.93 & 4.93 & 62.5 & 24.1 & 17.7 &
    {\tt F2} & {\tt nessai} & 1000 &
    $14.3_{-3.0}^{+1.0}$ & $2.07_{-0.15}^{+0.57}$ &
    $125_{-42}^{+30}$ &
    $0.55_{-0.05}^{+0.10}$ & $0.00_{-0.52}^{+0.49}$ & $0.48_{-0.04}^{+0.06}$ & -- & $1.22_{-1.05}^{+1.20}$ & $632_{-577}^{+1211}$ & 22.6 & 17.0 &
    (\checkmark,\checkmark) & 450 \\
    {\tt A17-d12-p180} & 4.93 & 4.93 & 62.5 & 25.2 & 17.9 &
    {\tt XP} & {\tt nessai} & 1000 &
    $4.73_{-0.18}^{+0.39}$ & $4.35_{-0.35}^{+0.17}$ &
    $85.1_{-26.9}^{+13.7}$ &
    -- & -- & -- & -- & -- & -- & 22.6 & 17.0 &
    (\checkmark,\checkmark) & 328 \\
    {\tt A17-d12-p180} & 4.93 & 4.93 & 62.5 & 25.2 & 17.9 &
    {\tt XP} & {\tt nessai} & 1000 &
    $7.43_{-0.99}^{+1.82}$ & $4.11_{-0.76}^{+0.62}$ &
    $76.3_{-24.6}^{+20.5}$ &
    $0.87_{-0.14}^{+0.10}$ & $0.86_{-0.29}^{+0.12}$ & $0.68_{-0.07}^{+0.07}$ & $0.45_{-0.20}^{+0.18}$ & -- & -- & 24.5 & 17.9 &
    (\checkmark,\checkmark) & 405 \\
    {\tt A17-d12-p180} & 36.2 & 36.2 & 500 & 35.8 & 25.3 &
    {\tt XP} & {\tt nessai} & 1000 &
    $31.6_{-1.9}^{+3.2}$ & $28.3_{-2.6}^{+2.1}$ &
    $756_{-300}^{+195}$ &
    --& -- & -- & -- & -- & -- & 29.2 & 22.1 &
    (\checkmark,\checkmark) & 792 \\
    {\tt A17-d12-p180} & 36.2 & 36.2 & 500 & 35.8 & 25.3 &
    {\tt XP} & {\tt nessai} & 1000 &
    $45.3_{-4.9}^{+7.0}$ & $29.4_{-5.2}^{+4.4}$ &
    $808_{-257}^{+272}$ &
    $0.81_{-0.25}^{+0.16}$ & $0.75_{-0.51}^{+0.21}$ & $0.51_{-0.06}^{+0.06}$ & $0.44_{-0.27}^{+0.20}$ & -- & -- & 34.5 & 25.2 &
    (\checkmark,\checkmark) & 906 \\
    {\tt A17-d12-p180} & 36.2 & 36.2 & 500 & 35.8 & 25.3 &
    {\tt XPHM} & {\tt dynesty} & 500 &
    $32.9_{-2.7}^{+2.5}$ & $28.8_{-2.6}^{+2.0}$ &
    $623_{-239}^{+325}$ &
    -- & -- & -- & -- & -- & -- & 29.9 & 21.9 &
    (\checkmark,\checkmark) & 744 \\
    {\tt A17-d12-p180} & 36.2 & 36.2 & 500 & 35.8 & 25.3 &
    {\tt XPHM} & {\tt dynesty} & 500 &
    $46.7_{-4.7}^{+4.2}$ & $24.8_{-2.2}^{+3.7}$ &
    $1168_{-154}^{+119}$ &
    $0.65_{-0.21}^{+0.17}$ & $0.63_{-0.55}^{+0.33}$ & $0.48_{-0.06}^{+0.06}$ & $0.32_{-0.19}^{+0.21}$ & -- & -- & 33.6 & 24.8 &
    (\checkmark,\checkmark) & 885 \\
    \hline 
    %
    {\tt A17-d15-p180} & 2.48& 2.48& 30 & 23.7 & 16.8 &
    {\tt XP} & {\tt nessai} & 1000 &
    $ 2.52^{+0.38}_{-0.20}$ & $ 2.11^{+0.18}_{-0.29}$ & $ 42.6^{+7.0}_{-14.8}$ &
    -- & -- &
    -- & -- &
    -- & -- & 21.9 & 16.4 &
    (\checkmark,\checkmark) &  348 \\
    %
    {\tt A17-d15-p180} & 2.48& 2.48& 30 & 23.7 & 16.8 &
    {\tt XP} & {\tt nessai} & 1000 &
    $2.94^{+0.83}_{-0.52}$ & $1.86^{+0.36}_{-0.35}$ & $41.7^{+7.9}_{-13.4}$ &
    $0.27^{+0.28}_{-0.23}$ & $0.55^{+0.39}_{-0.48}$ &
    $0.09^{+0.15}_{-0.11}$ & $0.32^{+0.27}_{-0.23}$ &
    -- & -- & 23.9 & 14.4 &
    (\checkmark,\checkmark) & 347 \\
    %
    {\tt A17-d15-p180} & 4.93& 4.93& 60 & 33.3 & 23.6 &
    {\tt XP} & {\tt nessai} & 1000 &
    $ 4.77^{+0.40}_{-0.20}$ & $ 4.34^{+0.19}_{-0.35}$ & $ 97.6^{+22.1}_{-38.5}$&
    -- & -- &
    -- & -- &
    -- & -- & 31.9 & 21.1 &
    (\checkmark,\checkmark) &  697 \\
    %
    {\tt A17-d15-p180} & 4.93& 4.93& 60 & 33.3 & 23.6 &
    {\tt XP} & {\tt nessai} & 1000 &
    $7.77^{+1.72}_{-0.92}$ & $3.05^{+0.36}_{-0.44}$ & $95.8^{+13.0}_{-26.3}$ &
    $0.75^{+0.16}_{-0.12}$ & $0.84^{+0.14}_{-0.43}$ &
    $0.31^{+0.10}_{-0.07}$ & $0.55^{+0.24}_{-0.21}$ &
    -- & -- & 31.6 & 24.5 &
    (\checkmark,\checkmark) & 800 \\
    %
    {\tt A17-d15-p180} & 9.73& 9.73&125 & 39.5 & 28.0 &
    {\tt XP} & {\tt nessai} & 1000 &
    $ 9.03^{+0.41}_{-0.20}$ & $ 8.66^{+0.21}_{-0.36}$ & $ 201^{+43}_{-73}$ &
    -- & -- &
    -- & -- &
    -- & -- & 36.8 & 28.3 &
    (\checkmark,\checkmark) &  949 \\
    %
    {\tt A17-d15-p180} & 9.73& 9.73&125 & 39.5 & 28.0 &
    {\tt XP} & {\tt nessai} & 1000 &
    $13.6^{+1.8 }_{-2.0 }$ & $6.84^{+1.08}_{-0.90}$ & $ 127^{+81}_{-14}$ &
    $0.60^{+0.17}_{-0.18}$ & $0.30^{+0.55}_{-0.26}$ &
    $0.29^{+0.05}_{-0.05}$ & $0.40^{+0.22}_{-0.19}$ &
    -- & -- & 36.9 & 28.4 &
    (\checkmark,\checkmark) &1070 \\
    %
    {\tt A17-d15-p180} & 19.0& 19.0&250 & 42.3 & 30.0 &
    {\tt XP} & {\tt nessai} & 1000 &
    $ 17.3^{+0.7 }_{-0.5 }$ & $16.6^{+0.6 }_{-0.6 }$ & $ 406^{+87}_{-166}$ &
    -- & -- &
    -- & -- &
    -- & -- & 39.3 & 26.3 &
    (\checkmark,\checkmark) & 1054 \\
    %
    {\tt A17-d15-p180} & 19.0& 19.0&250 & 42.3 & 30.0 &
    {\tt XP} & {\tt nessai} & 1000 &
    $20.0^{+0.5 }_{-0.4 }$ & $16.4^{+0.5 }_{-0.6 }$ & $ 452^{+38}_{-54}$ &
    $0.97^{+0.02}_{-0.08}$ & $0.88^{+0.10}_{-0.21}$ &
    $0.27^{+0.04}_{-0.05}$ & $0.72^{+0.15}_{-0.15}$ &
    -- & -- & 43.6 & 30.6 &
    (\checkmark,\checkmark) &1322 \\
    %
    {\tt A17-d15-p180} & 28.2& 28.2&300 & 50.4 & 35.8 &
    {\tt XP} & {\tt nessai} & 1000 &
    $ 25.4^{+1.2 }_{-0.8 }$ & $24.3^{+1.0 }_{-1.0 }$ & $ 474^{+103}_{-187}$ &
    -- & -- &
    -- & -- &
    -- & -- & 45.6 & 33.5 &
    (\checkmark,\checkmark) & 1523 \\
    %
    {\tt A17-d15-p180} & 28.2& 28.2&300 & 50.4 & 35.8 &
    {\tt XP} & {\tt nessai} & 1000 &
    $37.2^{+2.4 }_{-2.6 }$ & $20.4^{+1.9 }_{-1.8 }$ & $ 589^{+44}_{-122}$ &
    $0.68^{+0.15}_{-0.13}$ & $0.42^{+0.52}_{-0.37}$ &
    $0.38^{+0.04}_{-0.12}$ & $0.22^{+0.21}_{-0.13}$ &
    -- & -- & 50.2 & 35.1 &
    (\checkmark,\checkmark) &1860 \\
    %
    {\tt A17-d15-p180} & 36.2& 36.2& 500& 37.6 & 26.7 &
    {\tt XP} & {\tt nessai} & 1000 &
    $ 33.7^{+4.2 }_{-3.1 }$ & $27.8^{+2.9 }_{-3.2 }$ & $ 718^{+212 }_{-298 }$ &
    -- & -- &
    -- & -- &
    -- & -- & 34.2 & 23.9 &
    (\checkmark,\checkmark) &  850 \\
    %
    {\tt A17-d15-p180} & 36.2& 36.2&500 & 37.6 & 26.7 &
    {\tt XP} & {\tt nessai} & 1000 &
    $49.5^{+3.6 }_{-3.8 }$ & $30.9^{+2.7 }_{-2.5 }$ & $ 446^{+193 }_{-101 }$ &
    $0.91^{+0.07}_{-0.13}$ & $0.73^{+0.22}_{-0.26}$ &
    $0.47^{+0.04}_{-0.04}$ & $0.41^{+0.14}_{-0.13}$ &
    -- & -- & 38.9 & 28.0 &
    (\checkmark,\checkmark) &1045 \\
    %
    {\tt A17-d15-p180} & 44.3& 44.3& 625& 35.5 & 25.2 &
    {\tt XP} & {\tt nessai} & 1000 &
    $ 53.9^{+3.7 }_{-3.1 }$ & $23.4^{+2.0 }_{-1.6 }$ & $ 762^{+180 }_{-295 }$ &
    -- & -- &
    -- & -- &
    -- & -- & 34.5 & 24.2 &
    (\checkmark,\checkmark) &  870 \\
    %
    {\tt A17-d15-p180} & 44.3& 44.3&625 & 35.5 & 25.2 &
    {\tt XP} & {\tt nessai} & 1000 &
    $60.3^{+5.4 }_{-6.7 }$ & $32.7^{+5.8 }_{-3.4 }$ & $ 875^{+307 }_{-242 }$ &
    $0.71^{+0.17}_{-0.17}$ & $0.83^{+0.15}_{-0.41}$ &
    $0.43^{+0.08}_{-0.06}$ & $0.39^{+0.17}_{-0.19}$ &
    -- & -- & 37.1 & 22.8 &
    (\checkmark,\checkmark) & 940 \\
    %
    {\tt A17-d15-p180} & 52.1& 52.1& 750& 33.7 & 24.0 &
    {\tt XP} & {\tt nessai} & 1000 &
    $ 68.9^{+4.9 }_{-4.1 }$ & $22.4^{+2.0 }_{-1.7 }$ & $ 808^{+203 }_{-326 }$ &
    -- & -- &
    -- & -- &
    -- & -- & 32.6 & 23.1 &
    (\checkmark,\checkmark) &  758 \\
    %
    {\tt A17-d15-p180} & 52.1& 52.1&750 & 33.7 & 24.0 &
    {\tt XP} & {\tt nessai} & 1000 &
    $69.3^{+5.7 }_{-5.7 }$ & $32.5^{+4.9 }_{-3.6 }$ & $1112^{+238 }_{-358 }$ &
    $0.49^{+0.18}_{-0.20}$ & $0.57^{+0.37}_{-0.48}$ &
    $0.33^{+0.08}_{-0.08}$ & $0.27^{+0.19}_{-0.17}$ &
    -- & -- & 34.8 & 25.1 &
    (\checkmark,\checkmark) & 884 \\
    \hline \hline
    \end{tabular}
\end{sidewaystable}
\endgroup

\begingroup
\squeezetable
\begin{sidewaystable}
\centering

    \begin{tabular}{l|ccccc|ccc|ccccccccccc|cc}

    \hline \hline
    \multicolumn{1}{c}{} &
    \multicolumn{5}{|c}{Injection} &
    \multicolumn{3}{|c}{Analysis} &
    \multicolumn{11}{|c}{Recovery} &
    \multicolumn{2}{|c}{Consistency} \\

    \hline
    Family & $\frac{m_1}{M_{\odot}}$ & $\frac{m_2}{M_{\odot}}$ & $\frac{d_L}{\rm Mpc}$ & $\rho^{\rm H1}$ & $\rho^{\rm L1}$ &
    Appr. & Sampler & $N_s$ &
    $\frac{m_1}{M_{\odot}}$ & $\frac{m_2}{M_{\odot}}$ & $\frac{d_L}{\rm Mpc}$ & $a_1$ & $a_2$ & $\chi_{\rm eff}$ & $\chi_{\rm p}$ & $\Lambda_1$ & $\Lambda_2$ & $\rho^{\rm H1}$ & $\rho^{\rm L1}$ &
    Res.(H,L) & $\log(\mathcal{B}^{\rm{S}}_{\rm{N}})$ \\
    \hline
    %
    %
    {\tt A17-d15-p090 } & 2.48& 2.48& 30 & 23.8 & 16.9 &
    {\tt XP} & {\tt nessai} &1000 &
    $2.69^{+0.47}_{-0.33}$ & $2.00^{+0.29}_{-0.32}$ & $43.3^{+6.5}_{-14.3}$ &
    -- & -- &
    -- & -- &
    -- & -- & 22.3 & 15.7 &
    (\checkmark,\checkmark) & 342 \\
    %
    {\tt A17-d15-p090 } & 2.48& 2.48& 30 & 23.8 & 16.9 &
    {\tt XP} & {\tt nessai} &1000 &
    $3.00^{+0.57}_{-0.29}$ & $1.82^{+0.16}_{-0.26}$ & $33.6^{+12.3}_{-8.5}$ &
    $0.43^{+0.14}_{-0.12}$ & $0.79^{+0.17}_{-0.32}$ &
    $0.06^{+0.10}_{-0.08}$ & $0.45^{+0.14}_{-0.15}$ &
    -- & --- & 22.9 & 18.1 &
    (\checkmark,\checkmark) & 420 \\
    {\tt A17-d15-p090 } & 2.48 & 2.48 & 31.25 & 22.8 & 16.2 &
    {\tt Pv2T} & {\tt nessai} & 1000 &
    $2.53_{-0.15}^{+0.37}$ & $2.21_{-0.31}^{+0.15}$ & $53.6_{-19.2}^{+11.0}$ &
    $0.40_{-0.36}^{+0.48}$ & $0.48_{-0.42}^{+0.43}$ &
    $0.22_{-0.18}^{+0.22}$ & $0.40_{-0.27}^{+0.35}$ &
    $114_{-110}^{+288}$ & $190_{-175}^{+366}$ & 21.6 & 16.7 & 
    (\checkmark,\checkmark) & 342 \\
    %
    {\tt A17-d15-p090} & 4.93& 4.93& 60 & 33.2 & 23.6 &
    {\tt XP} & {\tt nessai} & 1000 &
    $ 4.77^{+0.33}_{-0.16}$ & $4.41^{+0.16}_{-0.29}$ & $96.1^{+21.6}_{-34.9}$ &
    -- & -- &
    -- & -- &
    -- & -- & 31.6 & 22.4 &
    (\checkmark,\checkmark) &  722 \\
    %
    {\tt A17-d15-p090} & 4.93& 4.93& 60 & 33.2 & 23.6 &
    {\tt XP} & {\tt nessai} & 1000 &
    $5.04^{+1.38}_{-0.11}$ & $4.28^{+0.09}_{-0.90}$ & $99.6^{+17.1}_{-24.3}$ &
    $0.86^{+0.10}_{-0.34}$ & $0.92^{+0.07}_{-0.24}$ &
    $0.04^{+0.06}_{-0.05}$ & $0.83^{+0.12}_{-0.37}$ &
    -- & -- & 32.4 & 22.3 &
    (\checkmark,\checkmark) & 750 \\
    {\tt A17-d15-p090 } & 4.93 & 4.93 & 62.5 & 31.9 & 22.7 &
    {\tt F2} & {\tt nessai} & 1000 &
    $14.1_{-0.6}^{+0.2}$ & $1.82_{-0.03}^{+0.08}$ & $104_{-32}^{+24}$ &
    $0.42_{-0.03}^{+0.04}$ & $-0.04_{-0.57}^{+0.44}$ &
    $0.37_{-0.03}^{+0.03}$ & -- &
    $0.13_{-0.12}^{+0.29}$ & $146_{-131}^{+338}$ & 31.0 & 24.3 & 
    (\checkmark,\checkmark) & 730 \\
    %
    {\tt A17-d15-p090 } & 4.93 & 4.93 & 62.5 & 31.9 & 22.7 &
    {\tt XP} & {\tt nessai} & 1000 &
    $4.82^{+0.42}_{-0.21}$ & $4.35^{+0.21}_{-0.36}$ & $100^{+23}_{-36}$ &
     -- & -- & -- & -- &
     -- & -- & 31.6 & 22.4 &.
    (\checkmark,\checkmark) & 698 \\
    %
    {\tt A17-d15-p090} & 9.73& 9.73&125 & 39.1 & 27.8 &
    {\tt XP} & {\tt nessai} & 1000 &
    $ 9.26^{+0.57}_{-0.31}$ & $8.66^{+0.31}_{-0.50}$ & $ 208^{+46}_{-85 }$ &
    -- & -- &.
    -- & -- &
    -- & -- & 38.6 & 26.4 &
    (\checkmark,\checkmark) & 1037 \\
    %
    {\tt A17-d15-p090} & 9.73& 9.73&125 & 39.1 & 27.8 &
    {\tt XP} & {\tt nessai} & 1000 &
    $9.87^{+2.56}_{-0.26}$ & $8.45^{+0.28}_{-1.60}$ & $ 206^{+26}_{-41}$ &
    $0.76^{+0.14}_{-0.25}$ & $0.77^{+0.18}_{-0.33}$ &
    $0.09^{+0.06}_{-0.03}$ & $0.74^{+0.15}_{-0.29}$ &
    -- & -- & 38.6 & 29.5 &
    (\checkmark,\checkmark) &1174 \\
    %
    {\tt A17-d15-p090} & 19.0 & 19.0 & 250 & 41.8 & 29.7 &
    {\tt XP} & {\tt nessai} & 1000 &
    $17.9^{+1.2}_{-0.7}$ & $16.7^{+0.8}_{-1.0}$ & $387^{+96}_{-174}$ &
    -- & -- &
    -- & -- &
    -- & -- & 41.0 & 28.0 &
    (\checkmark,\checkmark) & 1200 \\
    %
    {\tt A17-d15-p090} & 19.0 & 19.0 & 250 & 41.8 & 29.7 &
    {\tt XP} & {\tt nessai} & 1000 &
    $24.4^{+1.9}_{-2.3}$ & $13.0^{+1.4}_{-0.9}$ & $389^{+49}_{-93}$ &
    $0.38^{+0.28}_{-0.19}$ & $0.56^{+0.37}_{-0.45}$ &
    $0.12^{+0.05}_{-0.05}$ & $0.32^{+0.25}_{-0.17}$ &
    -- & -- & 39.8 & 31.6 &
    (\checkmark,\checkmark) & 1309 \\
    %
    {\tt A17-d15-p090} & 28.2 & 28.2 & 300 & 49.3 & 35.0 &
    {\tt XP} & {\tt nessai} & 1000 &
    $26.4^{+1.7}_{-1.1}$ & $24.7^{+1.2}_{-1.5}$ & $445^{+115}_{-201}$ &
    -- & -- &
    -- & -- &
    -- & -- & 48.5 & 34.9 &
    (\checkmark,\checkmark) & 1698 \\
    %
    {\tt A17-d15-p090} & 28.2 & 28.2 & 300 & 49.3 & 35.0 &
    {\tt XP} & {\tt nessai} & 1000 &
    $38.3^{+2.9}_{-3.1}$ & $18.0^{+1.7}_{-1.4}$ & $529^{+58}_{-131}$ &
    $0.28^{+0.19}_{-0.19}$ & $0.63^{+0.32}_{-0.52}$ &
    $0.19^{+0.05}_{-0.05}$ & $0.26^{+0.16}_{-0.15}$ &
    -- & -- & 46.0 & 34.2 &
    (\checkmark,\checkmark) & 1599 \\
    %
    {\tt A17-d15-p090} & 36.2 & 36.2 & 500 & 36.8 & 26.2 &
    {\tt XP} & {\tt nessai} & 1000 &
    $39.1^{+3.4}_{-4.4}$ & $26.3^{+3.7}_{-2.3}$ & $710^{+156}_{-291}$ &
    -- & -- &
    -- & -- &
    -- & -- & 37.8 & 27.5 &
    (\checkmark,\checkmark) & 1019 \\
    %
    {\tt A17-d15-p090} & 36.2 & 36.2 & 500 & 36.8 & 26.2 &
    {\tt XP} & {\tt nessai} & 1000 &
    $40.6^{+5.4}_{-5.0}$ & $30.1^{+4.6}_{-4.0}$ & $935^{+111}_{-275}$ &
    $0.47^{+0.32}_{-0.39}$ & $0.49^{+0.42}_{-0.43}$ &
    $0.29^{+0.05}_{-0.05}$ & $0.34^{+0.31}_{-0.21}$ &
    -- & -- & 37.4 & 27.6 &
    (\checkmark,\checkmark) & 1010 \\
    %
    {\tt A17-d15-p090} & 36.2& 36.2&500 & 36.8 & 26.2 &
    {\tt XPHM} & {\tt cpnest} &  500 &
    $38.9^{+7.5 }_{-3.1 }$ & $33.2^{+4.1 }_{-8.6 }$ & $ 851^{+292 }_{-382 }$ &
    $0.43^{+0.38}_{-0.37}$ & $0.55^{+0.38}_{-0.44}$ &
    $0.30^{+0.06}_{-0.09}$ & $0.34^{+0.40}_{-0.20}$ &
    -- & -- & 34.6 & 23.9 &
    (\checkmark,\checkmark) & 952 \\
    %
    {\tt A17-d15-p090} & 36.2& 36.2&500 & 36.8 & 26.2 &
    {\tt XPHM} & {\tt cpnest} & 1000 &
    $41.2^{+7.0 }_{-5.8 }$ & $29.0^{+6.4 }_{-5.4 }$ & $ 818^{+215 }_{-285 }$ &
    $0.41^{+0.36}_{-0.35}$ & $0.54^{+0.38}_{-0.42}$ &
    $0.24^{+0.09}_{-0.08}$ & $0.36^{+0.30}_{-0.23}$ &
    -- & -- & 39.5 & 29.7 &
    (\checkmark,\checkmark) &1047 \\
    %
    {\tt A17-d15-p090} & 44.3 & 44.3 & 625 & 34.7 & 24.7 &
    {\tt XP} & {\tt nessai} & 1000 &
    $54.0^{+4.1}_{-4.0}$ & $26.1^{+2.6}_{-2.1}$ & $824^{+202}_{-307}$ &
    -- & -- &
    -- & -- &
    -- & -- & 34.0 & 23.2 &
    (\checkmark,\checkmark) &  802 \\
    %
    {\tt A17-d15-p090} & 44.3 & 44.3 & 625 & 34.7 & 24.7 &
    {\tt XP} & {\tt nessai} & 1000 &
    $50.0^{+6.3}_{-6.0}$ & $37.3^{+5.2}_{-5.4}$ & $ 932^{+255}_{-322}$ &
    $0.54^{+0.32}_{-0.39}$ & $0.56^{+0.37}_{-0.46}$ &
    $0.27^{+0.07}_{-0.07}$ & $0.43^{+0.29}_{-0.26}$ &
    -- & -- & 34.7 & 25.8 &
    (\checkmark,\checkmark) & 936 \\
    %
    {\tt A17-d15-p090} & 52.1 & 52.1 & 750 & 33.0 & 23.5 &
    {\tt XP} & {\tt nessai} & 1000 &
    $69.1^{+4.7}_{-3.9}$ & $26.0^{+2.4}_{-2.0}$ & $844^{+203}_{-323}$ &
    -- & -- &
    -- & -- &
    -- & -- & 33.8 & 24.2 &
    (\checkmark,\checkmark) &  817 \\
    %
    {\tt A17-d15-p090} & 52.1 & 52.1 & 750 & 33.0 & 23.5 &
    {\tt XP} & {\tt nessai} & 1000 &
    $59.1^{+7.1}_{-8.5}$ & $36.5^{+9.1}_{-6.0}$ & $1230^{+269}_{-423}$ &
    $0.40^{+0.34}_{-0.32}$ & $0.47^{+0.44}_{-0.42}$ &.
    $0.21^{+0.09}_{-0.10}$ & $0.32^{+0.31}_{-0.21}$ &
    -- & -- & 32.2 & 24.3 &
    (\checkmark,\checkmark) & 759 \\
    \hline
    %
    %
    %
    %
    {\tt A17-d12-e1} & 2.48 & 2.48 & 31.25 & 16.4 & 11.6 & 
    {\tt Pv2T} & {\tt nessai} & 1000 &
    $2.62_{-0.19}^{+0.36}$ & $2.32_{-0.30}^{+0.18}$ &
    $61.0_{-24}^{+14}$ &
    $0.47_{-0.41}^{+0.43}$ & $0.49_{-0.43}^{+0.43}$ & $0.26_{-0.19}^{+0.18}$ & $0.41_{-0.28}^{+0.37}$ & $25.1_{-23.0}^{+76.0}$ & $39.1_{-35.4}^{+100.0}$ & 16.9 & 11.1 &
    (\checkmark,\checkmark) & 177 \\
    {\tt A17-d12-e1} & 2.49 & 2.49 & 17.0 & 30.2 & 21.4 & 
    {\tt XP} & {\tt nessai} & 1000 &
    $3.28_{-0.65}^{+1.25}$ & $1.94_{-0.43}^{+0.45}$ &
    $30.3_{-10.2}^{+7.0}$ &
    $0.60_{-0.37}^{+0.29}$ & $0.63_{-0.51}^{+0.31}$ & $0.30_{-0.13}^{+0.15}$ & $0.49_{-0.25}^{+0.32}$ & -- & -- & 30.1 & 21.7 & 
    (\checkmark,\checkmark) & 646 \\ 
    {\tt A17-d12-e1} & 2.48 & 2.48 & 31.25 & 16.4 & 11.6 & 
    {\tt XP} & {\tt nessai} & 1000 &
    $2.66_{-0.26}^{+0.49}$ & $2.07_{-0.35}^{+0.23}$ &
    $49.0_{-21.2}^{+15.6}$ &
    -- & -- & -- & -- & -- & -- & 16.5 & 11.9 & 
    (\checkmark,\checkmark) & 174 \\ 
    {\tt A17-d12-e1} & 2.48 & 2.48 & 31.25 & 16.4 & 11.6 & 
    {\tt XP} & {\tt nessai} & 1000 &
    $3.32_{-0.79}^{+2.24}$ & $1.88_{-0.62}^{+0.52}$ &
    $54.5_{-19.8}^{+14.5}$ &
    $0.68_{-0.51}^{+0.27}$ & $0.52_{-0.46}^{+0.42}$ & $0.29_{-0.23}^{+0.24}$ & $0.52_{-0.31}^{+0.34}$ & -- & -- & 17.1 & 12.1 & 
    (\checkmark,\checkmark) & 177 \\
    {\tt A17-d12-e1} & 4.93 & 4.93 & 62.5 & 24.6 & 17.4 &
    {\tt F2} & {\tt nessai} & 1000 &
    $13.4_{-2.2}^{+0.9}$ & $1.95_{-0.14}^{+0.40}$ &
    $118_{-44}^{+34}$ &
    $0.43_{-0.04}^{+0.06}$ & $0.00_{-0.52}^{+0.53}$ & $0.37_{-0.04}^{+0.04}$ & -- & $0.31_{-0.28}^{+0.60}$ & $205_{-187}^{+593}$ & 23.4 & 18.1 &
    (\checkmark,\checkmark) & 376 \\
    {\tt A17-d12-e1} & 4.93 & 4.93 & 62.5 & 24.6 & 17.4 & 
    {\tt XP} & {\tt nessai} & 1000 &
    $4.98_{-0.30}^{+0.65}$ & $4.33_{-0.53}^{+0.28}$ &
    $81.8_{-29.0}^{+16.4}$ &
    -- & -- & -- & -- & -- & -- & 24.4 & 15.8 &
    (\checkmark,\checkmark) & 407 \\
    {\tt A17-d12-e1} & 4.93 & 4.93 & 62.5 & 24.6 & 17.4 & 
    {\tt XP} & {\tt nessai} & 1000 &
    $5.65_{-0.71}^{+1.46}$ & $4.19_{-0.82}^{+0.60}$ &
    $117_{-47}^{+22}$
    & $0.49_{-0.39}^{+0.40}$ & $0.48_{-0.42}^{+0.44}$ & $0.19_{-0.10}^{+0.10}$ & $0.44_{-0.29}^{+0.41}$ & -- & -- & 24.8 & 15.6 &
    (\checkmark,\checkmark) & 409 \\
    {\tt A17-d12-e1} & 9.73 & 9.73 & 125 & 31.6 & 22.4 & 
    {\tt XP} & {\tt nessai} & 1000 &
    $9.63_{-0.52}^{+1.04}$ & $8.54_{-0.88}^{+0.49}$ &
    $202_{-95}^{+52}$
    & -- & -- & -- & -- & -- & -- & 31.7 & 23.5 &
    (\checkmark,\checkmark) & 724 \\
    {\tt A17-d12-e1} & 9.73 & 9.73 & 125 & 31.6 & 22.4 & 
    {\tt XP} & {\tt nessai} & 1000 &
    $11.7_{-1.8}^{+1.5}$ & $7.65_{-0.89}^{+1.23}$ &
    $178_{-64}^{+66}$
    & $0.36_{-0.20}^{+0.24}$ & $0.48_{-0.40}^{+0.37}$ & $0.15_{-0.07}^{+0.07}$ & $0.34_{-0.17}^{+0.24}$ & -- & -- & 32.9 & 24.0 &
    (\checkmark,\checkmark) & 729 \\
    {\tt A17-d12-e1} & 19.0 & 19.0 & 250 & 35.9 & 25.5 & 
    {\tt XP} & {\tt nessai} & 1000 &
    $19.8_{-1.1}^{+2.1}$ & $17.3_{-1.8}^{+1.1}$ &
    $87.8_{-16.1}^{+10.8}$
    & -- & -- & -- & -- & -- & -- & 37.7 & 23.8 &
    (\checkmark,\checkmark) & 915 \\
    {\tt A17-d12-e1} & 19.0 & 19.0 & 250 & 35.9 & 25.5 & 
    {\tt XP} & {\tt nessai} & 1000 &
    $20.9_{-2.3}^{+2.3}$ & $16.0_{-1.7}^{+1.9}$ &
    $180_{-39}^{+17}$
    & $0.17_{-0.15}^{+0.25}$ & $0.23_{-0.21}^{+0.39}$ & $0.04_{-0.05}^{+0.06}$ & $0.16_{-0.13}^{+0.24}$ & -- & -- & 36.6 & 22.8 &
    (\checkmark,\checkmark) & 914 \\
    {\tt A17-d12-e1} & 27.8 & 27.8 & 375 & 35.6 & 25.2 & 
    {\tt XP} & {\tt nessai} & 1000 &
    $29.5_{-3.0}^{+3.2}$ & $22.7_{-2.5}^{+2.7}$ &
    $431_{-117}^{+62}$
    & -- & -- & -- & -- & -- & -- & 36.8 & 26.9 &
    (\checkmark,\checkmark) & 957 \\
    {\tt A17-d12-e1} & 27.8 & 27.8 & 375 & 35.6 & 25.2 & 
    {\tt XP} & {\tt nessai} & 1000 &
    $31.2_{-3.9}^{+3.0}$ & $22.1_{-2.01}^{+3.12}$ &
    $422_{-98}^{+70}$
    & $0.20_{-0.18}^{+0.40}$ & $0.28_{-0.25}^{+0.55}$ & $0.05_{-0.05}^{+0.06}$ & $0.21_{-0.16}^{+0.37}$ & -- & -- & 32.1 & 21.6 &
    (\checkmark,\checkmark) & 955 \\
    {\tt A17-d12-e1} & 36.2 & 36.2 & 500 & 33.6 & 23.7 & 
    {\tt XPHM} & {\tt dynesty} & 500 &
    $35.8_{-3.5}^{+4.8}$ & $27.8_{-4.5}^{+4.4}$ &
    $971_{-307}^{+156}$ &
    -- & -- & -- & -- & -- & -- & 32.4 & 23.3 &
    (\checkmark,\checkmark) & 817 \\
    {\tt A17-d12-e1} & 36.2 & 36.2 & 500 & 33.6 & 23.7 & 
    {\tt XPHM} & {\tt dynesty} & 500 &
    $37.1_{-4.1}^{+4.5}$ & $27.2_{-3.9}^{+4.8}$ &
    $999_{-247}^{+129}$ &
    $0.21_{-0.19}^{+0.43}$ & $0.31_{-0.28}^{+0.56}$ & $0.04_{-0.07}^{+0.07}$ & $0.26_{-0.21}^{+0.36}$ & -- & -- & 33.1 & 24.8 &
    (\checkmark,\checkmark) & 816 \\
    {\tt A17-d12-e1} & 36.2 & 36.2 & 500 & 33.6 & 23.7 & 
    {\tt XP} & {\tt nessai} & 1000 &
    $38.1_{-5.02}^{+4.59}$ & $27.2_{-3.14}^{+4.38}$ &
    $826_{-306}^{+204}$ &
    -- & -- & -- & -- & -- & -- & 34.0 & 24.5 &
    (\checkmark,\checkmark) & 820 \\
    {\tt A17-d12-e1} & 36.2 & 36.2 & 500 & 33.2 & 23.8 & 
    {\tt XP} & {\tt nessai} & 500 &
    $38.2_{-4.86}^{+4.54}$ & 
    $27.1_{-3.3}^{+4.4}$ &
    $901_{-317}^{+168}$ &
    $0.15_{-0.13}^{+0.32}$ & $0.23_{-0.20}^{+0.51}$ & $0.02_{-0.06}^{+0.07}$ & $0.19_{-0.15}^{+0.32}$ & -- & -- & 33.1 & 24.8 &
    (\checkmark,\checkmark) & 817 \\
    {\tt A17-d12-e1} & 44.3 & 44.3 & 625 & 31.7 & 22.3 & 
    {\tt XP} & {\tt nessai} & 1000 &
    $54.6_{-6.4}^{+4.6}$ & $33.2_{-3.5}^{+5.6}$ &
    $411_{-127}^{+81}$ &
    -- & -- & -- & -- & -- & -- & 31.3 & 23.1 &
    (\checkmark,\checkmark) & 725 \\
    {\tt A17-d12-e1} & 44.3 & 44.3 & 625 & 31.7 & 22.3 & 
    {\tt XP} & {\tt nessai} & 500 &
    $53.1_{-6.6}^{+5.3}$ & $35.0_{-5.0}^{+7.0}$ &
    $434_{-119}^{+60}$ &
    $0.13_{-0.11}^{+0.27}$ & $0.23_{-0.20}^{+0.42}$ & $0.03_{-0.07}^{+0.08}$ & $0.16_{-0.11}^{+0.27}$ & -- & -- & 30.6 & 23.4 &
    (\checkmark,\checkmark) & 722 \\
    {\tt A17-d12-e1} & 52.1 & 52.1 & 750 & 30.0 & 21.2 & 
    {\tt XP} & {\tt nessai} & 1000 &
    $67.0_{-8.6}^{+6.0}$ & $39.2_{-4.8}^{+8.2}$ &
    $449_{-155}^{+47}$ &
    -- & -- & -- & -- & -- & -- & 29.9 & 21.6 &
    (\checkmark,\checkmark) & 652 \\
    {\tt A17-d12-e1} & 52.1 & 52.1 & 750 & 30.0 & 21.2 & 
    {\tt XP} & {\tt nessai} & 500 &
    $68.1_{-6.2}^{+5.3}$ & $34.5_{-6.5}^{+8.3}$ &
    $455_{-109}^{+40}$ &
    $0.16_{-0.14}^{+0.26}$ & $0.27_{-0.24}^{+0.48}$ & $-0.05_{-0.13}^{+0.11}$ & $0.17_{-0.12}^{+0.22}$ & -- & -- & 30.1 & 22.3 &
    (\checkmark,\checkmark) & 652 \\
    \hline \hline
    \end{tabular}
\end{sidewaystable}
\endgroup

\twocolumngrid
%


\begin{thebibliography}{63}%
\makeatletter
\providecommand \@ifxundefined [1]{%
 \@ifx{#1\undefined}
}%
\providecommand \@ifnum [1]{%
 \ifnum #1\expandafter \@firstoftwo
 \else \expandafter \@secondoftwo
 \fi
}%
\providecommand \@ifx [1]{%
 \ifx #1\expandafter \@firstoftwo
 \else \expandafter \@secondoftwo
 \fi
}%
\providecommand \natexlab [1]{#1}%
\providecommand \enquote  [1]{``#1''}%
\providecommand \bibnamefont  [1]{#1}%
\providecommand \bibfnamefont [1]{#1}%
\providecommand \citenamefont [1]{#1}%
\providecommand \href@noop [0]{\@secondoftwo}%
\providecommand \href [0]{\begingroup \@sanitize@url \@href}%
\providecommand \@href[1]{\@@startlink{#1}\@@href}%
\providecommand \@@href[1]{\endgroup#1\@@endlink}%
\providecommand \@sanitize@url [0]{\catcode `\\12\catcode `\$12\catcode
  `\&12\catcode `\#12\catcode `\^12\catcode `\_12\catcode `\%12\relax}%
\providecommand \@@startlink[1]{}%
\providecommand \@@endlink[0]{}%
\providecommand \url  [0]{\begingroup\@sanitize@url \@url }%
\providecommand \@url [1]{\endgroup\@href {#1}{\urlprefix }}%
\providecommand \urlprefix  [0]{URL }%
\providecommand \Eprint [0]{\href }%
\providecommand \doibase [0]{https://doi.org/}%
\providecommand \selectlanguage [0]{\@gobble}%
\providecommand \bibinfo  [0]{\@secondoftwo}%
\providecommand \bibfield  [0]{\@secondoftwo}%
\providecommand \translation [1]{[#1]}%
\providecommand \BibitemOpen [0]{}%
\providecommand \bibitemStop [0]{}%
\providecommand \bibitemNoStop [0]{.\EOS\space}%
\providecommand \EOS [0]{\spacefactor3000\relax}%
\providecommand \BibitemShut  [1]{\csname bibitem#1\endcsname}%
\let\auto@bib@innerbib\@empty
\bibitem [{\citenamefont {Abbott}\ \emph {et~al.}(2016)\citenamefont {Abbott}
  \emph {et~al.}}]{Abbott:2016blz}%
  \BibitemOpen
  \bibfield  {author} {\bibinfo {author} {\bibfnamefont {B.~P.}\ \bibnamefont
  {Abbott}} \emph {et~al.},\ }\bibfield  {title} {\bibinfo {title}
  {{Observation of Gravitational Waves from a Binary Black Hole Merger}},\
  }\href {https://doi.org/10.1103/PhysRevLett.116.061102} {\bibfield  {journal}
  {\bibinfo  {journal} {Phys. Rev. Lett.}\ }\textbf {\bibinfo {volume} {116}},\
  \bibinfo {pages} {061102} (\bibinfo {year} {2016})},\ \Eprint
  {https://arxiv.org/abs/1602.03837} {arXiv:1602.03837 [gr-qc]} \BibitemShut
  {NoStop}%
\bibitem [{\citenamefont {Abbott}\ \emph {et~al.}(2018)\citenamefont {Abbott}
  \emph {et~al.}}]{AdvancedDetectorsLVK2018}%
  \BibitemOpen
  \bibfield  {author} {\bibinfo {author} {\bibfnamefont {B.~P.}\ \bibnamefont
  {Abbott}} \emph {et~al.} (\bibinfo {collaboration} {KAGRA, LIGO Scientific,
  and Virgo}),\ }\bibfield  {title} {\bibinfo {title} {{Prospects for Observing
  and Localizing Gravitational-Wave Transients with Advanced LIGO, Advanced
  Virgo and KAGRA}},\ }\href {https://doi.org/10.1007/s41114-018-0012-9,
  10.1007/lrr-2016-1} {\bibfield  {journal} {\bibinfo  {journal} {Living Rev.
  Rel.}\ }\textbf {\bibinfo {volume} {21}},\ \bibinfo {pages} {3} (\bibinfo
  {year} {2018})},\ \Eprint {https://arxiv.org/abs/1304.0670} {arXiv:1304.0670
  [gr-qc]} \BibitemShut {NoStop}%
\bibitem [{\citenamefont {Abbott}\ \emph {et~al.}(2023)\citenamefont {Abbott}
  \emph {et~al.}}]{KAGRA:2021vkt}%
  \BibitemOpen
  \bibfield  {author} {\bibinfo {author} {\bibfnamefont {R.}~\bibnamefont
  {Abbott}} \emph {et~al.} (\bibinfo {collaboration} {KAGRA, Virgo, and LIGO
  Scientific}),\ }\bibfield  {title} {\bibinfo {title} {{GWTC-3: Compact Binary
  Coalescences Observed by LIGO and Virgo during the Second Part of the Third
  Observing Run}},\ }\href {https://doi.org/10.1103/PhysRevX.13.041039}
  {\bibfield  {journal} {\bibinfo  {journal} {Phys. Rev. X}\ }\textbf {\bibinfo
  {volume} {13}},\ \bibinfo {pages} {041039} (\bibinfo {year} {2023})},\
  \Eprint {https://arxiv.org/abs/2111.03606} {arXiv:2111.03606 [gr-qc]}
  \BibitemShut {NoStop}%
\bibitem [{\citenamefont {Nitz}\ \emph {et~al.}(2023)\citenamefont {Nitz},
  \citenamefont {Kumar}, \citenamefont {Wang}, \citenamefont {Kastha},
  \citenamefont {Wu}, \citenamefont {Sch\"afer}, \citenamefont {Dhurkunde},\
  and\ \citenamefont {Capano}}]{Nitz:2021zwj}%
  \BibitemOpen
  \bibfield  {author} {\bibinfo {author} {\bibfnamefont {A.~H.}\ \bibnamefont
  {Nitz}}, \bibinfo {author} {\bibfnamefont {S.}~\bibnamefont {Kumar}},
  \bibinfo {author} {\bibfnamefont {Y.-F.}\ \bibnamefont {Wang}}, \bibinfo
  {author} {\bibfnamefont {S.}~\bibnamefont {Kastha}}, \bibinfo {author}
  {\bibfnamefont {S.}~\bibnamefont {Wu}}, \bibinfo {author} {\bibfnamefont
  {M.}~\bibnamefont {Sch\"afer}}, \bibinfo {author} {\bibfnamefont
  {R.}~\bibnamefont {Dhurkunde}},\ and\ \bibinfo {author} {\bibfnamefont
  {C.~D.}\ \bibnamefont {Capano}},\ }\bibfield  {title} {\bibinfo {title}
  {{4-OGC: Catalog of Gravitational Waves from Compact Binary Mergers}},\
  }\href {https://doi.org/10.3847/1538-4357/aca591} {\bibfield  {journal}
  {\bibinfo  {journal} {Astrophys. J.}\ }\textbf {\bibinfo {volume} {946}},\
  \bibinfo {pages} {59} (\bibinfo {year} {2023})},\ \Eprint
  {https://arxiv.org/abs/2112.06878} {arXiv:2112.06878 [astro-ph.HE]}
  \BibitemShut {NoStop}%
\bibitem [{\citenamefont {Cardoso}\ and\ \citenamefont
  {Pani}(2019)}]{Cardoso:2019rvt}%
  \BibitemOpen
  \bibfield  {author} {\bibinfo {author} {\bibfnamefont {V.}~\bibnamefont
  {Cardoso}}\ and\ \bibinfo {author} {\bibfnamefont {P.}~\bibnamefont {Pani}},\
  }\bibfield  {title} {\bibinfo {title} {{Testing the nature of dark compact
  objects: a status report}},\ }\href
  {https://doi.org/10.1007/s41114-019-0020-4} {\bibfield  {journal} {\bibinfo
  {journal} {Living Rev. Rel.}\ }\textbf {\bibinfo {volume} {22}},\ \bibinfo
  {pages} {4} (\bibinfo {year} {2019})},\ \Eprint
  {https://arxiv.org/abs/arXiv:1904.05363 [gr-qc]} {arXiv:1904.05363 [gr-qc]}
  \BibitemShut {NoStop}%
\bibitem [{\citenamefont {Kaup}(1968)}]{Kaup:1968zz}%
  \BibitemOpen
  \bibfield  {author} {\bibinfo {author} {\bibfnamefont {D.~J.}\ \bibnamefont
  {Kaup}},\ }\bibfield  {title} {\bibinfo {title} {{Klein-Gordon Geon}},\
  }\href {https://doi.org/10.1103/PhysRev.172.1331} {\bibfield  {journal}
  {\bibinfo  {journal} {Phys. Rev.}\ }\textbf {\bibinfo {volume} {172}},\
  \bibinfo {pages} {1331} (\bibinfo {year} {1968})}\BibitemShut {NoStop}%
\bibitem [{\citenamefont {Ruffini}\ and\ \citenamefont
  {Bonazzola}(1969)}]{Ruffini:1969qy}%
  \BibitemOpen
  \bibfield  {author} {\bibinfo {author} {\bibfnamefont {R.}~\bibnamefont
  {Ruffini}}\ and\ \bibinfo {author} {\bibfnamefont {S.}~\bibnamefont
  {Bonazzola}},\ }\bibfield  {title} {\bibinfo {title} {{Systems of
  selfgravitating particles in general relativity and the concept of an
  equation of state}},\ }\href {https://doi.org/10.1103/PhysRev.187.1767}
  {\bibfield  {journal} {\bibinfo  {journal} {Phys. Rev.}\ }\textbf {\bibinfo
  {volume} {187}},\ \bibinfo {pages} {1767} (\bibinfo {year}
  {1969})}\BibitemShut {NoStop}%
\bibitem [{\citenamefont {Liebling}\ and\ \citenamefont
  {Palenzuela}(2023)}]{Liebling:2012fv}%
  \BibitemOpen
  \bibfield  {author} {\bibinfo {author} {\bibfnamefont {S.~L.}\ \bibnamefont
  {Liebling}}\ and\ \bibinfo {author} {\bibfnamefont {C.}~\bibnamefont
  {Palenzuela}},\ }\bibfield  {title} {\bibinfo {title} {{Dynamical boson
  stars}},\ }\href {https://doi.org/10.1007/s41114-023-00043-4} {\bibfield
  {journal} {\bibinfo  {journal} {Living Rev. Rel.}\ }\textbf {\bibinfo
  {volume} {26}},\ \bibinfo {pages} {1} (\bibinfo {year} {2023})},\ \Eprint
  {https://arxiv.org/abs/1202.5809} {arXiv:1202.5809 [gr-qc]} \BibitemShut
  {NoStop}%
\bibitem [{\citenamefont {Sharma}\ \emph {et~al.}(2008)\citenamefont {Sharma},
  \citenamefont {Karmakar},\ and\ \citenamefont {Mukherjee}}]{Sharma:2008sc}%
  \BibitemOpen
  \bibfield  {author} {\bibinfo {author} {\bibfnamefont {R.}~\bibnamefont
  {Sharma}}, \bibinfo {author} {\bibfnamefont {S.}~\bibnamefont {Karmakar}},\
  and\ \bibinfo {author} {\bibfnamefont {S.}~\bibnamefont {Mukherjee}},\
  }\bibfield  {title} {\bibinfo {title} {{Boson star and dark matter}},\
  }\href@noop {} {\  (\bibinfo {year} {2008})},\ \Eprint
  {https://arxiv.org/abs/0812.3470} {arXiv:0812.3470 [gr-qc]} \BibitemShut
  {NoStop}%
\bibitem [{\citenamefont {Marsh}(2016)}]{Marsh:2015xka}%
  \BibitemOpen
  \bibfield  {author} {\bibinfo {author} {\bibfnamefont {D.~J.~E.}\
  \bibnamefont {Marsh}},\ }\bibfield  {title} {\bibinfo {title} {{Axion
  Cosmology}},\ }\href {https://doi.org/10.1016/j.physrep.2016.06.005}
  {\bibfield  {journal} {\bibinfo  {journal} {Phys. Rept.}\ }\textbf {\bibinfo
  {volume} {643}},\ \bibinfo {pages} {1} (\bibinfo {year} {2016})},\ \Eprint
  {https://arxiv.org/abs/1510.07633} {arXiv:1510.07633 [astro-ph.CO]}
  \BibitemShut {NoStop}%
\bibitem [{\citenamefont {Vincent}\ \emph {et~al.}(2016)\citenamefont
  {Vincent}, \citenamefont {Meliani}, \citenamefont {Grandclement},
  \citenamefont {Gourgoulhon},\ and\ \citenamefont {Straub}}]{Vincent:2015xta}%
  \BibitemOpen
  \bibfield  {author} {\bibinfo {author} {\bibfnamefont {F.~H.}\ \bibnamefont
  {Vincent}}, \bibinfo {author} {\bibfnamefont {Z.}~\bibnamefont {Meliani}},
  \bibinfo {author} {\bibfnamefont {P.}~\bibnamefont {Grandclement}}, \bibinfo
  {author} {\bibfnamefont {E.}~\bibnamefont {Gourgoulhon}},\ and\ \bibinfo
  {author} {\bibfnamefont {O.}~\bibnamefont {Straub}},\ }\bibfield  {title}
  {\bibinfo {title} {{Imaging a boson star at the Galactic center}},\ }\href
  {https://doi.org/10.1088/0264-9381/33/10/105015} {\bibfield  {journal}
  {\bibinfo  {journal} {Class. Quant. Grav.}\ }\textbf {\bibinfo {volume}
  {33}},\ \bibinfo {pages} {105015} (\bibinfo {year} {2016})},\ \Eprint
  {https://arxiv.org/abs/1510.04170} {arXiv:1510.04170 [gr-qc]} \BibitemShut
  {NoStop}%
\bibitem [{\citenamefont {Rosa}\ and\ \citenamefont
  {Rubiera-Garcia}(2022)}]{Rosa:2022tfv}%
  \BibitemOpen
  \bibfield  {author} {\bibinfo {author} {\bibfnamefont {J.~L.}\ \bibnamefont
  {Rosa}}\ and\ \bibinfo {author} {\bibfnamefont {D.}~\bibnamefont
  {Rubiera-Garcia}},\ }\bibfield  {title} {\bibinfo {title} {{Shadows of boson
  and Proca stars with thin accretion disks}},\ }\href
  {https://doi.org/10.1103/PhysRevD.106.084004} {\bibfield  {journal} {\bibinfo
   {journal} {Phys. Rev. D}\ }\textbf {\bibinfo {volume} {106}},\ \bibinfo
  {pages} {084004} (\bibinfo {year} {2022})},\ \Eprint
  {https://arxiv.org/abs/2204.12949} {arXiv:2204.12949 [gr-qc]} \BibitemShut
  {NoStop}%
\bibitem [{\citenamefont {Grandcl\'ement}(2017)}]{Grandclement:2016eng}%
  \BibitemOpen
  \bibfield  {author} {\bibinfo {author} {\bibfnamefont {P.}~\bibnamefont
  {Grandcl\'ement}},\ }\bibfield  {title} {\bibinfo {title} {{Light rings and
  light points of boson stars}},\ }\href
  {https://doi.org/10.1103/PhysRevD.95.084011} {\bibfield  {journal} {\bibinfo
  {journal} {Phys. Rev. D}\ }\textbf {\bibinfo {volume} {95}},\ \bibinfo
  {pages} {084011} (\bibinfo {year} {2017})},\ \Eprint
  {https://arxiv.org/abs/1612.07507} {arXiv:1612.07507 [gr-qc]} \BibitemShut
  {NoStop}%
\bibitem [{\citenamefont {Cunha}\ \emph {et~al.}(2017)\citenamefont {Cunha},
  \citenamefont {Berti},\ and\ \citenamefont {Herdeiro}}]{Cunha:2017qtt}%
  \BibitemOpen
  \bibfield  {author} {\bibinfo {author} {\bibfnamefont {P.~V.~P.}\
  \bibnamefont {Cunha}}, \bibinfo {author} {\bibfnamefont {E.}~\bibnamefont
  {Berti}},\ and\ \bibinfo {author} {\bibfnamefont {C.~A.~R.}\ \bibnamefont
  {Herdeiro}},\ }\bibfield  {title} {\bibinfo {title} {{Light-Ring Stability
  for Ultracompact Objects}},\ }\href
  {https://doi.org/10.1103/PhysRevLett.119.251102} {\bibfield  {journal}
  {\bibinfo  {journal} {Phys. Rev. Lett.}\ }\textbf {\bibinfo {volume} {119}},\
  \bibinfo {pages} {251102} (\bibinfo {year} {2017})},\ \Eprint
  {https://arxiv.org/abs/1708.04211} {arXiv:1708.04211 [gr-qc]} \BibitemShut
  {NoStop}%
\bibitem [{\citenamefont {Cunha}\ \emph {et~al.}(2023)\citenamefont {Cunha},
  \citenamefont {Herdeiro}, \citenamefont {Radu},\ and\ \citenamefont
  {Sanchis-Gual}}]{Cunha:2022gde}%
  \BibitemOpen
  \bibfield  {author} {\bibinfo {author} {\bibfnamefont {P.~V.~P.}\
  \bibnamefont {Cunha}}, \bibinfo {author} {\bibfnamefont {C.}~\bibnamefont
  {Herdeiro}}, \bibinfo {author} {\bibfnamefont {E.}~\bibnamefont {Radu}},\
  and\ \bibinfo {author} {\bibfnamefont {N.}~\bibnamefont {Sanchis-Gual}},\
  }\bibfield  {title} {\bibinfo {title} {{Exotic Compact Objects and the Fate
  of the Light-Ring Instability}},\ }\href
  {https://doi.org/10.1103/PhysRevLett.130.061401} {\bibfield  {journal}
  {\bibinfo  {journal} {Phys. Rev. Lett.}\ }\textbf {\bibinfo {volume} {130}},\
  \bibinfo {pages} {061401} (\bibinfo {year} {2023})},\ \Eprint
  {https://arxiv.org/abs/2207.13713} {arXiv:2207.13713 [gr-qc]} \BibitemShut
  {NoStop}%
\bibitem [{\citenamefont {Sennett}\ \emph {et~al.}(2017)\citenamefont
  {Sennett}, \citenamefont {Hinderer}, \citenamefont {Steinhoff}, \citenamefont
  {Buonanno},\ and\ \citenamefont {Ossokine}}]{Sennett:2017etc}%
  \BibitemOpen
  \bibfield  {author} {\bibinfo {author} {\bibfnamefont {N.}~\bibnamefont
  {Sennett}}, \bibinfo {author} {\bibfnamefont {T.}~\bibnamefont {Hinderer}},
  \bibinfo {author} {\bibfnamefont {J.}~\bibnamefont {Steinhoff}}, \bibinfo
  {author} {\bibfnamefont {A.}~\bibnamefont {Buonanno}},\ and\ \bibinfo
  {author} {\bibfnamefont {S.}~\bibnamefont {Ossokine}},\ }\bibfield  {title}
  {\bibinfo {title} {{Distinguishing Boson Stars from Black Holes and Neutron
  Stars from Tidal Interactions in Inspiraling Binary Systems}},\ }\href
  {https://doi.org/10.1103/PhysRevD.96.024002} {\bibfield  {journal} {\bibinfo
  {journal} {Phys. Rev. D}\ }\textbf {\bibinfo {volume} {96}},\ \bibinfo
  {pages} {024002} (\bibinfo {year} {2017})},\ \Eprint
  {https://arxiv.org/abs/arXiv:1704.08651 [gr-qc]} {arXiv:1704.08651 [gr-qc]}
  \BibitemShut {NoStop}%
\bibitem [{\citenamefont {Palenzuela}\ \emph {et~al.}(2017)\citenamefont
  {Palenzuela}, \citenamefont {Pani}, \citenamefont {Bezares}, \citenamefont
  {Cardoso}, \citenamefont {Lehner},\ and\ \citenamefont
  {Liebling}}]{Palenzuela:2017kcg}%
  \BibitemOpen
  \bibfield  {author} {\bibinfo {author} {\bibfnamefont {C.}~\bibnamefont
  {Palenzuela}}, \bibinfo {author} {\bibfnamefont {P.}~\bibnamefont {Pani}},
  \bibinfo {author} {\bibfnamefont {M.}~\bibnamefont {Bezares}}, \bibinfo
  {author} {\bibfnamefont {V.}~\bibnamefont {Cardoso}}, \bibinfo {author}
  {\bibfnamefont {L.}~\bibnamefont {Lehner}},\ and\ \bibinfo {author}
  {\bibfnamefont {S.}~\bibnamefont {Liebling}},\ }\bibfield  {title} {\bibinfo
  {title} {{Gravitational Wave Signatures of Highly Compact Boson Star
  Binaries}},\ }\href {https://doi.org/10.1103/PhysRevD.96.104058} {\bibfield
  {journal} {\bibinfo  {journal} {Phys. Rev. D}\ }\textbf {\bibinfo {volume}
  {96}},\ \bibinfo {pages} {104058} (\bibinfo {year} {2017})},\ \Eprint
  {https://arxiv.org/abs/arXiv:1710.09432 [gr-qc]} {arXiv:1710.09432 [gr-qc]}
  \BibitemShut {NoStop}%
\bibitem [{\citenamefont {Bezares}\ \emph {et~al.}(2022)\citenamefont
  {Bezares}, \citenamefont {Bo\v{s}kovi\'c}, \citenamefont {Liebling},
  \citenamefont {Palenzuela}, \citenamefont {Pani},\ and\ \citenamefont
  {Barausse}}]{Bezares:2022obu}%
  \BibitemOpen
  \bibfield  {author} {\bibinfo {author} {\bibfnamefont {M.}~\bibnamefont
  {Bezares}}, \bibinfo {author} {\bibfnamefont {M.}~\bibnamefont
  {Bo\v{s}kovi\'c}}, \bibinfo {author} {\bibfnamefont {S.}~\bibnamefont
  {Liebling}}, \bibinfo {author} {\bibfnamefont {C.}~\bibnamefont
  {Palenzuela}}, \bibinfo {author} {\bibfnamefont {P.}~\bibnamefont {Pani}},\
  and\ \bibinfo {author} {\bibfnamefont {E.}~\bibnamefont {Barausse}},\
  }\bibfield  {title} {\bibinfo {title} {{Gravitational waves and kicks from
  the merger of unequal mass, highly compact boson stars}},\ }\href
  {https://doi.org/10.1103/PhysRevD.105.064067} {\bibfield  {journal} {\bibinfo
   {journal} {Phys. Rev. D}\ }\textbf {\bibinfo {volume} {105}},\ \bibinfo
  {pages} {064067} (\bibinfo {year} {2022})},\ \Eprint
  {https://arxiv.org/abs/2201.06113} {arXiv:2201.06113 [gr-qc]} \BibitemShut
  {NoStop}%
\bibitem [{\citenamefont {Croft}\ \emph {et~al.}(2023)\citenamefont {Croft},
  \citenamefont {Helfer}, \citenamefont {Ge}, \citenamefont {Radia},
  \citenamefont {Evstafyeva}, \citenamefont {Lim}, \citenamefont {Sperhake},\
  and\ \citenamefont {Clough}}]{Croft:2022bxq}%
  \BibitemOpen
  \bibfield  {author} {\bibinfo {author} {\bibfnamefont {R.}~\bibnamefont
  {Croft}}, \bibinfo {author} {\bibfnamefont {T.}~\bibnamefont {Helfer}},
  \bibinfo {author} {\bibfnamefont {B.-X.}\ \bibnamefont {Ge}}, \bibinfo
  {author} {\bibfnamefont {M.}~\bibnamefont {Radia}}, \bibinfo {author}
  {\bibfnamefont {T.}~\bibnamefont {Evstafyeva}}, \bibinfo {author}
  {\bibfnamefont {E.~A.}\ \bibnamefont {Lim}}, \bibinfo {author} {\bibfnamefont
  {U.}~\bibnamefont {Sperhake}},\ and\ \bibinfo {author} {\bibfnamefont
  {K.}~\bibnamefont {Clough}},\ }\bibfield  {title} {\bibinfo {title} {{The
  gravitational afterglow of boson stars}},\ }\href
  {https://doi.org/10.1088/1361-6382/acace4} {\bibfield  {journal} {\bibinfo
  {journal} {Class. Quant. Grav.}\ }\textbf {\bibinfo {volume} {40}},\ \bibinfo
  {pages} {065001} (\bibinfo {year} {2023})},\ \Eprint
  {https://arxiv.org/abs/2207.05690} {arXiv:2207.05690 [gr-qc]} \BibitemShut
  {NoStop}%
\bibitem [{\citenamefont {Siemonsen}\ and\ \citenamefont
  {East}(2023{\natexlab{a}})}]{Siemonsen:2023age}%
  \BibitemOpen
  \bibfield  {author} {\bibinfo {author} {\bibfnamefont {N.}~\bibnamefont
  {Siemonsen}}\ and\ \bibinfo {author} {\bibfnamefont {W.~E.}\ \bibnamefont
  {East}},\ }\bibfield  {title} {\bibinfo {title} {{Generic initial data for
  binary boson stars}},\ }\href {https://doi.org/10.1103/PhysRevD.108.124015}
  {\bibfield  {journal} {\bibinfo  {journal} {Phys. Rev. D}\ }\textbf {\bibinfo
  {volume} {108}},\ \bibinfo {pages} {124015} (\bibinfo {year}
  {2023}{\natexlab{a}})},\ \Eprint {https://arxiv.org/abs/2306.17265}
  {arXiv:2306.17265 [gr-qc]} \BibitemShut {NoStop}%
\bibitem [{\citenamefont {Siemonsen}\ and\ \citenamefont
  {East}(2023{\natexlab{b}})}]{Siemonsen:2023hko}%
  \BibitemOpen
  \bibfield  {author} {\bibinfo {author} {\bibfnamefont {N.}~\bibnamefont
  {Siemonsen}}\ and\ \bibinfo {author} {\bibfnamefont {W.~E.}\ \bibnamefont
  {East}},\ }\bibfield  {title} {\bibinfo {title} {{Binary boson stars: Merger
  dynamics and formation of rotating remnant stars}},\ }\href
  {https://doi.org/10.1103/PhysRevD.107.124018} {\bibfield  {journal} {\bibinfo
   {journal} {Phys. Rev. D}\ }\textbf {\bibinfo {volume} {107}},\ \bibinfo
  {pages} {124018} (\bibinfo {year} {2023}{\natexlab{b}})},\ \Eprint
  {https://arxiv.org/abs/2302.06627} {arXiv:2302.06627 [gr-qc]} \BibitemShut
  {NoStop}%
\bibitem [{\citenamefont {Siemonsen}(2024)}]{Siemonsen:2024snb}%
  \BibitemOpen
  \bibfield  {author} {\bibinfo {author} {\bibfnamefont {N.}~\bibnamefont
  {Siemonsen}},\ }\bibfield  {title} {\bibinfo {title} {{Nonlinear Treatment of
  a Black Hole Mimicker Ringdown}},\ }\href
  {https://doi.org/10.1103/PhysRevLett.133.031401} {\bibfield  {journal}
  {\bibinfo  {journal} {Phys. Rev. Lett.}\ }\textbf {\bibinfo {volume} {133}},\
  \bibinfo {pages} {031401} (\bibinfo {year} {2024})},\ \Eprint
  {https://arxiv.org/abs/2404.14536} {arXiv:2404.14536 [gr-qc]} \BibitemShut
  {NoStop}%
\bibitem [{\citenamefont {Pacilio}\ \emph {et~al.}(2020)\citenamefont
  {Pacilio}, \citenamefont {Vaglio}, \citenamefont {Maselli},\ and\
  \citenamefont {Pani}}]{Pacilio:2020jza}%
  \BibitemOpen
  \bibfield  {author} {\bibinfo {author} {\bibfnamefont {C.}~\bibnamefont
  {Pacilio}}, \bibinfo {author} {\bibfnamefont {M.}~\bibnamefont {Vaglio}},
  \bibinfo {author} {\bibfnamefont {A.}~\bibnamefont {Maselli}},\ and\ \bibinfo
  {author} {\bibfnamefont {P.}~\bibnamefont {Pani}},\ }\bibfield  {title}
  {\bibinfo {title} {{Gravitational-wave detectors as particle-physics
  laboratories: Constraining scalar interactions with a coherent inspiral model
  of boson-star binaries}},\ }\href
  {https://doi.org/10.1103/PhysRevD.102.083002} {\bibfield  {journal} {\bibinfo
   {journal} {Phys. Rev. D}\ }\textbf {\bibinfo {volume} {102}},\ \bibinfo
  {pages} {083002} (\bibinfo {year} {2020})},\ \Eprint
  {https://arxiv.org/abs/2007.05264} {arXiv:2007.05264 [gr-qc]} \BibitemShut
  {NoStop}%
\bibitem [{\citenamefont {Vaglio}\ \emph {et~al.}(2023)\citenamefont {Vaglio},
  \citenamefont {Pacilio}, \citenamefont {Maselli},\ and\ \citenamefont
  {Pani}}]{Vaglio:2023lrd}%
  \BibitemOpen
  \bibfield  {author} {\bibinfo {author} {\bibfnamefont {M.}~\bibnamefont
  {Vaglio}}, \bibinfo {author} {\bibfnamefont {C.}~\bibnamefont {Pacilio}},
  \bibinfo {author} {\bibfnamefont {A.}~\bibnamefont {Maselli}},\ and\ \bibinfo
  {author} {\bibfnamefont {P.}~\bibnamefont {Pani}},\ }\bibfield  {title}
  {\bibinfo {title} {{Bayesian parameter estimation on boson-star binary
  signals with a coherent inspiral template and spin-dependent quadrupolar
  corrections}},\ }\href {https://doi.org/10.1103/PhysRevD.108.023021}
  {\bibfield  {journal} {\bibinfo  {journal} {Phys. Rev. D}\ }\textbf {\bibinfo
  {volume} {108}},\ \bibinfo {pages} {023021} (\bibinfo {year} {2023})},\
  \Eprint {https://arxiv.org/abs/2302.13954} {arXiv:2302.13954 [gr-qc]}
  \BibitemShut {NoStop}%
\bibitem [{\citenamefont {Guo}\ \emph {et~al.}(2019)\citenamefont {Guo},
  \citenamefont {Sinha},\ and\ \citenamefont {Sun}}]{Guo:2019sns}%
  \BibitemOpen
  \bibfield  {author} {\bibinfo {author} {\bibfnamefont {H.-K.}\ \bibnamefont
  {Guo}}, \bibinfo {author} {\bibfnamefont {K.}~\bibnamefont {Sinha}},\ and\
  \bibinfo {author} {\bibfnamefont {C.}~\bibnamefont {Sun}},\ }\bibfield
  {title} {\bibinfo {title} {{Probing Boson Stars with Extreme Mass Ratio
  Inspirals}},\ }\href {https://doi.org/10.1088/1475-7516/2019/09/032}
  {\bibfield  {journal} {\bibinfo  {journal} {JCAP}\ }\textbf {\bibinfo
  {volume} {09}},\ \bibinfo {pages} {032}},\ \Eprint
  {https://arxiv.org/abs/1904.07871} {arXiv:1904.07871 [hep-ph]} \BibitemShut
  {NoStop}%
\bibitem [{\citenamefont {Calder\'on~Bustillo}\ \emph
  {et~al.}(2021)\citenamefont {Calder\'on~Bustillo}, \citenamefont
  {Sanchis-Gual}, \citenamefont {Torres-Forn\'e}, \citenamefont {Font},
  \citenamefont {Vajpeyi}, \citenamefont {Smith}, \citenamefont {Herdeiro},
  \citenamefont {Radu},\ and\ \citenamefont
  {Leong}}]{CalderonBustillo:2020fyi}%
  \BibitemOpen
  \bibfield  {author} {\bibinfo {author} {\bibfnamefont {J.}~\bibnamefont
  {Calder\'on~Bustillo}}, \bibinfo {author} {\bibfnamefont {N.}~\bibnamefont
  {Sanchis-Gual}}, \bibinfo {author} {\bibfnamefont {A.}~\bibnamefont
  {Torres-Forn\'e}}, \bibinfo {author} {\bibfnamefont {J.~A.}\ \bibnamefont
  {Font}}, \bibinfo {author} {\bibfnamefont {A.}~\bibnamefont {Vajpeyi}},
  \bibinfo {author} {\bibfnamefont {R.}~\bibnamefont {Smith}}, \bibinfo
  {author} {\bibfnamefont {C.}~\bibnamefont {Herdeiro}}, \bibinfo {author}
  {\bibfnamefont {E.}~\bibnamefont {Radu}},\ and\ \bibinfo {author}
  {\bibfnamefont {S.~H.~W.}\ \bibnamefont {Leong}},\ }\bibfield  {title}
  {\bibinfo {title} {{GW190521 as a Merger of Proca Stars: A Potential New
  Vector Boson of $8.7\times 10^{-13}$ eV}},\ }\href
  {https://doi.org/10.1103/PhysRevLett.126.081101} {\bibfield  {journal}
  {\bibinfo  {journal} {Phys. Rev. Lett.}\ }\textbf {\bibinfo {volume} {126}},\
  \bibinfo {pages} {081101} (\bibinfo {year} {2021})},\ \Eprint
  {https://arxiv.org/abs/2009.05376} {arXiv:2009.05376 [gr-qc]} \BibitemShut
  {NoStop}%
\bibitem [{\citenamefont {Calderon~Bustillo}\ \emph {et~al.}(2023)\citenamefont
  {Calderon~Bustillo}, \citenamefont {Sanchis-Gual}, \citenamefont {Leong},
  \citenamefont {Chandra}, \citenamefont {Torres-Forne}, \citenamefont {Font},
  \citenamefont {Herdeiro}, \citenamefont {Radu}, \citenamefont {Wong},\ and\
  \citenamefont {Li}}]{CalderonBustillo:2022cja}%
  \BibitemOpen
  \bibfield  {author} {\bibinfo {author} {\bibfnamefont {J.}~\bibnamefont
  {Calderon~Bustillo}}, \bibinfo {author} {\bibfnamefont {N.}~\bibnamefont
  {Sanchis-Gual}}, \bibinfo {author} {\bibfnamefont {S.~H.~W.}\ \bibnamefont
  {Leong}}, \bibinfo {author} {\bibfnamefont {K.}~\bibnamefont {Chandra}},
  \bibinfo {author} {\bibfnamefont {A.}~\bibnamefont {Torres-Forne}}, \bibinfo
  {author} {\bibfnamefont {J.~A.}\ \bibnamefont {Font}}, \bibinfo {author}
  {\bibfnamefont {C.}~\bibnamefont {Herdeiro}}, \bibinfo {author}
  {\bibfnamefont {E.}~\bibnamefont {Radu}}, \bibinfo {author} {\bibfnamefont
  {I.~C.~F.}\ \bibnamefont {Wong}},\ and\ \bibinfo {author} {\bibfnamefont
  {T.~G.~F.}\ \bibnamefont {Li}},\ }\bibfield  {title} {\bibinfo {title}
  {{Searching for vector boson-star mergers within LIGO-Virgo intermediate-mass
  black-hole merger candidates}},\ }\href
  {https://doi.org/10.1103/PhysRevD.108.123020} {\bibfield  {journal} {\bibinfo
   {journal} {Phys. Rev. D}\ }\textbf {\bibinfo {volume} {108}},\ \bibinfo
  {pages} {123020} (\bibinfo {year} {2023})},\ \Eprint
  {https://arxiv.org/abs/2206.02551} {arXiv:2206.02551 [gr-qc]} \BibitemShut
  {NoStop}%
\bibitem [{\citenamefont {Lee}\ and\ \citenamefont {Pang}(1992)}]{Lee:1991ax}%
  \BibitemOpen
  \bibfield  {author} {\bibinfo {author} {\bibfnamefont {T.~D.}\ \bibnamefont
  {Lee}}\ and\ \bibinfo {author} {\bibfnamefont {Y.}~\bibnamefont {Pang}},\
  }\bibfield  {title} {\bibinfo {title} {{Nontopological solitons}},\ }\href
  {https://doi.org/10.1016/0370-1573(92)90064-7} {\bibfield  {journal}
  {\bibinfo  {journal} {Phys. Rept.}\ }\textbf {\bibinfo {volume} {221}},\
  \bibinfo {pages} {251} (\bibinfo {year} {1992})}\BibitemShut {NoStop}%
\bibitem [{\citenamefont {Lee}(1987)}]{Lee:1986ts}%
  \BibitemOpen
  \bibfield  {author} {\bibinfo {author} {\bibfnamefont {T.~D.}\ \bibnamefont
  {Lee}},\ }\bibfield  {title} {\bibinfo {title} {{Soliton Stars and the
  Critical Masses of Black Holes}},\ }\href
  {https://doi.org/10.1103/PhysRevD.35.3637} {\bibfield  {journal} {\bibinfo
  {journal} {Phys. Rev. D}\ }\textbf {\bibinfo {volume} {35}},\ \bibinfo
  {pages} {3637} (\bibinfo {year} {1987})}\BibitemShut {NoStop}%
\bibitem [{\citenamefont {Evstafyeva}\ \emph
  {et~al.}(2023{\natexlab{a}})\citenamefont {Evstafyeva}, \citenamefont
  {Rosca-Mead}, \citenamefont {Sperhake},\ and\ \citenamefont
  {Brugmann}}]{Evstafyeva:2023kfg}%
  \BibitemOpen
  \bibfield  {author} {\bibinfo {author} {\bibfnamefont {T.}~\bibnamefont
  {Evstafyeva}}, \bibinfo {author} {\bibfnamefont {R.}~\bibnamefont
  {Rosca-Mead}}, \bibinfo {author} {\bibfnamefont {U.}~\bibnamefont
  {Sperhake}},\ and\ \bibinfo {author} {\bibfnamefont {B.}~\bibnamefont
  {Brugmann}},\ }\bibfield  {title} {\bibinfo {title} {{Boson stars in massless
  and massive scalar-tensor gravity}},\ }\href
  {https://doi.org/10.1103/PhysRevD.108.104064} {\bibfield  {journal} {\bibinfo
   {journal} {Phys. Rev. D}\ }\textbf {\bibinfo {volume} {108}},\ \bibinfo
  {pages} {104064} (\bibinfo {year} {2023}{\natexlab{a}})},\ \Eprint
  {https://arxiv.org/abs/2310.05200} {arXiv:2310.05200 [gr-qc]} \BibitemShut
  {NoStop}%
\bibitem [{\citenamefont {Helfer}\ \emph {et~al.}(2022)\citenamefont {Helfer},
  \citenamefont {Sperhake}, \citenamefont {Croft}, \citenamefont {Radia},
  \citenamefont {Ge},\ and\ \citenamefont {Lim}}]{Helfer:2021brt}%
  \BibitemOpen
  \bibfield  {author} {\bibinfo {author} {\bibfnamefont {T.}~\bibnamefont
  {Helfer}}, \bibinfo {author} {\bibfnamefont {U.}~\bibnamefont {Sperhake}},
  \bibinfo {author} {\bibfnamefont {R.}~\bibnamefont {Croft}}, \bibinfo
  {author} {\bibfnamefont {M.}~\bibnamefont {Radia}}, \bibinfo {author}
  {\bibfnamefont {B.-X.}\ \bibnamefont {Ge}},\ and\ \bibinfo {author}
  {\bibfnamefont {E.~A.}\ \bibnamefont {Lim}},\ }\bibfield  {title} {\bibinfo
  {title} {{Malaise and remedy of binary boson-star initial data}},\ }\href
  {https://doi.org/10.1088/1361-6382/ac53b7} {\bibfield  {journal} {\bibinfo
  {journal} {Class. Quant. Grav.}\ }\textbf {\bibinfo {volume} {39}},\ \bibinfo
  {pages} {074001} (\bibinfo {year} {2022})},\ \Eprint
  {https://arxiv.org/abs/2108.11995} {arXiv:2108.11995 [gr-qc]} \BibitemShut
  {NoStop}%
\bibitem [{\citenamefont {Clough}\ \emph {et~al.}(2015)\citenamefont {Clough},
  \citenamefont {Figueras}, \citenamefont {Finkel}, \citenamefont {Kunesch},
  \citenamefont {Lim},\ and\ \citenamefont {Tunyasuvunakool}}]{Clough:2015sqa}%
  \BibitemOpen
  \bibfield  {author} {\bibinfo {author} {\bibfnamefont {K.}~\bibnamefont
  {Clough}}, \bibinfo {author} {\bibfnamefont {P.}~\bibnamefont {Figueras}},
  \bibinfo {author} {\bibfnamefont {H.}~\bibnamefont {Finkel}}, \bibinfo
  {author} {\bibfnamefont {M.}~\bibnamefont {Kunesch}}, \bibinfo {author}
  {\bibfnamefont {E.~A.}\ \bibnamefont {Lim}},\ and\ \bibinfo {author}
  {\bibfnamefont {S.}~\bibnamefont {Tunyasuvunakool}},\ }\bibfield  {title}
  {\bibinfo {title} {{GRChombo : Numerical Relativity with Adaptive Mesh
  Refinement}},\ }\href {https://doi.org/10.1088/0264-9381/32/24/245011}
  {\bibfield  {journal} {\bibinfo  {journal} {Class. Quant. Grav.}\ }\textbf
  {\bibinfo {volume} {32}},\ \bibinfo {pages} {245011} (\bibinfo {year}
  {2015})},\ \Eprint {https://arxiv.org/abs/1503.03436} {arXiv:1503.03436
  [gr-qc]} \BibitemShut {NoStop}%
\bibitem [{\citenamefont {Radia}\ \emph {et~al.}(2022)\citenamefont {Radia},
  \citenamefont {Sperhake}, \citenamefont {Drew}, \citenamefont {Clough},
  \citenamefont {Figueras}, \citenamefont {Lim}, \citenamefont {Ripley},
  \citenamefont {Aurrekoetxea}, \citenamefont {Fran\c{c}a},\ and\ \citenamefont
  {Helfer}}]{Radia:2021smk}%
  \BibitemOpen
  \bibfield  {author} {\bibinfo {author} {\bibfnamefont {M.}~\bibnamefont
  {Radia}}, \bibinfo {author} {\bibfnamefont {U.}~\bibnamefont {Sperhake}},
  \bibinfo {author} {\bibfnamefont {A.}~\bibnamefont {Drew}}, \bibinfo {author}
  {\bibfnamefont {K.}~\bibnamefont {Clough}}, \bibinfo {author} {\bibfnamefont
  {P.}~\bibnamefont {Figueras}}, \bibinfo {author} {\bibfnamefont {E.~A.}\
  \bibnamefont {Lim}}, \bibinfo {author} {\bibfnamefont {J.~L.}\ \bibnamefont
  {Ripley}}, \bibinfo {author} {\bibfnamefont {J.~C.}\ \bibnamefont
  {Aurrekoetxea}}, \bibinfo {author} {\bibfnamefont {T.}~\bibnamefont
  {Fran\c{c}a}},\ and\ \bibinfo {author} {\bibfnamefont {T.}~\bibnamefont
  {Helfer}},\ }\bibfield  {title} {\bibinfo {title} {{Lessons for adaptive mesh
  refinement in numerical relativity}},\ }\href
  {https://doi.org/10.1088/1361-6382/ac6fa9} {\bibfield  {journal} {\bibinfo
  {journal} {Class. Quant. Grav.}\ }\textbf {\bibinfo {volume} {39}},\ \bibinfo
  {pages} {135006} (\bibinfo {year} {2022})},\ \Eprint
  {https://arxiv.org/abs/2112.10567} {arXiv:2112.10567 [gr-qc]} \BibitemShut
  {NoStop}%
\bibitem [{\citenamefont {Andrade}\ \emph {et~al.}(2021)\citenamefont {Andrade}
  \emph {et~al.}}]{Andrade:2021rbd}%
  \BibitemOpen
  \bibfield  {author} {\bibinfo {author} {\bibfnamefont {T.}~\bibnamefont
  {Andrade}} \emph {et~al.},\ }\bibfield  {title} {\bibinfo {title} {{GRChombo:
  An adaptable numerical relativity code for fundamental physics}},\ }\href
  {https://doi.org/10.21105/joss.03703} {\bibfield  {journal} {\bibinfo
  {journal} {J. Open Source Softw.}\ }\textbf {\bibinfo {volume} {6}},\
  \bibinfo {pages} {3703} (\bibinfo {year} {2021})},\ \Eprint
  {https://arxiv.org/abs/2201.03458} {arXiv:2201.03458 [gr-qc]} \BibitemShut
  {NoStop}%
\bibitem [{\citenamefont {Sperhake}(2007)}]{Sperhake:2006cy}%
  \BibitemOpen
  \bibfield  {author} {\bibinfo {author} {\bibfnamefont {U.}~\bibnamefont
  {Sperhake}},\ }\bibfield  {title} {\bibinfo {title} {Binary black-hole
  evolutions of excision and puncture data},\ }\href
  {https://doi.org/10.1103/PhysRevD.76.104015} {\bibfield  {journal} {\bibinfo
  {journal} {Phys. Rev. D}\ }\textbf {\bibinfo {volume} {76}},\ \bibinfo
  {pages} {104015} (\bibinfo {year} {2007})},\ \bibinfo {note}
  {gr-qc/0606079}\BibitemShut {NoStop}%
\bibitem [{\citenamefont {Alic}\ \emph {et~al.}(2012)\citenamefont {Alic},
  \citenamefont {Bona-Casas}, \citenamefont {Bona}, \citenamefont {Rezzolla},\
  and\ \citenamefont {Palenzuela}}]{Alic:2011gg}%
  \BibitemOpen
  \bibfield  {author} {\bibinfo {author} {\bibfnamefont {D.}~\bibnamefont
  {Alic}}, \bibinfo {author} {\bibfnamefont {C.}~\bibnamefont {Bona-Casas}},
  \bibinfo {author} {\bibfnamefont {C.}~\bibnamefont {Bona}}, \bibinfo {author}
  {\bibfnamefont {L.}~\bibnamefont {Rezzolla}},\ and\ \bibinfo {author}
  {\bibfnamefont {C.}~\bibnamefont {Palenzuela}},\ }\bibfield  {title}
  {\bibinfo {title} {{Conformal and covariant formulation of the Z4 system with
  constraint-violation damping}},\ }\href
  {https://doi.org/10.1103/PhysRevD.85.064040} {\bibfield  {journal} {\bibinfo
  {journal} {Phys. Rev. D}\ }\textbf {\bibinfo {volume} {85}},\ \bibinfo
  {pages} {064040} (\bibinfo {year} {2012})},\ \Eprint
  {https://arxiv.org/abs/1106.2254} {arXiv:1106.2254 [gr-qc]} \BibitemShut
  {NoStop}%
\bibitem [{\citenamefont {Campanelli}\ \emph
  {et~al.}(2006{\natexlab{a}})\citenamefont {Campanelli}, \citenamefont
  {Lousto}, \citenamefont {Marronetti},\ and\ \citenamefont
  {Zlochower}}]{Campanelli:2005dd}%
  \BibitemOpen
  \bibfield  {author} {\bibinfo {author} {\bibfnamefont {M.}~\bibnamefont
  {Campanelli}}, \bibinfo {author} {\bibfnamefont {C.~O.}\ \bibnamefont
  {Lousto}}, \bibinfo {author} {\bibfnamefont {P.}~\bibnamefont {Marronetti}},\
  and\ \bibinfo {author} {\bibfnamefont {Y.}~\bibnamefont {Zlochower}},\
  }\bibfield  {title} {\bibinfo {title} {{Accurate evolutions of orbiting
  black-hole binaries without excision}},\ }\href
  {https://doi.org/10.1103/PhysRevLett.96.111101} {\bibfield  {journal}
  {\bibinfo  {journal} {Phys. Rev. Lett.}\ }\textbf {\bibinfo {volume} {96}},\
  \bibinfo {pages} {111101} (\bibinfo {year} {2006}{\natexlab{a}})},\ \Eprint
  {https://arxiv.org/abs/gr-qc/0511048} {arXiv:gr-qc/0511048} \BibitemShut
  {NoStop}%
\bibitem [{\citenamefont {Baker}\ \emph {et~al.}(2006)\citenamefont {Baker},
  \citenamefont {Centrella}, \citenamefont {Choi}, \citenamefont {Koppitz},\
  and\ \citenamefont {van Meter}}]{Baker:2005vv}%
  \BibitemOpen
  \bibfield  {author} {\bibinfo {author} {\bibfnamefont {J.~G.}\ \bibnamefont
  {Baker}}, \bibinfo {author} {\bibfnamefont {J.}~\bibnamefont {Centrella}},
  \bibinfo {author} {\bibfnamefont {D.-I.}\ \bibnamefont {Choi}}, \bibinfo
  {author} {\bibfnamefont {M.}~\bibnamefont {Koppitz}},\ and\ \bibinfo {author}
  {\bibfnamefont {J.}~\bibnamefont {van Meter}},\ }\bibfield  {title} {\bibinfo
  {title} {{Gravitational wave extraction from an inspiraling configuration of
  merging black holes}},\ }\href
  {https://doi.org/10.1103/PhysRevLett.96.111102} {\bibfield  {journal}
  {\bibinfo  {journal} {Phys. Rev. Lett.}\ }\textbf {\bibinfo {volume} {96}},\
  \bibinfo {pages} {111102} (\bibinfo {year} {2006})},\ \Eprint
  {https://arxiv.org/abs/gr-qc/0511103} {arXiv:gr-qc/0511103} \BibitemShut
  {NoStop}%
\bibitem [{\citenamefont {Adams}\ \emph {et~al.}(2019)\citenamefont {Adams}
  \emph {et~al.}}]{chombo}%
  \BibitemOpen
  \bibfield  {author} {\bibinfo {author} {\bibfnamefont {M.}~\bibnamefont
  {Adams}} \emph {et~al.},\ }\href@noop {} {\emph {\bibinfo {title} {{Chombo
  Software Package for AMR Applications - Design Document}}}},\ \bibinfo {type}
  {Tech. Rep.}\ \bibinfo {number} {{LBNL}-6616{E}}\ (\bibinfo  {institution}
  {{Lawrence Berkeley National Laboratory}},\ \bibinfo {year}
  {{2019}})\BibitemShut {NoStop}%
\bibitem [{\citenamefont {{Allen, G. and Goodale, T. and Mass\'o, J. and
  Seidel, E.}}(1999)}]{Allen:1999}%
  \BibitemOpen
  \bibfield  {author} {\bibinfo {author} {\bibnamefont {{Allen, G. and Goodale,
  T. and Mass\'o, J. and Seidel, E.}}},\ }\bibfield  {title} {\bibinfo {title}
  {{The Cactus Computational Toolkit and Using Distributed Computing to Collide
  Neutron Stars}},\ }in\ \href@noop {} {\emph {\bibinfo {booktitle}
  {{Proceedings of Eighth IEEE International Symposium on High Performance
  Distributed Computing, HPDC-8, Redondo Beach, 1999}}}}\ (\bibinfo
  {publisher} {{IEEE Press}},\ \bibinfo {address} {{Piscataway, New Jersey,
  United States}},\ \bibinfo {year} {1999})\BibitemShut {NoStop}%
\bibitem [{\citenamefont {Schnetter}\ \emph {et~al.}(2004)\citenamefont
  {Schnetter}, \citenamefont {Hawley},\ and\ \citenamefont
  {Hawke}}]{Schnetter:2003rb}%
  \BibitemOpen
  \bibfield  {author} {\bibinfo {author} {\bibfnamefont {E.}~\bibnamefont
  {Schnetter}}, \bibinfo {author} {\bibfnamefont {S.~H.}\ \bibnamefont
  {Hawley}},\ and\ \bibinfo {author} {\bibfnamefont {I.}~\bibnamefont
  {Hawke}},\ }\bibfield  {title} {\bibinfo {title} {{Evolutions in 3-D
  numerical relativity using fixed mesh refinement}},\ }\href
  {https://doi.org/10.1088/0264-9381/21/6/014} {\bibfield  {journal} {\bibinfo
  {journal} {Class. Quant. Grav.}\ }\textbf {\bibinfo {volume} {21}},\ \bibinfo
  {pages} {1465} (\bibinfo {year} {2004})},\ \Eprint
  {https://arxiv.org/abs/gr-qc/0310042} {arXiv:gr-qc/0310042} \BibitemShut
  {NoStop}%
\bibitem [{\citenamefont {Thornburg}(2004)}]{Thornburg:2003sf}%
  \BibitemOpen
  \bibfield  {author} {\bibinfo {author} {\bibfnamefont {J.}~\bibnamefont
  {Thornburg}},\ }\bibfield  {title} {\bibinfo {title} {{A Fast apparent
  horizon finder for three-dimensional Cartesian grids in numerical
  relativity}},\ }\href {https://doi.org/10.1088/0264-9381/21/2/026} {\bibfield
   {journal} {\bibinfo  {journal} {Class. Quant. Grav.}\ }\textbf {\bibinfo
  {volume} {21}},\ \bibinfo {pages} {743} (\bibinfo {year} {2004})},\ \bibinfo
  {note} {gr-qc/0306056}\BibitemShut {NoStop}%
\bibitem [{\citenamefont {Evstafyeva}\ \emph
  {et~al.}(2023{\natexlab{b}})\citenamefont {Evstafyeva}, \citenamefont
  {Sperhake}, \citenamefont {Helfer}, \citenamefont {Croft}, \citenamefont
  {Radia}, \citenamefont {Ge},\ and\ \citenamefont {Lim}}]{Evstafyeva:2022bpr}%
  \BibitemOpen
  \bibfield  {author} {\bibinfo {author} {\bibfnamefont {T.}~\bibnamefont
  {Evstafyeva}}, \bibinfo {author} {\bibfnamefont {U.}~\bibnamefont
  {Sperhake}}, \bibinfo {author} {\bibfnamefont {T.}~\bibnamefont {Helfer}},
  \bibinfo {author} {\bibfnamefont {R.}~\bibnamefont {Croft}}, \bibinfo
  {author} {\bibfnamefont {M.}~\bibnamefont {Radia}}, \bibinfo {author}
  {\bibfnamefont {B.-X.}\ \bibnamefont {Ge}},\ and\ \bibinfo {author}
  {\bibfnamefont {E.~A.}\ \bibnamefont {Lim}},\ }\bibfield  {title} {\bibinfo
  {title} {{Unequal-mass boson-star binaries: initial data and merger
  dynamics}},\ }\href {https://doi.org/10.1088/1361-6382/acc2a8} {\bibfield
  {journal} {\bibinfo  {journal} {Class. Quant. Grav.}\ }\textbf {\bibinfo
  {volume} {40}},\ \bibinfo {pages} {085009} (\bibinfo {year}
  {2023}{\natexlab{b}})},\ \Eprint {https://arxiv.org/abs/2212.08023}
  {arXiv:2212.08023 [gr-qc]} \BibitemShut {NoStop}%
\bibitem [{\citenamefont {Mroue}\ \emph {et~al.}(2010)\citenamefont {Mroue},
  \citenamefont {Pfeiffer}, \citenamefont {Kidder},\ and\ \citenamefont
  {Teukolsky}}]{Mroue:2010re}%
  \BibitemOpen
  \bibfield  {author} {\bibinfo {author} {\bibfnamefont {A.~H.}\ \bibnamefont
  {Mroue}}, \bibinfo {author} {\bibfnamefont {H.~P.}\ \bibnamefont {Pfeiffer}},
  \bibinfo {author} {\bibfnamefont {L.~E.}\ \bibnamefont {Kidder}},\ and\
  \bibinfo {author} {\bibfnamefont {S.~A.}\ \bibnamefont {Teukolsky}},\
  }\bibfield  {title} {\bibinfo {title} {{Measuring orbital eccentricity and
  periastron advance in quasi-circular black hole simulations}},\ }\href
  {https://doi.org/10.1103/PhysRevD.82.124016} {\bibfield  {journal} {\bibinfo
  {journal} {Phys. Rev. D}\ }\textbf {\bibinfo {volume} {82}},\ \bibinfo
  {pages} {124016} (\bibinfo {year} {2010})},\ \Eprint
  {https://arxiv.org/abs/1004.4697} {arXiv:1004.4697 [gr-qc]} \BibitemShut
  {NoStop}%
\bibitem [{\citenamefont {Hinder}\ \emph {et~al.}(2014)\citenamefont {Hinder}
  \emph {et~al.}}]{Hinder:2013oqa}%
  \BibitemOpen
  \bibfield  {author} {\bibinfo {author} {\bibfnamefont {I.}~\bibnamefont
  {Hinder}} \emph {et~al.},\ }\bibfield  {title} {\bibinfo {title}
  {{Error-analysis and comparison to analytical models of numerical waveforms
  produced by the NRAR Collaboration}},\ }\href
  {https://doi.org/10.1088/0264-9381/31/2/025012} {\bibfield  {journal}
  {\bibinfo  {journal} {Class. Quant. Grav.}\ }\textbf {\bibinfo {volume}
  {31}},\ \bibinfo {pages} {025012} (\bibinfo {year} {2014})},\ \bibinfo {note}
  {arXiv:1307.5307 [gr-qc]}\BibitemShut {NoStop}%
\bibitem [{\citenamefont {Palenzuela}\ \emph {et~al.}(2007)\citenamefont
  {Palenzuela}, \citenamefont {Olabarrieta}, \citenamefont {Lehner},\ and\
  \citenamefont {Liebling}}]{Palenzuela:2006wp}%
  \BibitemOpen
  \bibfield  {author} {\bibinfo {author} {\bibfnamefont {C.}~\bibnamefont
  {Palenzuela}}, \bibinfo {author} {\bibfnamefont {I.}~\bibnamefont
  {Olabarrieta}}, \bibinfo {author} {\bibfnamefont {L.}~\bibnamefont
  {Lehner}},\ and\ \bibinfo {author} {\bibfnamefont {S.~L.}\ \bibnamefont
  {Liebling}},\ }\bibfield  {title} {\bibinfo {title} {{Head-on collisions of
  boson stars}},\ }\href {https://doi.org/10.1103/PhysRevD.75.064005}
  {\bibfield  {journal} {\bibinfo  {journal} {Phys. Rev. D}\ }\textbf {\bibinfo
  {volume} {75}},\ \bibinfo {pages} {064005} (\bibinfo {year} {2007})},\
  \Eprint {https://arxiv.org/abs/gr-qc/0612067} {arXiv:gr-qc/0612067}
  \BibitemShut {NoStop}%
\bibitem [{\citenamefont {Sanchis-Gual}\ \emph {et~al.}(2022)\citenamefont
  {Sanchis-Gual}, \citenamefont {Calder\'on~Bustillo}, \citenamefont
  {Herdeiro}, \citenamefont {Radu}, \citenamefont {Font}, \citenamefont
  {Leong},\ and\ \citenamefont {Torres-Forn\'e}}]{Sanchis-Gual:2022mkk}%
  \BibitemOpen
  \bibfield  {author} {\bibinfo {author} {\bibfnamefont {N.}~\bibnamefont
  {Sanchis-Gual}}, \bibinfo {author} {\bibfnamefont {J.}~\bibnamefont
  {Calder\'on~Bustillo}}, \bibinfo {author} {\bibfnamefont {C.}~\bibnamefont
  {Herdeiro}}, \bibinfo {author} {\bibfnamefont {E.}~\bibnamefont {Radu}},
  \bibinfo {author} {\bibfnamefont {J.~A.}\ \bibnamefont {Font}}, \bibinfo
  {author} {\bibfnamefont {S.~H.~W.}\ \bibnamefont {Leong}},\ and\ \bibinfo
  {author} {\bibfnamefont {A.}~\bibnamefont {Torres-Forn\'e}},\ }\bibfield
  {title} {\bibinfo {title} {{Impact of the wavelike nature of Proca stars on
  their gravitational-wave emission}},\ }\href
  {https://doi.org/10.1103/PhysRevD.106.124011} {\bibfield  {journal} {\bibinfo
   {journal} {Phys. Rev. D}\ }\textbf {\bibinfo {volume} {106}},\ \bibinfo
  {pages} {124011} (\bibinfo {year} {2022})},\ \Eprint
  {https://arxiv.org/abs/2208.11717} {arXiv:2208.11717 [gr-qc]} \BibitemShut
  {NoStop}%
\bibitem [{\citenamefont {Ashton}\ \emph {et~al.}(2019)\citenamefont {Ashton}
  \emph {et~al.}}]{Ashton:2018jfp}%
  \BibitemOpen
  \bibfield  {author} {\bibinfo {author} {\bibfnamefont {G.}~\bibnamefont
  {Ashton}} \emph {et~al.},\ }\bibfield  {title} {\bibinfo {title} {{BILBY: A
  user-friendly Bayesian inference library for gravitational-wave astronomy}},\
  }\href {https://doi.org/10.3847/1538-4365/ab06fc} {\bibfield  {journal}
  {\bibinfo  {journal} {Astrophys. J. Suppl.}\ }\textbf {\bibinfo {volume}
  {241}},\ \bibinfo {pages} {27} (\bibinfo {year} {2019})},\ \Eprint
  {https://arxiv.org/abs/1811.02042} {arXiv:1811.02042 [astro-ph.IM]}
  \BibitemShut {NoStop}%
\bibitem [{\citenamefont {{Romero-Shaw}}\ \emph {et~al.}(2020)\citenamefont
  {{Romero-Shaw}} \emph {et~al.}}]{Romero-Shaw:2020:Bilby}%
  \BibitemOpen
  \bibfield  {author} {\bibinfo {author} {\bibfnamefont {I.~M.}\ \bibnamefont
  {{Romero-Shaw}}} \emph {et~al.},\ }\bibfield  {title} {\bibinfo {title}
  {{Bayesian inference for compact binary coalescences with bilby: validation
  and application to the first LIGO\textendash{}Virgo gravitational-wave
  transient catalogue}},\ }\href {https://doi.org/10.1093/mnras/staa2850}
  {\bibfield  {journal} {\bibinfo  {journal} {Mon. Not. Roy. Astron. Soc.}\
  }\textbf {\bibinfo {volume} {499}},\ \bibinfo {pages} {3295} (\bibinfo {year}
  {2020})},\ \Eprint {https://arxiv.org/abs/2006.00714} {arXiv:2006.00714
  [astro-ph.IM]} \BibitemShut {NoStop}%
\bibitem [{\citenamefont {Husa}\ \emph {et~al.}(2016)\citenamefont {Husa},
  \citenamefont {Khan}, \citenamefont {Hannam}, \citenamefont {P\"urrer},
  \citenamefont {Ohme}, \citenamefont {Jim\'enez~Forteza},\ and\ \citenamefont
  {Boh\'e}}]{Husa:2015iqa}%
  \BibitemOpen
  \bibfield  {author} {\bibinfo {author} {\bibfnamefont {S.}~\bibnamefont
  {Husa}}, \bibinfo {author} {\bibfnamefont {S.}~\bibnamefont {Khan}}, \bibinfo
  {author} {\bibfnamefont {M.}~\bibnamefont {Hannam}}, \bibinfo {author}
  {\bibfnamefont {M.}~\bibnamefont {P\"urrer}}, \bibinfo {author}
  {\bibfnamefont {F.}~\bibnamefont {Ohme}}, \bibinfo {author} {\bibfnamefont
  {X.}~\bibnamefont {Jim\'enez~Forteza}},\ and\ \bibinfo {author}
  {\bibfnamefont {A.}~\bibnamefont {Boh\'e}},\ }\bibfield  {title} {\bibinfo
  {title} {{Frequency-domain gravitational waves from nonprecessing black-hole
  binaries. I. New numerical waveforms and anatomy of the signal}},\ }\href
  {https://doi.org/10.1103/PhysRevD.93.044006} {\bibfield  {journal} {\bibinfo
  {journal} {Phys. Rev. D}\ }\textbf {\bibinfo {volume} {93}},\ \bibinfo
  {pages} {044006} (\bibinfo {year} {2016})},\ \Eprint
  {https://arxiv.org/abs/1508.07250} {arXiv:1508.07250 [gr-qc]} \BibitemShut
  {NoStop}%
\bibitem [{\citenamefont {Khan}\ \emph {et~al.}(2016)\citenamefont {Khan},
  \citenamefont {Husa}, \citenamefont {Hannam}, \citenamefont {Ohme},
  \citenamefont {P\"urrer}, \citenamefont {Jim\'enez~Forteza},\ and\
  \citenamefont {Boh\'e}}]{Khan:2015jqa}%
  \BibitemOpen
  \bibfield  {author} {\bibinfo {author} {\bibfnamefont {S.}~\bibnamefont
  {Khan}}, \bibinfo {author} {\bibfnamefont {S.}~\bibnamefont {Husa}}, \bibinfo
  {author} {\bibfnamefont {M.}~\bibnamefont {Hannam}}, \bibinfo {author}
  {\bibfnamefont {F.}~\bibnamefont {Ohme}}, \bibinfo {author} {\bibfnamefont
  {M.}~\bibnamefont {P\"urrer}}, \bibinfo {author} {\bibfnamefont
  {X.}~\bibnamefont {Jim\'enez~Forteza}},\ and\ \bibinfo {author}
  {\bibfnamefont {A.}~\bibnamefont {Boh\'e}},\ }\bibfield  {title} {\bibinfo
  {title} {{Frequency-domain gravitational waves from nonprecessing black-hole
  binaries. II. A phenomenological model for the advanced detector era}},\
  }\href {https://doi.org/10.1103/PhysRevD.93.044007} {\bibfield  {journal}
  {\bibinfo  {journal} {Phys. Rev. D}\ }\textbf {\bibinfo {volume} {93}},\
  \bibinfo {pages} {044007} (\bibinfo {year} {2016})},\ \Eprint
  {https://arxiv.org/abs/1508.07253} {arXiv:1508.07253 [gr-qc]} \BibitemShut
  {NoStop}%
\bibitem [{\citenamefont {Khan}\ \emph {et~al.}(2019)\citenamefont {Khan},
  \citenamefont {Chatziioannou}, \citenamefont {Hannam},\ and\ \citenamefont
  {Ohme}}]{Khan:2018fmp}%
  \BibitemOpen
  \bibfield  {author} {\bibinfo {author} {\bibfnamefont {S.}~\bibnamefont
  {Khan}}, \bibinfo {author} {\bibfnamefont {K.}~\bibnamefont {Chatziioannou}},
  \bibinfo {author} {\bibfnamefont {M.}~\bibnamefont {Hannam}},\ and\ \bibinfo
  {author} {\bibfnamefont {F.}~\bibnamefont {Ohme}},\ }\bibfield  {title}
  {\bibinfo {title} {{Phenomenological model for the gravitational-wave signal
  from precessing binary black holes with two-spin effects}},\ }\href
  {https://doi.org/10.1103/PhysRevD.100.024059} {\bibfield  {journal} {\bibinfo
   {journal} {Phys. Rev. D}\ }\textbf {\bibinfo {volume} {100}},\ \bibinfo
  {pages} {024059} (\bibinfo {year} {2019})},\ \Eprint
  {https://arxiv.org/abs/1809.10113} {arXiv:1809.10113 [gr-qc]} \BibitemShut
  {NoStop}%
\bibitem [{\citenamefont {Pratten}\ \emph {et~al.}(2021)\citenamefont {Pratten}
  \emph {et~al.}}]{Pratten:2020ceb}%
  \BibitemOpen
  \bibfield  {author} {\bibinfo {author} {\bibfnamefont {G.}~\bibnamefont
  {Pratten}} \emph {et~al.},\ }\bibfield  {title} {\bibinfo {title}
  {{Computationally efficient models for the dominant and subdominant harmonic
  modes of precessing binary black holes}},\ }\href
  {https://doi.org/10.1103/PhysRevD.103.104056} {\bibfield  {journal} {\bibinfo
   {journal} {Phys. Rev. D}\ }\textbf {\bibinfo {volume} {103}},\ \bibinfo
  {pages} {104056} (\bibinfo {year} {2021})},\ \Eprint
  {https://arxiv.org/abs/2004.06503} {arXiv:2004.06503 [gr-qc]} \BibitemShut
  {NoStop}%
\bibitem [{\citenamefont {Messina}\ and\ \citenamefont
  {Nagar}(2017)}]{Messina:2017yjg}%
  \BibitemOpen
  \bibfield  {author} {\bibinfo {author} {\bibfnamefont {F.}~\bibnamefont
  {Messina}}\ and\ \bibinfo {author} {\bibfnamefont {A.}~\bibnamefont
  {Nagar}},\ }\bibfield  {title} {\bibinfo {title} {{Parametrized-4.5PN
  TaylorF2 approximants and tail effects to quartic nonlinear order from the
  effective one body formalism}},\ }\href
  {https://doi.org/10.1103/PhysRevD.95.124001} {\bibfield  {journal} {\bibinfo
  {journal} {Phys. Rev. D}\ }\textbf {\bibinfo {volume} {95}},\ \bibinfo
  {pages} {124001} (\bibinfo {year} {2017})},\ \bibinfo {note} {[Erratum:
  Phys.Rev.D 96, 049907(E) (2017)]},\ \Eprint
  {https://arxiv.org/abs/1703.08107} {arXiv:1703.08107 [gr-qc]} \BibitemShut
  {NoStop}%
\bibitem [{\citenamefont {Dietrich}\ \emph {et~al.}(2017)\citenamefont
  {Dietrich}, \citenamefont {Bernuzzi},\ and\ \citenamefont
  {Tichy}}]{Dietrich:2017aum}%
  \BibitemOpen
  \bibfield  {author} {\bibinfo {author} {\bibfnamefont {T.}~\bibnamefont
  {Dietrich}}, \bibinfo {author} {\bibfnamefont {S.}~\bibnamefont {Bernuzzi}},\
  and\ \bibinfo {author} {\bibfnamefont {W.}~\bibnamefont {Tichy}},\ }\bibfield
   {title} {\bibinfo {title} {{Closed-form tidal approximants for binary
  neutron star gravitational waveforms constructed from high-resolution
  numerical relativity simulations}},\ }\href
  {https://doi.org/10.1103/PhysRevD.96.121501} {\bibfield  {journal} {\bibinfo
  {journal} {Phys. Rev. D}\ }\textbf {\bibinfo {volume} {96}},\ \bibinfo
  {pages} {121501(R)} (\bibinfo {year} {2017})},\ \Eprint
  {https://arxiv.org/abs/1706.02969} {arXiv:1706.02969 [gr-qc]} \BibitemShut
  {NoStop}%
\bibitem [{\citenamefont {Dietrich}\ \emph {et~al.}(2019)\citenamefont
  {Dietrich}, \citenamefont {Samajdar}, \citenamefont {Khan}, \citenamefont
  {Johnson-McDaniel}, \citenamefont {Dudi},\ and\ \citenamefont
  {Tichy}}]{Dietrich:2019kaq}%
  \BibitemOpen
  \bibfield  {author} {\bibinfo {author} {\bibfnamefont {T.}~\bibnamefont
  {Dietrich}}, \bibinfo {author} {\bibfnamefont {A.}~\bibnamefont {Samajdar}},
  \bibinfo {author} {\bibfnamefont {S.}~\bibnamefont {Khan}}, \bibinfo {author}
  {\bibfnamefont {N.~K.}\ \bibnamefont {Johnson-McDaniel}}, \bibinfo {author}
  {\bibfnamefont {R.}~\bibnamefont {Dudi}},\ and\ \bibinfo {author}
  {\bibfnamefont {W.}~\bibnamefont {Tichy}},\ }\bibfield  {title} {\bibinfo
  {title} {{Improving the NRTidal model for binary neutron star systems}},\
  }\href {https://doi.org/10.1103/PhysRevD.100.044003} {\bibfield  {journal}
  {\bibinfo  {journal} {Phys. Rev. D}\ }\textbf {\bibinfo {volume} {100}},\
  \bibinfo {pages} {044003} (\bibinfo {year} {2019})},\ \Eprint
  {https://arxiv.org/abs/1905.06011} {arXiv:1905.06011 [gr-qc]} \BibitemShut
  {NoStop}%
\bibitem [{\citenamefont {Nagar}\ \emph {et~al.}(2018)\citenamefont {Nagar}
  \emph {et~al.}}]{Nagar:2018zoe}%
  \BibitemOpen
  \bibfield  {author} {\bibinfo {author} {\bibfnamefont {A.}~\bibnamefont
  {Nagar}} \emph {et~al.},\ }\bibfield  {title} {\bibinfo {title} {{Time-domain
  effective-one-body gravitational waveforms for coalescing compact binaries
  with nonprecessing spins, tides and self-spin effects}},\ }\href
  {https://doi.org/10.1103/PhysRevD.98.104052} {\bibfield  {journal} {\bibinfo
  {journal} {Phys. Rev. D}\ }\textbf {\bibinfo {volume} {98}},\ \bibinfo
  {pages} {104052} (\bibinfo {year} {2018})},\ \Eprint
  {https://arxiv.org/abs/1806.01772} {arXiv:1806.01772 [gr-qc]} \BibitemShut
  {NoStop}%
\bibitem [{\citenamefont {Wade}\ \emph {et~al.}(2014)\citenamefont {Wade},
  \citenamefont {Creighton}, \citenamefont {Ochsner}, \citenamefont {Lackey},
  \citenamefont {Farr}, \citenamefont {Littenberg},\ and\ \citenamefont
  {Raymond}}]{Wade:2014vqa}%
  \BibitemOpen
  \bibfield  {author} {\bibinfo {author} {\bibfnamefont {L.}~\bibnamefont
  {Wade}}, \bibinfo {author} {\bibfnamefont {J.~D.~E.}\ \bibnamefont
  {Creighton}}, \bibinfo {author} {\bibfnamefont {E.}~\bibnamefont {Ochsner}},
  \bibinfo {author} {\bibfnamefont {B.~D.}\ \bibnamefont {Lackey}}, \bibinfo
  {author} {\bibfnamefont {B.~F.}\ \bibnamefont {Farr}}, \bibinfo {author}
  {\bibfnamefont {T.~B.}\ \bibnamefont {Littenberg}},\ and\ \bibinfo {author}
  {\bibfnamefont {V.}~\bibnamefont {Raymond}},\ }\bibfield  {title} {\bibinfo
  {title} {{Systematic and statistical errors in a Bayesian approach to the
  estimation of the neutron-star equation of state using advanced gravitational
  wave detectors}},\ }\href {https://doi.org/10.1103/PhysRevD.89.103012}
  {\bibfield  {journal} {\bibinfo  {journal} {Phys. Rev. D}\ }\textbf {\bibinfo
  {volume} {89}},\ \bibinfo {pages} {103012} (\bibinfo {year} {2014})},\
  \Eprint {https://arxiv.org/abs/1402.5156} {arXiv:1402.5156 [gr-qc]}
  \BibitemShut {NoStop}%
\bibitem [{\citenamefont {Campanelli}\ \emph
  {et~al.}(2006{\natexlab{b}})\citenamefont {Campanelli}, \citenamefont
  {Lousto},\ and\ \citenamefont {Zlochower}}]{Campanelli:2006uy}%
  \BibitemOpen
  \bibfield  {author} {\bibinfo {author} {\bibfnamefont {M.}~\bibnamefont
  {Campanelli}}, \bibinfo {author} {\bibfnamefont {C.~O.}\ \bibnamefont
  {Lousto}},\ and\ \bibinfo {author} {\bibfnamefont {Y.}~\bibnamefont
  {Zlochower}},\ }\bibfield  {title} {\bibinfo {title} {{Spinning-black-hole
  binaries: The orbital hang up}},\ }\href
  {https://doi.org/10.1103/PhysRevD.74.041501} {\bibfield  {journal} {\bibinfo
  {journal} {Phys. Rev. D}\ }\textbf {\bibinfo {volume} {74}},\ \bibinfo
  {pages} {041501(R)} (\bibinfo {year} {2006}{\natexlab{b}})},\ \bibinfo {note}
  {gr-qc/0604012}\BibitemShut {NoStop}%
\bibitem [{Note1()}]{Note1}%
  \BibitemOpen
  \bibinfo {note} {\protect \href
  {https://github.com/GRTLCollaboration}{https://github.com/GRTLCollaboration}}\BibitemShut
  {NoStop}%
\bibitem [{\citenamefont {Evstafyeva}\ and\ \citenamefont
  {Sperhake}()}]{Evstafyeva_Boson_star_waveforms}%
  \BibitemOpen
  \bibfield  {author} {\bibinfo {author} {\bibfnamefont {T.}~\bibnamefont
  {Evstafyeva}}\ and\ \bibinfo {author} {\bibfnamefont {U.}~\bibnamefont
  {Sperhake}},\ }\href {https://github.com/tamaraevst/Boson-star-waveforms}
  {\bibinfo {title} {{Boson star waveforms}}}\BibitemShut {NoStop}%
\bibitem [{\citenamefont {Evstafyeva}\ \emph
  {et~al.}(2024{\natexlab{a}})\citenamefont {Evstafyeva}, \citenamefont
  {Sperhake}, \citenamefont {Romero-Shaw},\ and\ \citenamefont
  {Agathos}}]{video1}%
  \BibitemOpen
  \bibfield  {author} {\bibinfo {author} {\bibfnamefont {T.}~\bibnamefont
  {Evstafyeva}}, \bibinfo {author} {\bibfnamefont {U.}~\bibnamefont
  {Sperhake}}, \bibinfo {author} {\bibfnamefont {I.}~\bibnamefont
  {Romero-Shaw}},\ and\ \bibinfo {author} {\bibfnamefont {M.}~\bibnamefont
  {Agathos}},\ }\href {https://www.youtube.com/watch?v=QrbJKKAOerY} {\bibinfo
  {title} {{Inspiral and merger of equal-mass boson star binary forming a black
  hole post-merger}}} (\bibinfo {year} {2024}{\natexlab{a}})\BibitemShut
  {NoStop}%
\bibitem [{\citenamefont {Evstafyeva}\ \emph
  {et~al.}(2024{\natexlab{b}})\citenamefont {Evstafyeva}, \citenamefont
  {Sperhake}, \citenamefont {Romero-Shaw},\ and\ \citenamefont
  {Agathos}}]{video2}%
  \BibitemOpen
  \bibfield  {author} {\bibinfo {author} {\bibfnamefont {T.}~\bibnamefont
  {Evstafyeva}}, \bibinfo {author} {\bibfnamefont {U.}~\bibnamefont
  {Sperhake}}, \bibinfo {author} {\bibfnamefont {I.}~\bibnamefont
  {Romero-Shaw}},\ and\ \bibinfo {author} {\bibfnamefont {M.}~\bibnamefont
  {Agathos}},\ }\href {https://www.youtube.com/watch?v=hj1WzNPpeS0} {\bibinfo
  {title} {{Inspiral and merger of equal-mass boson star binary forming a boson
  star post-merger}}} (\bibinfo {year} {2024}{\natexlab{b}})\BibitemShut
  {NoStop}%
\end{thebibliography}

\begin{thebibliography}{12}%
\makeatletter
\providecommand \@ifxundefined [1]{%
 \@ifx{#1\undefined}
}%
\providecommand \@ifnum [1]{%
 \ifnum #1\expandafter \@firstoftwo
 \else \expandafter \@secondoftwo
 \fi
}%
\providecommand \@ifx [1]{%
 \ifx #1\expandafter \@firstoftwo
 \else \expandafter \@secondoftwo
 \fi
}%
\providecommand \natexlab [1]{#1}%
\providecommand \enquote  [1]{``#1''}%
\providecommand \bibnamefont  [1]{#1}%
\providecommand \bibfnamefont [1]{#1}%
\providecommand \citenamefont [1]{#1}%
\providecommand \href@noop [0]{\@secondoftwo}%
\providecommand \href [0]{\begingroup \@sanitize@url \@href}%
\providecommand \@href[1]{\@@startlink{#1}\@@href}%
\providecommand \@@href[1]{\endgroup#1\@@endlink}%
\providecommand \@sanitize@url [0]{\catcode `\\12\catcode `\$12\catcode
  `\&12\catcode `\#12\catcode `\^12\catcode `\_12\catcode `\%12\relax}%
\providecommand \@@startlink[1]{}%
\providecommand \@@endlink[0]{}%
\providecommand \url  [0]{\begingroup\@sanitize@url \@url }%
\providecommand \@url [1]{\endgroup\@href {#1}{\urlprefix }}%
\providecommand \urlprefix  [0]{URL }%
\providecommand \Eprint [0]{\href }%
\providecommand \doibase [0]{https://doi.org/}%
\providecommand \selectlanguage [0]{\@gobble}%
\providecommand \bibinfo  [0]{\@secondoftwo}%
\providecommand \bibfield  [0]{\@secondoftwo}%
\providecommand \translation [1]{[#1]}%
\providecommand \BibitemOpen [0]{}%
\providecommand \bibitemStop [0]{}%
\providecommand \bibitemNoStop [0]{.\EOS\space}%
\providecommand \EOS [0]{\spacefactor3000\relax}%
\providecommand \BibitemShut  [1]{\csname bibitem#1\endcsname}%
\let\auto@bib@innerbib\@empty
\bibitem [{\citenamefont {{LIGO Scientific Collaboration}}\ \emph
  {et~al.}(2018)\citenamefont {{LIGO Scientific Collaboration}}, \citenamefont
  {{Virgo Collaboration}},\ and\ \citenamefont {{KAGRA
  Collaboration}}}]{lalsuite}%
  \BibitemOpen
  \bibfield  {author} {\bibinfo {author} {\bibnamefont {{LIGO Scientific}}},
  \bibinfo {author} {\bibnamefont {{Virgo}}},\
  and\ \bibinfo {author} {\bibnamefont {{KAGRA Collaborations}}},\ }\href
  {https://doi.org/10.7935/GT1W-FZ16} {\bibinfo {title} {{LVK} {A}lgorithm
  {L}ibrary - {LALS}uite}},\ \bibinfo {howpublished} {Free software (GPL)}
  (\bibinfo {year} {2018})\BibitemShut {NoStop}%
\bibitem [{\citenamefont {Wette}(2020)}]{swiglal}%
  \BibitemOpen
  \bibfield  {author} {\bibinfo {author} {\bibfnamefont {K.}~\bibnamefont
  {Wette}},\ }\bibfield  {title} {\bibinfo {title} {{SWIGLAL: Python and Octave
  interfaces to the LALSuite gravitational-wave data analysis libraries}},\
  }\href {https://doi.org/10.1016/j.softx.2020.100634} {\bibfield  {journal}
  {\bibinfo  {journal} {SoftwareX}\ }\textbf {\bibinfo {volume} {12}},\
  \bibinfo {pages} {100634} (\bibinfo {year} {2020})}\BibitemShut {NoStop}%
\bibitem [{\citenamefont {Schmidt}\ \emph {et~al.}(2017)\citenamefont
  {Schmidt}, \citenamefont {Harry},\ and\ \citenamefont
  {Pfeiffer}}]{Schmidt2017numerical}%
  \BibitemOpen
  \bibfield  {author} {\bibinfo {author} {\bibfnamefont {P.}~\bibnamefont
  {Schmidt}}, \bibinfo {author} {\bibfnamefont {I.~W.}\ \bibnamefont {Harry}},\
  and\ \bibinfo {author} {\bibfnamefont {H.~P.}\ \bibnamefont {Pfeiffer}},\
  }\href@noop {} {\bibinfo {title} {Numerical relativity injection
  infrastructure}} (\bibinfo {year} {2017}),\ \Eprint
  {https://arxiv.org/abs/1703.01076} {arXiv:1703.01076 [gr-qc]} \BibitemShut
  {NoStop}%
\bibitem [{\citenamefont {Abbott}\ \emph {et~al.}(2018)\citenamefont {Abbott}
  \emph {et~al.}}]{AdvancedDetectorsLVK2018S}%
  \BibitemOpen
  \bibfield  {author} {\bibinfo {author} {\bibfnamefont {B.~P.}\ \bibnamefont
  {Abbott}} \emph {et~al.} (\bibinfo {collaboration} {KAGRA, LIGO Scientific,
  VIRGO}),\ }\bibfield  {title} {\bibinfo {title} {{Prospects for Observing and
  Localizing Gravitational-Wave Transients with Advanced LIGO, Advanced Virgo
  and KAGRA}},\ }\href {https://doi.org/10.1007/s41114-018-0012-9,
  10.1007/lrr-2016-1} {\bibfield  {journal} {\bibinfo  {journal} {Living Rev.
  Rel.}\ }\textbf {\bibinfo {volume} {21}},\ \bibinfo {pages} {3} (\bibinfo
  {year} {2018})},\ \Eprint {https://arxiv.org/abs/1304.0670} {arXiv:1304.0670
  [gr-qc]} \BibitemShut {NoStop}%
\bibitem [{\citenamefont {Speagle}(2020)}]{Speagle_2020}%
  \BibitemOpen
  \bibfield  {author} {\bibinfo {author} {\bibfnamefont {J.~S.}\ \bibnamefont
  {Speagle}},\ }\bibfield  {title} {\bibinfo {title} {dynesty: a dynamic nested
  sampling package for estimating {B}ayesian posteriors and evidences},\ }\href
  {https://doi.org/10.1093/mnras/staa278} {\bibfield  {journal} {\bibinfo
  {journal} {Mon. Not. R. Astron. Soc.}\ }\textbf
  {\bibinfo {volume} {493}},\ \bibinfo {pages} {3132–3158} (\bibinfo {year}
  {2020})}\BibitemShut {NoStop}%
\bibitem [{\citenamefont {{Del Pozzo}}\ and\ \citenamefont
  {{Veitch}}(2022)}]{2022ascl.soft05021D}%
  \BibitemOpen
  \bibfield  {author} {\bibinfo {author} {\bibfnamefont {W.}~\bibnamefont {{Del
  Pozzo}}}\ and\ \bibinfo {author} {\bibfnamefont {J.}~\bibnamefont
  {{Veitch}}},\ }\href {https://ui.adsabs.harvard.edu/abs/2022ascl.soft05021D}
  {\bibinfo {title} {{CPNest: Parallel nested sampling}}},\ \bibinfo
  {howpublished} {Astrophysics Source Code Library, record ascl:2205.021}
  (\bibinfo {year} {2022})\BibitemShut {NoStop}%
\bibitem [{\citenamefont {Williams}(2021)}]{nessai}%
  \BibitemOpen
  \bibfield  {author} {\bibinfo {author} {\bibfnamefont {M.~J.}\ \bibnamefont
  {Williams}},\ }\href {https://doi.org/10.5281/zenodo.4550693} {\bibinfo
  {title} {\\nessai: Nested sampling with artificial intelligence}} (\bibinfo
  {year} {2021}),\ \bibinfo {note}
  {https://doi.org/10.5281/zenodo.4550693}\BibitemShut {NoStop}%
\bibitem [{\citenamefont {Williams}\ \emph {et~al.}(2021)\citenamefont
  {Williams}, \citenamefont {Veitch},\ and\ \citenamefont
  {Messenger}}]{Williams:2021qyt}%
  \BibitemOpen
  \bibfield  {author} {\bibinfo {author} {\bibfnamefont {M.~J.}\ \bibnamefont
  {Williams}}, \bibinfo {author} {\bibfnamefont {J.}~\bibnamefont {Veitch}},\
  and\ \bibinfo {author} {\bibfnamefont {C.}~\bibnamefont {Messenger}},\
  }\bibfield  {title} {\bibinfo {title} {{Nested sampling with normalizing
  flows for gravitational-wave inference}},\ }\href
  {https://doi.org/10.1103/PhysRevD.103.103006} {\bibfield  {journal} {\bibinfo
   {journal} {Phys. Rev. D}\ }\textbf {\bibinfo {volume} {103}},\ \bibinfo
  {pages} {103006} (\bibinfo {year} {2021})},\ \Eprint
  {https://arxiv.org/abs/2102.11056} {arXiv:2102.11056 [gr-qc]} \BibitemShut
  {NoStop}%
\bibitem [{\citenamefont {Williams}\ \emph {et~al.}(2023)\citenamefont
  {Williams}, \citenamefont {Veitch},\ and\ \citenamefont
  {Messenger}}]{Williams:2023ppp}%
  \BibitemOpen
  \bibfield  {author} {\bibinfo {author} {\bibfnamefont {M.~J.}\ \bibnamefont
  {Williams}}, \bibinfo {author} {\bibfnamefont {J.}~\bibnamefont {Veitch}},\
  and\ \bibinfo {author} {\bibfnamefont {C.}~\bibnamefont {Messenger}},\
  }\bibfield  {title} {\bibinfo {title} {{Importance nested sampling with
  normalising flows}},\ }\href {https://doi.org/10.1088/2632-2153/acd5aa}
  {\bibfield  {journal} {\bibinfo  {journal} {Mach. Learn. Sci. Tech.}\
  }\textbf {\bibinfo {volume} {4}},\ \bibinfo {pages} {035011} (\bibinfo {year}
  {2023})},\ \Eprint {https://arxiv.org/abs/2302.08526} {arXiv:2302.08526
  [astro-ph.IM]} \BibitemShut {NoStop}%
\bibitem [{\citenamefont {Ashton}\ \emph {et~al.}(2019)\citenamefont {Ashton}
  \emph {et~al.}}]{Ashton:S2018jfp}%
  \BibitemOpen
  \bibfield  {author} {\bibinfo {author} {\bibfnamefont {G.}~\bibnamefont
  {Ashton}} \emph {et~al.},\ }\bibfield  {title} {\bibinfo {title} {{BILBY: A
  user-friendly Bayesian inference library for gravitational-wave astronomy}},\
  }\href {https://doi.org/10.3847/1538-4365/ab06fc} {\bibfield  {journal}
  {\bibinfo  {journal} {Astrophys. J. Suppl.}\ }\textbf {\bibinfo {volume}
  {241}},\ \bibinfo {pages} {27} (\bibinfo {year} {2019})},\ \Eprint
  {https://arxiv.org/abs/1811.02042} {arXiv:1811.02042 [astro-ph.IM]}
  \BibitemShut {NoStop}%
\bibitem [{\citenamefont {Palenzuela}\ \emph {et~al.}(2007)\citenamefont
  {Palenzuela}, \citenamefont {Olabarrieta}, \citenamefont {Lehner},\ and\
  \citenamefont {Liebling}}]{Palenzuela:S2006wp}%
  \BibitemOpen
  \bibfield  {author} {\bibinfo {author} {\bibfnamefont {C.}~\bibnamefont
  {Palenzuela}}, \bibinfo {author} {\bibfnamefont {I.}~\bibnamefont
  {Olabarrieta}}, \bibinfo {author} {\bibfnamefont {L.}~\bibnamefont
  {Lehner}},\ and\ \bibinfo {author} {\bibfnamefont {S.~L.}\ \bibnamefont
  {Liebling}},\ }\bibfield  {title} {\bibinfo {title} {{Head-on collisions of
  boson stars}},\ }\href {https://doi.org/10.1103/PhysRevD.75.064005}
  {\bibfield  {journal} {\bibinfo  {journal} {Phys. Rev. D}\ }\textbf {\bibinfo
  {volume} {75}},\ \bibinfo {pages} {064005} (\bibinfo {year} {2007})},\
  \Eprint {https://arxiv.org/abs/gr-qc/0612067} {arXiv:gr-qc/0612067}
  \BibitemShut {NoStop}%
\bibitem [{\citenamefont {Radia}\ \emph {et~al.}(2022)\citenamefont {Radia},
  \citenamefont {Sperhake}, \citenamefont {Drew}, \citenamefont {Clough},
  \citenamefont {Figueras}, \citenamefont {Lim}, \citenamefont {Ripley},
  \citenamefont {Aurrekoetxea}, \citenamefont {Fran\c{c}a},\ and\ \citenamefont
  {Helfer}}]{Radia:S2021smk}%
  \BibitemOpen
  \bibfield  {author} {\bibinfo {author} {\bibfnamefont {M.}~\bibnamefont
  {Radia}}, \bibinfo {author} {\bibfnamefont {U.}~\bibnamefont {Sperhake}},
  \bibinfo {author} {\bibfnamefont {A.}~\bibnamefont {Drew}}, \bibinfo {author}
  {\bibfnamefont {K.}~\bibnamefont {Clough}}, \bibinfo {author} {\bibfnamefont
  {P.}~\bibnamefont {Figueras}}, \bibinfo {author} {\bibfnamefont {E.~A.}\
  \bibnamefont {Lim}}, \bibinfo {author} {\bibfnamefont {J.~L.}\ \bibnamefont
  {Ripley}}, \bibinfo {author} {\bibfnamefont {J.~C.}\ \bibnamefont
  {Aurrekoetxea}}, \bibinfo {author} {\bibfnamefont {T.}~\bibnamefont
  {Fran\c{c}a}},\ and\ \bibinfo {author} {\bibfnamefont {T.}~\bibnamefont
  {Helfer}},\ }\bibfield  {title} {\bibinfo {title} {{Lessons for adaptive mesh
  refinement in numerical relativity}},\ }\href
  {https://doi.org/10.1088/1361-6382/ac6fa9} {\bibfield  {journal} {\bibinfo
  {journal} {Class. Quant. Grav.}\ }\textbf {\bibinfo {volume} {39}},\ \bibinfo
  {pages} {135006} (\bibinfo {year} {2022})},\ \Eprint
  {https://arxiv.org/abs/2112.10567} {arXiv:2112.10567 [gr-qc]} \BibitemShut
  {NoStop}%
\bibitem [{Note2()}]{Note2}%
  \BibitemOpen
  \bibinfo {note} {For nonprecessing models, we report projections of the
  dimensionless spins onto the orbital angular momentum, $\chi_1$, $\chi_2$,
  which are allowed to be negative.}\BibitemShut {Stop}%
\end{thebibliography}
%

\end{document}